\documentclass[useAMS,usenatbib,10pt]{mnras}
\usepackage{listings}
\usepackage{times}  
\usepackage{graphics}
\usepackage{subfigure}
\usepackage{hyperref}
\usepackage{amssymb}
\usepackage{aas_macros}
\usepackage{lscape}
\usepackage{rotating}
\usepackage[export]{adjustbox}
\usepackage{graphicx}
\def\gsim{\ga}
\def\lsim{\la}

\def\crate{\dot M_{\rm cool}}

\def\kB{k_{\rm B}}

\def\surfs{{\sc surfs}}
\def\shark{{\sc Shark}}
\def\gsim{ \lower .75ex \hbox{$\sim$} \llap{\raise .27ex \hbox{$>$}} }
\def\lsim{ \lower .75ex \hbox{$\sim$} \llap{\raise .27ex \hbox{$<$}} }
\def\simprop{ \lower .75ex \hbox{$\sim$} \llap{\raise .27ex \hbox{$\propto$}} }

\title[Shark]{
Shark: introducing an open source, free and flexible semi-analytic model of galaxy formation}
\author[C. Lagos et al.]{
\parbox[t]{\textwidth}{
\vspace{-1.0cm}
Claudia del P. Lagos$^{1,2}$\thanks{E-mail: claudia.lagos@icrar.org}, Rodrigo J. Tobar$^{1}$, Aaron S. G. Robotham$^{1,2}$, Danail Obreschkow$^{1,2}$, Peter D. Mitchell$^{3}$, Chris Power$^{1,2}$, Pascal J. Elahi$^{1,2}$}
\vspace*{6pt} \\
$^{1}$International Centre for Radio Astronomy Research (ICRAR), M468, University of Western Australia, 35 Stirling Hwy, Crawley, WA 6009, Australia.\\
$^{2}$ARC Centre of Excellence for All Sky Astrophysics in 3 Dimensions (ASTRO 3D).\\
$^{3}$Univ Lyon, Univ Lyon1, Ens de Lyon, CNRS, Centre de Recherche Astrophysique de Lyon UMR5574, F-69230 Saint-Genis-Laval, France.
\vspace*{-0.5cm}}

\begin{document}


\pagerange{\pageref{firstpage}--\pageref{lastpage}} \pubyear{2018}

\maketitle

\label{firstpage}

\begin{abstract}
We present a new, open source, free semi-analytic model (SAM) of galaxy formation, \shark, designed to be highly flexible and modular, allowing easy exploration of different physical processes and ways of modelling them. We introduce the philosophy behind \shark and provide an overview of the physical processes included in the model. \shark\ is written in C++11 and has been parallelized with OpenMP.  In the released version ({\sc v1.1}), we implement several different models for gas cooling, active galactic nuclei, stellar and photo-ionisation feedback, and star formation (SF). We demonstrate the basic performance of \shark\ using the \citet{Planck15} cosmology SURFS simulations, by comparing against a large set of observations, including: the stellar mass function (SMF) and stellar-halo mass relation at $z=0-4$; the cosmic evolution of the star formation rate density (SFRD), stellar mass, atomic and molecular hydrogen; local gas scaling relations; and structural galaxy properties, finding excellent agreement. Significant improvements over previous SAMs are seen in the mass-size relation for disks/bulges, the gas-stellar mass and stellar mass-metallicity relations. To illustrate the power of \shark\ in exploring the systematic effects of the galaxy formation modelling,  we quantify how the scatter of the SF main sequence and the gas scaling relations changes with the adopted SF law, and the effect of the starbursts H$_2$ depletion timescale on the SFRD and $\Omega_{\rm H_2}$. We compare \shark\ with other SAMs and the hydrodynamical simulation EAGLE, and find that SAMs have a much higher halo baryon fractions due to large amounts of intra-halo gas, which in the case of EAGLE is in the intergalactic medium.
\end{abstract}


\section{Introduction}

Galaxy formation and cosmology are fundamentally intertwined. The growth of the large scale structure in the Universe is dominated by dark matter (DM), as 
the latter is the main contributor to the matter budget. Thus, the growth rate of density peaks is mostly set by the 
abundance and initial clustering of DM post-inflation, and 
 the rate at which baryons flow towards the density peaks is also expected to follow closely that of the DM \citep{White78}. 
This shows that any thorough study of galaxy formation and evolution must include realistic cosmological environments and effects (see \citealt{Somerville14} for a
recent review).

Two widely used techniques to study galaxy formation in a cosmological context are hydrodynamical simulations
and semi-analytical models. 
Briefly, {hydro-dynamical simulations} solve the equations of gravity and fluid dynamics
simultaneously, allowing a detailed view of how the gas
and DM influence the evolution of each other and the complex gas structures
that typically form in highly active regions of galaxy formation (i.e. in halos).
The drawback of this technique is that it is computationally expensive, preventing us from
producing large cosmological boxes at resolutions that are interesting for galaxy formation
 and that have been 
well calibrated to key local observational data (i.e. typically the stellar mass of galaxies and their star formation rates, SFRs).
Currently achievable volumes where calibration is possible are of $\approx (100\,\rm Mpc)^3$ (\citealt{Schaye14}; \citealt{Crain15}; 
\citealt{Pillepich18}).

{Semi-analytic models} (SAMs) describe the physical processes giving rise to the
formation and evolution of galaxies in a simpler way, and are run over DM-only
$N$-body simulations. SAMs run are computationally inexpensive, and thus it is possible to explore
the parameter space thoroughly through statistical techniques (e.g. Markov Chain Monte Carlo
or genetic algorithms; {e.g. \citealt{Henriques09, Lu11, Ruiz15}}). The drawback of this technique is that
galaxies are described in much simpler terms than in hydro-dynamical simulations, lacking
information of the detailed internal structure, {particularly non-axisymmetric features}. 
{In general SAMs choose as coarse-graining scale the galaxy scale. However, 
significant effort has gone into improving the internal structures of galaxies in SAMs by describing them 
as concentric rings (e.g. \citealt{Stringer07,Fu10,Stevens16}), and in some cases even modelling the distribution 
of molecular clouds \citep{Lagos13}.}
The primary advantage of SAMs
is the possibility of simulating much larger cosmological boxes (up to box lengths of $1~$Gpc), which allow
us to have much better statistics and diversity of environments, at the same time as accurately calibrating them 
to a set of observations of the galaxy population (see recent example from \citealt{Benson14,Popping14,Somerville15,Henriques15,Croton16,Lacey15,Xie17,Cora18}).

The major challenge for both techniques is the same; namely, that the least understood physics 
takes place in scales below the resolution of the simulations {in the case of hydrodynamical simulations, and that it cannot 
be modelled in an ab-initio way in the case of SAMs.} This physics includes: star formation, stellar feedback, black hole growth 
and active galactic nuclei (AGN) feedback, which happen on sub-pc scales, while 
the highest resolution available for cosmological hydro-dynamical simulations is a few $100$~pc. 
Given their running speed, SAMs are ideally placed to thoroughly explore different physical phenomena but also different ways of describing any one 
physical process. This has been a well exploited strategy in SAMs (see for example 
the star formation law and interstellar medium modelling in \citealt{Lagos10}, the gas reincorporation timescale 
in \citealt{Mitchell14}, the stellar population synthesis modelling in \citealt{Gonzalez-Perez13} and the 
stellar feedback in \citealt{Hirschmann16}, just to mention a few). 

Though simulations of galaxy formation
have converged to {produce} approximately the correct evolution of the stellar mass growth of
galaxies (see Fig.~$16$ in \citealt{Driver17}), 
the detailed description of the physical processes listed above 
 is very uncertain. As a result, simulations can predict the same 
stellar mass growth with various different baryon models.
A consequence of this is that the predicted gas content of galaxies and halos
vary widely among models. \citet{Mitchell18} showed that
two different cosmological simulations of galaxy formation, using the two 
techniques above, {\sc EAGLE} \citep{Schaye14,Crain15,McAlpine15} and GALFORM \citep{Lacey15},
predicted practically the same stellar mass growth but for {\it very different reasons.}
These models include in principle the same physics: gas cooling, star formation,
stellar and black hole feedback. However, because these processes happen on scales
we are unable to directly simulate (sub parsec), we cannot model them in an ab-initio way. 
We, therefore, need to decide how to best model these processes and what approximations to make. The result is that
simulations can predict vastly different baryon components in both abundance (mass, metals) and structure (internal kinematics,
density and temperature), driving the need to gain a in-depth understanding of how the modelling of those phenomena
affect, in detail, the baryon components of galaxies.

In the coming years major facilities
will come online, that in combination will allow us to measure the properties of the interstellar medium (ISM)
 of galaxies. The dense and diffuse gas will be observed through molecular and atomic emission from
the Atacama Large Millimetre Array (ALMA; \citealt{Wootten09}), the Australian Square Kilometre Array Pathfinder (ASKAP; \citealt{Johnston08}),
the Karoo Array Telescope (\citealt{Booth09}), the next generation Very Large Array \citep{Bolatto17}, 
and in the future the Square Kilometre Array (SKA; \citealt{Schilizzi08}). On the other hand, the new James Webb Space Telescope (JWST; \citealt{Gardner06}; see \citealt{Kalirai18} for a recent discussion of the science objectives of JWST) will reveal the properties of the warm ionised ISM in galaxies as well
as the gas around them (through absorption metal lines and Lyman alpha in emission).
These telescopes will measure masses, metal abundances, as well as the dynamics of the gas.
The information above will be available from the epoch of formation of the first galaxies
to today.
However, in order to use these observations to learn about the physics of galaxy
formation we need to have robust predictions of the expected features different physical processes
and models would imprint on the galaxy properties being observed.
Such predictions require connecting physical processes from {sub-galactic} 
to the large scale structure (Gpcs). SAMs are ideally placed to play this role in the future decade(s), 
{as the constraints from a wide range of different observations can be more readily connected in a coherent physical framework
 in semi-analytic models. That said, cosmological hydrodynamical simulations are expected to evolve towards 
 better exploration of the parameter and physical space in the next years.}

\begin{table*}
\setlength\tabcolsep{2pt}
\centering\footnotesize
\caption{Compilation of several SAM characteristics. 
Information was obtained through published papers (`p'), private communication (`p.c.') or 
knowledge from the authors of this manuscript acquired by having worked with the codes (`e'); `-' indicates that this does not apply. 
The references correspond to (1) \citet{Lacey15}, (2) \citet{Henriques15}, (3) \citet{Somerville15}, (4) \citet{Croton16}, (5) \citet{Benson12}, (6) \citet{Hirschmann16} and (7) \citet{Cora18}. 
Note that L-galaxies, SAGE, GAEA and SAG were built upon the early version of L-galaxies of \citet{Springel01} (refer to as S01), which in itself it was heavily influenced 
by \citet{Kauffmann99}. These codes have gone through 
independent developments in the last decade or so.}
\begin{tabular}{@{\extracolsep{\fill}}l|ccccccccc|p{0.45\textwidth}}
\hline
\hline
    Model & \shark & GALFORM & L-galaxies & Santa-Cruz & SAGE & Galacticus & GAEA & SAG\\
\hline
    Recent Reference & This paper & (1) & (2) & (3) & (4) & (5)& (6) & (7) \\
    Language & C++ & Fortran90 (e)& C & C++ (p.c.) & C & Fortran2003 & C (p.c.) & C (e)\\
    Based on other code & no & no & S01 &  no & S01 & no & S01 & S01\\
    License & GPLv3 & - & GPLv3 & - & MIT & GPLv3 & - & - \\
    source code available  & yes & no & one static version & no & yes & yes & no & no\\
                &      &    & released &                         &   &     &   &    \\
    version controller & git & git (e) & git & git (p.c.) & git & Mercurial & git (p.c.) & Mercurial (p.c.)\\
    available from & github & - & github & - & github & bitbucket & - & -\\
\hline
\end{tabular}
\label{tab:compmodels}
\end{table*}

\subsection{Why a new model? Mission and philosophy of Shark}

Semi-analytic models are very flexible and fast tools to explore galaxy formation physics, and as such, they have been widely used 
by both the theory and observational communities. In our opinion, desirable features of state-of-the-art SAMs include 
open source, portable and flexible code that can easily allow for a range of models and control over numerical convergence. 
In Table~\ref{tab:compmodels} we present a compilation of key information of well known semi-analytic models, including 
what language they are written in, license adopted, whether they are freely available and version controlled, etc. 
Existing SAMs fulfil some but not all of the desirable features above.

Current SAMs typically implement one set of physics models (i.e. one model for gas cooling, for 
angular momentum growth, star formation, etc.). This makes it challenging to explore how the modelling 
affects the properties of galaxies and, hence, to draw inferences on the detailed physics from observations. 
Exploring a wide range of models within the {\it same} computational framework (merger trees, time-steps, numerical integration schemes, etc.) 
would allow a more robust comparison of how different, non-linear physical processes interplay and determine the evolving 
galaxy population. For this to become feasible, the SAM would need to be flexible enough to allow models of different 
complexity (rather than just different parameters) and it should be modular enough for the user to have minimal 
interactions with the code.

Another hurdle we have faced in the SAM community is that code is rarely made publicly available. 
To our knowledge, the publicly available models are Galacticus \citep{Benson12}, SAGE \citep{Croton16}, a branch of SAGE called 
DARK SAGE \citep{Stevens16}, and a static version of 
the L-galaxies code \citet{Henriques15}. 
The latter does not allow 
 users access the latest improvements and to make contributions. 
Galacticus and SAGE, on the other hand, are aimed at solving these issues by being constantly updated and released in versions. 
Galacticus includes a very large range of physical models and implementations of any one physical process, and uses numerical 
solvers with adaptive and flexible stepsizes for the suite of differential equations describing the physics of the SAM, and as such it fulfils many of desirable criteria. 
However, Galacticus uses a complex, custom-made build system, making it not straightforward to compile the software or add support for different platforms.
SAGE is written in {\tt C}, it is easy to compile, making it 
 portable. 
However, SAGE's range of physical models and implementations are limited, and its numerical solver assumes fixed stepsizes 
for the suite of differential equations, 
which makes control of numerical precision, and consequently numerical convergence, challenging\footnote{SAGE's stepsize is fixed by the time span between snapshots and assumes 
the solutions to the equations to be linear with time.}.

In this paper we present a new SAM, named \shark. \shark\ has been designed in collaboration with computer scientists to be flexible and to allow for easy 
extension and modification of physical models of any complexity, solving their differential equations with adaptive stepsizes.
The code is aimed at being a {\it community} code, in which users can contribute to its development, distributing the work and the benefits this brings to a wider community. \shark\ is written in {\tt C++11}, using 
the open source {\tt GSL} libraries and a flexible compilation system using {\tt cmake}. The community aspect is a very important 
feature as it is, in our opinion, a key factor that can bring closer the observational and theory astrophysical communities, hopefully placing galaxy formation simulations 
in the backbone behind the planning and building of coming observational surveys and instruments. This is only achieved by easy and wide access to 
resources. Table~\ref{tab:compmodels} shows \shark's features in comparison to some well known semi-analytic models.
 
In this paper we present the design of \shark, the basic set of physical processes and models included in the first release of the code 
{\sc v1.1}, and its basic performance. The paper is organized as follows. $\S$~\ref{SURFSSec} presents the suite of $N$-body DM only simulations
which provide the basis for \shark. Note, however, that \shark\ is not limited to this suite of simulations. 
$\S$~\ref{sharkdesign}  presents the design of the code with comments on scalability and High Performance Computing (HPC) environments. $\S$~\ref{ModelDescription} 
describes the design of \shark\ and the suite of physical processes and model already implemented in {\sc v1.1}. $\S$~\ref{basicresults} 
presents a wide range of results of \shark, including those of the default, best-fitting model, and variations arising from using 
different parameters and models. Finally, we present our conclusions and future prospects in $\S$~\ref{conclusions}. 

\section{The SURFS simulation suite}\label{SURFSSec}

The \surfs\ suite consists of N-body simulations, most with cubic volumes of $210\,\rm cMpc/h$ on a side, and span a range in particle number, 
currently up to $8.5$ billion particles using a $\Lambda$CDM \citet{Planck15} cosmology\footnote{We adopt the \citet{Planck15} parameters which combine temperature, polarisation and lensing, without external data,  
given in the 2nd last column of Table~$4$ in \citet{Planck15}}. The simulation parameters are 
listed in Table~\ref{tab:sims}. Our simulations are split into moderate volume, high resolution simulations focused on galaxy formation 
for upcoming surveys like the Wide Area Vista Extragalactic Survey ({\sc waves}; \citealt{Driver16}) and {\sc wallaby}, the Australian Square Kilometre Array Pathfinder 
 HI All-Sky Survey \citep{Duffy12c}, and larger volume simulations designed for surveys focused on cosmological 
parameters like the Taipan survey \citep{daCunha17}. 
All simulations were run with a memory lean version of the {\sc gadget2} code on the Magnus supercomputer at the Pawsey Supercomputing Centre.

These simulations provide an excellent test-bed for numerical convergence, studies into the growth of halos and the evolution of subhalos 
down to DM halo masses of $\sim10^{9}\rm \, M_{\odot}$ (and galaxy stellar masses down to $\sim 10^{7}\rm \, M_{\odot}$). 
We produce $200$ snapshots and associated halo catalogues in evenly spaced logarithmic intervals in the growth factor starting at $z=24$ 
for our L210 and L40 simulations. This high cadence, with the time between snapshots being $\approx 6-80$~Myr, higher than was used in the Millennium simulations \citep{Springel05}, 
is necessary for halo merger trees that accurately capture the evolution of DM halos as each 
snapshot is separated by less than the freefall time of overdensities of $200\rho_{\rm crit}$, i.e., halos.
A full description of the simulation suite is presented in \citet{Elahi18}. 
Halo catalogs and merger trees for SURFS, described below, are available upon request\footnote{By emailing 
{\tt icrar-surfs@icrar.org}}. Throughout this paper we use the L210N1536 simulation (see Table~\ref{tab:sims} for details). 
In Appendix~\ref{ConvTests} we present \shark\ results 
based on the other SURFS run and analyse convergence in a subset of galaxy properties.
 
\subsection{Halo Catalogues}\label{sec:analysis}

We identify halos and subhalos, and calculate their properties using {\sc VELOCIraptor} (\citealt{Elahi11}, 
Elahi et al., in prep\footnote{\href{https://github.com/pelahi/VELOCIraptor-STF}{\url{https://github.com/pelahi/VELOCIraptor-STF}}}, 
\citealt{Canas18}). 
In {\sc VELOCIraptor}, subhalos correspond to all the bounded substructure 
in the 3D FOF, and thus, include the central subhalo. The halo corresponds to the 3D FOF structure.
This code first identifies halos 
using a 3D friends-of-friends (FOF) algorithm, also applying a 6D~FOF to each candidate FOF halo using the velocity dispersion of the candidate object to clean the 
halo catalogue of objects spuriously linked by artificial particle bridges, this is useful for disentangling early stage mergers.
The code then identifies substructures using a phase-space FOF algorithm on 
particles that appear to be dynamically distinct from the mean halo background, i.e. particles which have a local velocity distribution 
that differs significantly from the mean, i.e. smooth background halo. Since this approach is capable of not only finding subhalos, 
but also tidal streams surrounding subhalos as well as tidal streams from completely disrupted subhalos \citep{Elahi13}, 
for this analysis, we also ensure that a group is roughly self-bound, {allowing particles to have a ratio between the absolute value of the potential energy to 
kinetic energy, $\rm |U|/K$, of at least $0.95$}\footnote{{A common practice in configuration space finders, such as SUBFIND \citep{Springel01}, is to use 
$\rm |U|/K>1$. However, that choice is driven by the poor initial membership assignment of particles (i.e. numerous, unbound background particles are collected as part of halos). 
Thus, restricting the ratio of $\rm |U|/K$ to $>1$ 
avoids significant contamination. In {\sc VELOCIraptor} the background contamination is not so important, and thus one can keep particles 
with $\rm |U|/K<1$. The value of $0.95$ was chosen based on tests of subhalos orbiting halos in idealised simulations: 
DM particles in subhalos with $\rm |U|/K>0.95$ typically took more than an orbital time-scales to get stripped away.}}
{In SURFS, we consider all halos composed of $\ge 20$ dark matter particles.}

These halos/subhalos and trees are the backbone of our model. Specifically, the properties we 
use are their assembly histories, masses, and angular momentum.
The subhalo masses used by \shark\ correspond to the exclusive total mass in the 6D~FOF (i.e. including only particles that 
are uniquely tagged to that 6D~FOF structure). The halo mass is calculated as the sum of all its subhalos.

\subsection{Merger Trees}

The next step is the construction of a halo merger tree. We
use the halo merger tree code 
{\sc TreeFrog}\footnote{\href{https://github.com/pelahi/TreeFrog}{\url{https://github.com/pelahi/TreeFrog}}}, developed to work on {\sc VELOCIraptor} \citep{Elahi18}. At the simplest level, this code is a particle 
correlator and relies on particle IDs being continuous across
time (or halo catalogues). {\sc TreeFrog} makes the connections at the level of subhalos, and does this
by calculating a merit based on the fraction of particles shared by two subhalos $i$ and $j$.
There are instances where several matches are identified for one subhalo with similar merits. 
This can happen when several similar mass haloes merge at once, as
loosely bound particles can be readily exchanged between haloes. \citet{Elahi18} explained
 that {\sc TreeFrog} deals with these situations by ranking particles based on their 
binding energy. The latter is used to estimate a combined merit function that makes use of total number 
of particles shared and the information of the binding energy (see Eq.~$3$ in \citealt{Elahi18}). 

We produce a tree following haloes forward in time, identifying the optimal links between progenitors and descendants. We
rank progenitor/descendant link as primary and secondary. A primary link is the bijective one; that is, 
it is a positive match in two directions progenitor and descendant. The merit is maximum 
 both forward and backward.
All other connections are classified as secondary links.
 {\sc TreeFrog} searches for several snapshots to identify optimal links, and by default we search up to $4$ snapshots.

Poulton et al. (submitted) show that the treatment described here plus the superior behaviour of {\sc VELOCIraptor} 
at identifying structures (see also \citealt{Canas18}), lead to very well behaved merger trees, with orbits that 
are well reconstructed. \citet{Elahi18b} also show that these orbits reproduce the bias in the halo mass estimate 
obtained from using the peculiar velocities of galaxies. 

\section{Shark design}\label{sharkdesign}

\shark{} is written in C++11, and therefore
can be compiled with any C++11-enabled compiler
(gcc 5+, clang 3.3+, and others).
\shark{} uses the standard \texttt{cmake} compilation system,
and requires only the HDF5, GSL and boost libraries to build.
These can be commonly found in most Linux distributions,
MacOS package managers and HPC systems.
This ease to compile and install \shark{}
in a number of different machines and operating systems
is an important aspect to pay attention to
if wider adoption is sought.
A set of python modules (compatible with python $2.7$ and $3$) are also distributed with \shark\ to produce 
a set of standard plots, including those presented in this paper.
The code is hosted in GitHub\footnote{\href{https://github.com/ICRAR/shark}{\url{https://github.com/ICRAR/shark}}}
and is free for everyone to download and use. 

A continuous integration service has also been setup
on Travis\footnote{\href{https://travis-ci.org/ICRAR/shark}{\url{https://travis-ci.org/ICRAR/shark}}}
to ensure that after each change introduced in the code,
the code compiles using different compilers,
under different operating systems,
that \shark{} runs successfully against a test dataset,
and that all standard plots are successfully produced. 
These runs are generated using the parameter file used by the default \shark\ model 
analysed in $\S$~\ref{basicresults}, and a subset of merger trees of the L210N512 simulation 
(see Table~\ref{tab:sims} for details).
\shark\ adopts the GPLv3 license.

\subsection{Design}

\begin{figure*}
\begin{center}
\includegraphics[width=0.99\textwidth]{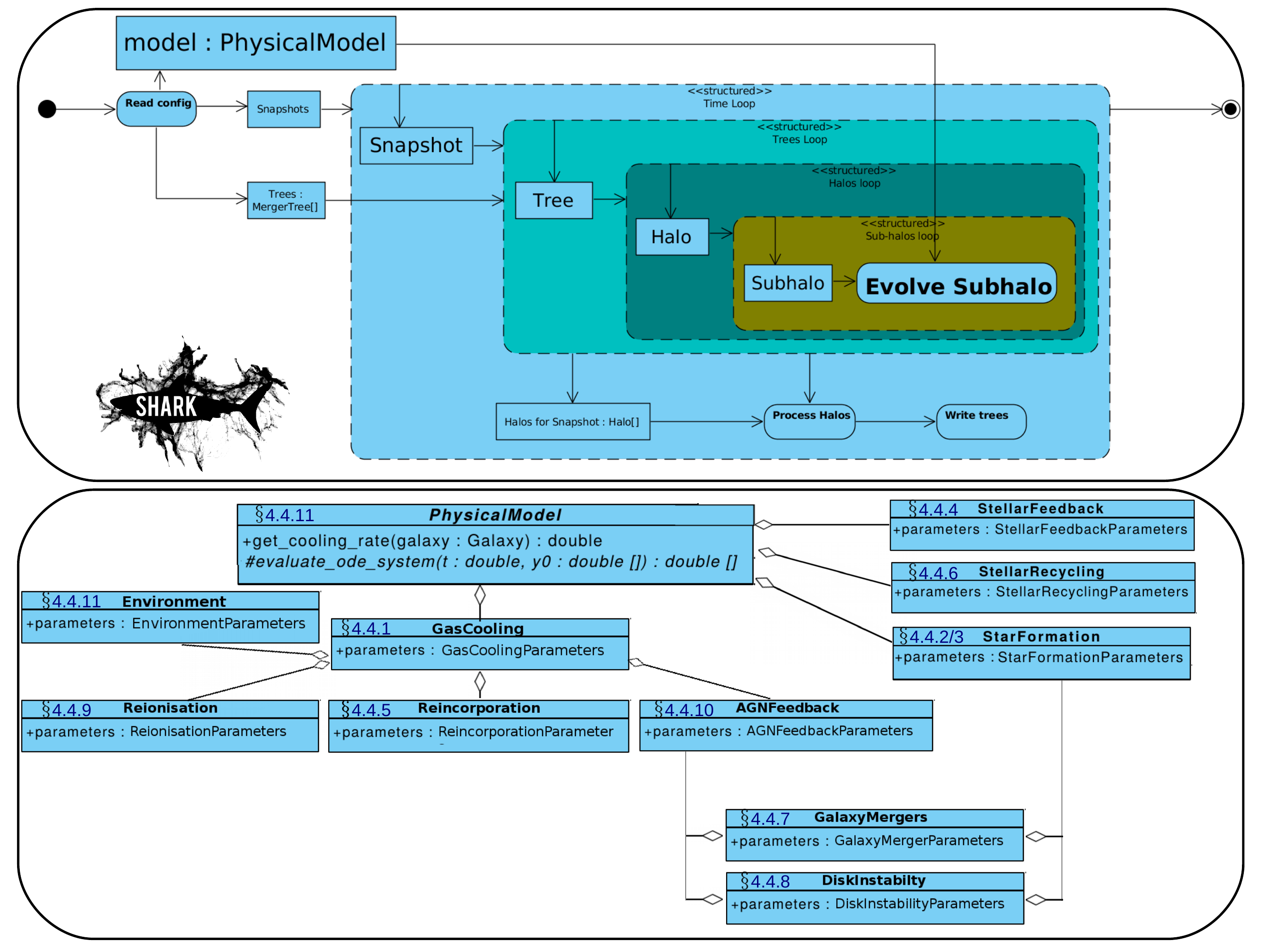}
\caption{{\it Top panel:} Design of \shark. The \emph{physical model} is the set of ODEs  
that are solved numerically by \shark\ (described in $\S$~\ref{sec:odes} for the current \shark\ model). The implementation 
of the \emph{physical model} is detached from the main evolution loop, which allows the code a lot of flexibility to change 
the description of the interplay between mass, metal and angular momentum components. Note that \shark\ also allows halos at any given 
snapshot to interact with each other, despite belonging to different merger trees. 
{\it Bottom panel:} zoom into the structure of the \emph{physical model}. Most of the relevant physical processes in galaxies respond directly to 
the \emph{physical model}. The only two important physical models that are called separately are galaxy mergers and disk instabilities 
only due to computational expense and simplicity of the code. The classes of AGN feedback and star formation are also linked with those 
of galaxy mergers and disk instabilities as in these two processes we expect black holes to grow and central starbursts to be driven. 
We show next 
to each physical process the section in this paper where we describe the details of the modelling.}
\label{design}
\end{center}
\end{figure*}

\shark\ evolves galaxies across snapshots
using a \emph{physical model}. 
{This \emph{physical model} describes the way in which the different physical processes included in the model 
interplay with each other. In the practice, this means the exact way in which the exchange of mass, metals and angular momentum 
take place (Eqs.~\ref{eqn:sff}-\ref{eqn:sff:jf} in the case of \shark).}
The particular physical model used by \shark\ 
is not hard-coded in the main evolution loop,
but implemented separately and provided as an input
to the main evolution routine. With `main evolution loop' we refer 
to as the loop over snapshots, which contains the loop over merger trees and halos 
(see top panel in Fig.~\ref{design}).
This design allows for different physical models
to be seamlessly exchanged.
We currently offer a single physical model
represented by a set of ordinary differential equations (ODEs) described in Eqs.~\ref{eqn:sff}-\ref{eqn:sff:jf}
to evolve each galaxy,
but other models
can be implemented.
This is shown in the schematic of Fig.~\ref{design}.

Following the same principle,
the individual physical processes
that participate in the physical model
are not hard-wired to the physical model itself,
but implemented as independent classes
and provided as inputs to the physical model.
These classes implement only the logic
associated to the particular physical process
they represent, exposing it to its callers.
If available,
different implementations of the same physical process
can also be chosen at runtime.

\subsection{Scalability}

\shark\ scales naturally with its input data.
Input volumes are usually divided into independent \emph{sub-volumes}
that can be individually processed.
On the other hand a single \shark\ execution 
can be commanded to process one or more sub-volumes.
This simple but flexible scheme
allows for easy parallelisation
based on input data,
where multiple \shark\ executions can be run 
to process a large number of sub-volumes in parallel.
This strategy does not require communication between executions,
reducing both the complexity of the software and its dependencies.

Depending on the size of its inputs,
\shark\ will usually be limited
by the amount of available memory.
Memory usually scales with the number of CPUs,
and therefore \shark\ will usually have
multiple CPUs at its disposal.
We take advantage of this
by further parallelising the execution of \shark\
using OpenMP.
During the main evolution loop, and for any snapshot,
the evolution of galaxies
belonging to different merger trees
is independent from each other.
This the place in the code in which 
most of the time is spent, and thus
we parallelise
the evolution of individual merger trees
so they take place in different threads.
The number of threads to use
is specified on the command-line,
and can be set to either a fixed number,
or to the default value provided by the OpenMP library.
In addition to this
other parts of the code use OpenMP
to parallelise their execution.

{In some important physical applications, for example to model 
the epoch of reionisation, halos belonging to different merger trees 
need to interact with each other \citep{Mutch16}. This in principle can be 
implemented in \shark, but in that case the user should run without OpenMP.}

\subsection{High Performance Computing environments}

\shark{} can also be efficiently and easily run
in HPC environments.
\shark{} comes with a \texttt{shark-submit} script
to spawn multiple \shark{} instances to an HPC cluster
running over a set of sub-volumes
and using a common configuration.
The script abstracts away the details
of the underlying queueing system
and takes care of using all resources optimally
(in terms of memory and CPUs),
while offering users flexibility over the submission parameters.
The script also creates well-organized, per-submission outputs,
making it easy to inspect them independently.
At the moment of writing only SLURM is supported,
but support for Torque/PBS will follow,
and more could be added in the future if required.

\subsection{Diagnostic plots}

\shark\ includes a set of python scripts (robust under python~2.7 and 3), which produce all the plots in this paper plus several other diagnostic 
plots. These scripts can be run automatically using {\tt shark-submit}. 

\section{Shark physics}\label{ModelDescription}

In this section we provide a description of the physics included in \shark, showing 
how the different models are referred to as in the code.

\subsection{Evolving galaxies through merger trees}\label{sec:evolvemergertree}

The merger trees and subhalo catalogue of {\sc VELOCIraptor+TreeFrog} provide a static skeleton within which we need to evolve galaxies. 
In \shark\ we make a postprocessing treatment of these merger trees before forming and evolving galaxies across the skeleton, described below.

\begin{itemize}
\item {\it Interpolating halos/subhalos}. Because {\sc TreeFrog} searches for primary links up to $4$ snapshots in the future, it can happen that a subhalo has 
as a descendant subhalo that is not necessarily at the next snapshot. This causes discontinuities at the moment of evolving galaxies. Thus, in \shark\  
we place subhalos between the snapshots of the current subhalo and its descendant, which we term `interpolated' subhalos. The properties of these interpolated subhalos 
are frozen to those of their progenitor subhalo. This measure ensures continuity to solve the equations of galaxy formation that we detail in $\S$~\ref{physics}.
It happens commonly with all available merger tree builders we are aware of, that subhalos disappear from the merger tree (i.e. have no identified descendant). 
In \shark\ one can choose to ignore those in the calculation by setting the execution parameter 
{\tt skip$_{-}$missing$_{-}$descendants} to {\tt true}. 
\item {\it Ensuring mass growth of halos.} Once merger trees are constructed, we navigate them to ensure that the mass of a halo is strictly equal or larger than the 
halo mass of its most massive progenitor. This is done to ensure that matter accretion onto halos is always $\ge 0$. $\S$~\ref{DMhalosAcc} describes how the gas accretion rate 
onto halos is calculated. Other SAMs follow a similar procedure \citep{Lacey15}, but with the aim of giving the user control over these decisions, \shark\ includes 
a boolean parameter, {\tt ensure$_{-}$mass$_{-}$growth}, which should be set to {\tt false} if the user does not wish to include this step. 
In that case, the negative accretion rates are ignored and set to $0$. 
{Note that Galacticus \citep{Benson12} also allows for these two options.}
In \shark, we find the results to be very mildly affected by this step 
due to {\sc VELOCIraptor+TreeFrog} providing high quality identifications and links that need little additional processing (Poulton et al. submitted).
\item {\it Defining the central subhalo.} In order to define the central subhalo of every halo in the catalogue, we step at $z=0$ and define the most massive subhalo 
of every existing halo as the central one. We subsequently make the main progenitor of those centrals as the centrals of their respective halo. We do this iteratively back 
in time. At every snapshot we find those halos that merge into another, and are not the main progenitors, and apply the same logic described above to designate their 
central subhalo.
\end{itemize}

Every subhalo/halo connects to its progenitor(s) and descendant subhalo/halo, which connect to the merger tree in which they have lived throughout 
their existence. Halos also point to their central subhalo and its satellite subhalos. 
Every subhalo points to the list of galaxies it may contain, but only central subhalos are allowed to have a central galaxy (which in turn 
is the central galaxy of the host halo). 

The merger trees are a static skeleton and we treat them as such in \shark. 
Thus, in order to form galaxies and subsequently evolve them, we identify all the halos that first appear in the catalogue (those with no 
progenitors) and initialize the galaxy pointer, so far composed of one galaxy with zero mass. 
The central subhalo of that halo is assigned a halo gas reservoir of mass $\Omega_{\rm b}/\Omega_{\rm m}\times M_{\rm halo}$. 
Having a halo gas mass $>0$ ignites gas cooling and the subsequent formation of a cold gas disk (as detailed in $\S$~\ref{physics}).
At the end of every snapshot, we transfer all the galaxies that are hosted by any one subhalo to its descendant and proceed to evolve them.
If subhalos appear for the first time in the merger tree as a satellite subhalo, and with the design above, it is defined as a dark subhalo with no 
allowed to form there.

\shark\ galaxies exist in $3$ different types: $\tt type = 0$ is the central galaxy of the central subhalo, while every other central galaxy of satellite 
subhalos are $\tt type=1$. If a subhalo merges onto another one and it is not the main progenitor, its given as defunct. All the galaxies 
of defunct subhalos are made $\tt type=2$ and transferred to the central subhalo of their descendant host halo. 
Galaxies $\tt type=2$ are widely referred to as `orphan galaxies' \citep{Guo15}.
Note that satellite subhalos can have subhalos themselves (i.e. subsubhalos), and {\sc VELOCIraptor+TreeFrog} allow for several levels of hierarchy 
in the subhalo population. However, in \shark\ all satellite subhalos are treated the same way (as containing satellite galaxies $\tt type=1$) regardless 
of their hierarchy. In this logic, central subhalos can have one central galaxy and any number of $\tt type=2$ galaxies, while satellite subhalos can only have 
one $\tt type=1$ galaxy. Any $\tt type=2$ galaxies a subhalo can have before becoming satellite, are transferred to the central subhalo once 
it becomes a satellite subhalo.

Table~\ref{tab:execparameters} lists the execution parameters in \shark\ with their possible values and those adopted by default. 
The parameter, {\tt ode$_{-}$solver$_{-}$precision}, determines the numerical precision to which the user wishes to solve the set of ODEs 
of $\S$~\ref{sec:odes}.

\begin{table}
\setlength\tabcolsep{2pt}
\centering\footnotesize
\caption{\shark\ execution parameters. Here we show the names these variables have in the code and the values chosen for our 
default \shark\ model in parenthesis in the middle column.}
\begin{tabular}{@{\extracolsep{\fill}}l|cc|p{0.45\textwidth}}
\hline
\hline
    Execution parameter & suggested value range\\
\hline
         {\tt ensure$_{-}$mass$_{-}$growth} & {\tt true} or {\tt false} ({\tt true})\\
         {\tt skip$_{-}$missing$_{-}$descendants} & {\tt true} or {\tt false} ({\tt true})\\
         {\tt ode$_{-}$solver$_{-}$precision} & $10^{-3}-0.1$ ($0.05$)\\
\hline
\end{tabular}
\label{tab:execparameters}
\end{table}

\subsection{Dark matter halos}\label{DMhalos}

When halos are formed, we assume them to have virial radii $r_{\rm vir} =
(3 M_{\rm halo}/(4\pi \Delta_{\rm vir} \rho_{\rm crit}))^{1/3}$, where $M_{\rm halo}$ is the
halo mass, $\rho_{\rm crit}$ is the cosmological critical density at that
redshift, and the overdensity $\Delta_{\rm vir}(\Omega_m,\Omega_v)=200$. 
Based on the spherical collapse model, \citet{Cole96} estimated 
 $\Delta_{\rm vir}(\Omega_m,\Omega_v)=178$, but a more widely adopted value 
based on $N$-body simulations is $200$. 
We assume DM halos to have a density profile that follow a 
 NFW profile:
\begin{equation}
\rho_{\rm DM}(r) \propto \frac{1}{(r/r_{\rm s})(1+r/r_{\rm s})^2} ,
\label{EqNFW}
\end{equation}
where $r_{\rm s}$ is the scale radius, related to the virial radius by
the concentration, $r_{\rm s} = r_{\rm vir}/c_{\rm NFW}$. In \shark\ we
 estimate concentrations using the \citet{Duffy08} relation between concentration, halo's virial mass and redshift. 
 {The user can choose to do this using instea the relation of \citet{Dutton14}. The latter is controlled by 
 the input parameter {\tt concentration$_{-}$model}.}

Halos grow via merging with other halos and by 
accretion. The properties $r_{\rm vir}$ and 
$M_{\rm halo}$ are calculated by {\sc VELOCIraptor} at each snapshot. 
In addition, the user can choose to either use the input halo's spin parameter, $\lambda_{\rm DM}$, calculated in 
{\sc VELOCIraptor} (which corresponds to the \citealt{Bullock01} spin parameter), 
or draw it from a log-normal distribution of mean $0.03$ and width $0.5$. 
This is controlled by the boolean parameter ${\tt lambda_{-}random}$.
The halo's angular momentum is then calculated from the mass and halo's spin 
parameters, adopting \citet{Mo98},

\begin{equation}
	J_{\rm h}= \frac{\sqrt{2}\,G^{2/3}}{(10\,H(z))^{1/3}} \lambda_{\rm DM}\, M^{5/3}_{\rm h},
\label{jhalo}
\end{equation}

\noindent where $G$ is Newton's gravity constant and $H(z)$ is the Hubble parameter. 
In the future we plan to add additional plausible profiles (e.g. \citealt{Einasto65})  for the users to decide which one 
they prefer. 

\subsection{Matter accretion onto halos}\label{DMhalosAcc}

When halos are first formed, we assume that a fraction $\Omega_{\rm b}$ 
is in the form of hot halo gas with a temperature 
$T_{\rm vir} = (\mu m_{\rm H}/2\kB) v_{\rm vir}^2$, where $v_{\rm vir} =
(G\,M_{\rm halo}/r_{\rm vir})^{1/2}$, and $\mu$ is the mean molecular weight.
We assume that this gas has a minimum fraction of metals $Z_{\rm min}$. 

Any subsequent gas accretion onto the halos from the cosmic web is calculated 
based on the DM mass a halo gains that does not come via mergers. We 
do this by adding up all the mass contributed by the progenitor halos, and taking the difference 
with the halo mass at the current timestep, $\Delta M= M_{\rm halo,curr} - M_{\rm halo,prog}$.
\citet{Contreras17} showed that halos can have sudden changes in their mass due to misidentification 
by halo/subhalo finders and by the construction of the merger tree. 
Analysing several snapshots to clean the subhalo catalogues as is done in several algorithms \citep{Behroozi13b,Jiang14,Elahi18}, helps 
to avoid this problem, but to some extent it can still be present. In order to avoid 
these sudden changes in mass to have a big effect on the accretion rates calculated here and introduce systematic problems, 
we limit the maximum baryon mass halos can have to the universal baryon fraction,
with baryons here 
including galaxy masses, halo gas and ejected gas. The latter is not strictly within halos, but makes up the 
intergalactic medium, which is formed by the effect of outflows (details presented in $\S$~\ref{sec:stellarfeed}~and~\ref{sec:reinco}).
We assume that the accreted matter brings a fraction $\Omega_{\rm b}$
of baryons with a metallicity $Z_{\rm min}$. \citet{Bromm04} argue that 
population III stars can lead to a nearly uniform enrichment of the universe to a
level of $Z~10^{-4}\rm\,Z_{\odot}$, and thus $Z_{\rm min}$ would typically take a value close to that. 

\subsection{Physical modelling of galaxy formation and evolution}\label{physics}

In \shark\ we include a large library of physical models describing gas cooling, star formation, 
stellar feedback and chemical enrichment, BH growth, AGN feebdack, galaxy mergers, disk instabilities, the development of 
galaxy sizes and environmental effects. Each of these mechanisms can be modelled in different ways, so \shark\ includes 
several different models for any one physical process. One of the missions of \shark\ is to be constantly updating the code 
to include more plausible models for all the different physical processes above, and possibly to add additional physical 
processes over time. 

Below we describe the models that are included in {\sc v1.1}.
At the end of this section we present all the parameters and models that can be included in \shark\ in 
Tables~\ref{tab:parameters}~and~\ref{tab:parameters2},
together with suggested ranges of values, and the values adopted in our default model.

\subsubsection{Gas in halos and cooling}\label{sec:cooling}

Gas in halos are assumed in \shark\ to have two phases: cold and hot. 
Cold halo gas is the gas that cools down within a simulation snapshot,
while the hot halo gas corresponds to the gas that is at the virial temperature 
and that has not had time to cool down yet. The halo gas settles into a spherically symmetric distribution
with some density profile. We implement an isothermal profile,

\begin{equation}
\rho_{\rm g}(r) = \frac{m_{\rm gas,h}}{4\pi\,r^2_{\rm vir}\,r},
\end{equation}
 
\noindent where $m_{\rm gas,h}$ is the total halo gas.

This halo gas then loses its
thermal energy by radiative cooling due to atomic processes, at a rate
per unit volume $\rho^2_{\rm g} \Lambda(T_{\rm vir}, z_{\rm gas,h})$, in which collisional
ionization equilibrium is assumed, and where $z_{\rm gas,h}$ is the metallicity of this
gas. We use {\sc CLOUDY} version 08 \citep{Ferland98} to 
produce tabulated cooling functions in a grid of $(T_{\rm vir}, z_{\rm gas,h})$,  
{assuming collisional ionization equilibrium and without considering the effects of dust}. 
Alternatively, the user can also choose to use instead the cooling tables of \citet{Sutherland93}.  
We interpolate over this large grid at each snapshot for each halo
to estimate $\Lambda$.
{Note that 
\shark\ uses the cooling rate to solve the ODEs of Eqs.~\ref{eqn:sff}-\ref{eqn:sff:jf}, and therefore 
it is not very sensitive to the simulation snapshots, even though the cold halo gas mass is.}

The cooling time is related to the density as

\begin{equation}
t_{\rm cool}(r) = \frac{3}{2} \frac{\mu m_{\rm H}\, \kappa_{\rm B}\, T_{\rm vir}}{\rho_{\rm g}(r)\,  \Lambda(T_{\rm vir}, z_{\rm gas,h})},
\label{tcool}
\end{equation}

\noindent where $\kappa_{\rm B}$ is Boltzmann's constant.
The cooled gas corresponds to that enclosed within the cooling radius, $r_{\rm cool}$. 
We refer to that gas as cold halo gas. 
In {\sc v1.1}, we calculate the cooling time and radius by employing two different models, 
\citet{Benson10} and \citet{Croton06}.
These two modes take different approaches 
to estimate the cooling time, which is then used to estimate $r_{\rm cool}$.
Below we summarize these two approaches.

\begin{itemize}
\item The {\tt Croton06} model. \citet{Croton06}, also adopted by \citet{Guo11} and \citet{Henriques15}, assume that the cooling time, $t_{\rm cool}$, 
at the cooling radius is of a similar magnitude to the dynamical timescale, and simply calculates it as 
$t_{\rm cool} \equiv r_{\rm vir}/v_{\rm vir}$. \citet{Croton06} then 
derive $r_{\rm cool}$ from Eq.~\ref{tcool}.  
The cooling rate is then calculated from the continuity equation

\begin{equation}
\dot{m}_{\rm cool} = 4\pi\,\rho_{\rm g}(r_{\rm cool})\,r^2_{\rm cool}\dot{r}_{\rm cool}.
\end{equation}

\noindent This is valid only if $r_{\rm cool}<r_{\rm vir}$ (referred to as `hot-halo mode'). 
In the case $r_{\rm cool}>r_{\rm vir}$, all the halo gas 
is accreted onto the galaxy in a dynamical timescale (referred to as `cold-halo mode'). 
\item The {\tt Benson10} model. 
\citet{Benson10} define a time available for cooling. In the case of an static halo, this 
time available for cooling is equivalent to the time since the halo came into existence. 
\citet{Benson10} assume $t_{\rm cool}\equiv t_{\rm avail}$, where 

\begin{equation}
t_{\rm avail} = \frac{\int_{0}^{t}\,[T_{\rm v}(t^{\prime})\,M_{\rm gas,h}(t^{\prime})/t_{\rm cool}(t^{\prime})]\, {\rm d}t^{\prime}}{T_{\rm v}(t)\,M_{\rm gas,h}(t)/t_{\rm cool}(t)},
\end{equation}

\noindent with $t$ corresponding to the current time. The cooling time above is computed at the mean density of the notional 
profile, $\rho_{\rm g}$. 
Having computed $t_{\rm avail}$ we solve for $r_{\rm cool}$ by 
$t_{\rm cool}(r_{\rm cool}) = t_{\rm avail}$. 
The current infall radius, $r_{\rm infall}$,  is then taken to be the smaller 
of the cooling and freefall radii. The cooled mass that is accreted 
onto the galaxy is simply that enclosed by $r_{\rm infall}$. 
\end{itemize}

In summary, the main difference between the {\tt Croton06} and {\tt Benson10} models, is what they assume for the time available for cooling. 
Which in the former, it is assumed to be equal to the dynamical time of the halo. 
The adopted model for cooling can be seen as `adjustable parameters' in the sense that using one or the other can change 
the predicted galaxy population. The distinction between `cold' and `hot' halo gas is reset at every snapshot, 
and is used when solving the ODEs described in Eqs.~\ref{eqn:sff}-\ref{eqn:sff:jf}. 

\subsubsection{Star formation in disks}\label{sec:sf}

The gas is assumed to follow 
an exponential profile of  
 half-mass radius $r_{\rm gas, disk}$ (see $\S$~\ref{sizes} for a definition).
We calculate the SFR surface density assuming a constant depletion time 
for the molecular gas,

\begin{equation}
\Sigma_{\rm SFR} = \nu_{\rm SF}\,f_{\rm mol}\,\Sigma_{\rm gas},
\label{SFLaw}
\end{equation}

\noindent where $\nu_{\rm SF}$ is the inverse of the H$_2$ depletion timescale, and 
$f_{\rm mol}\equiv \Sigma_{\rm mol}/\Sigma_{\rm gas}$, where $\Sigma_{\rm mol}$ is the molecular gas surface density and 
$\Sigma_{\rm gas}$ is the total gas surface density.
 Below we provide details on how we estimate $f_{\rm mol}$.
Some of the models included in \shark\ calculate a $\nu_{\rm SF}$ that depends on galaxy properties, while other 
models assume the observational value $\nu_{\rm SF} = 1/\tau_{\rm H_2}$, with $\tau_{\rm H_2}=2.2^{+2.1}_{-1.1}\,\rm Gyr$ \citep{Leroy13} being 
the observed molecular gas depletion timescale. The latter is the case for the {\tt BR06} and {\tt GD14} models.  

The HI surface densities cannot extend to infinitely small surface densities because the UV background can easily 
ionise very low density gas. Thus, we impose a limit in the minimum HI density allowed before the gas becomes ionised, 
$\Sigma_{\rm thresh}$, and adopt $\Sigma_{\rm thresh}=0.1\,\rm M_{\odot}\,pc^{-2}$ following the results of the hydrodynamical 
simulations of \citet{Gnedin12}. In reality this threshold should evolve with redshift, increasing at earlier epochs 
when the UV background is brighter \citep{Haardt12}.

We integrate $\Sigma_{\rm SFR}$ over the radii range $0-10\rm r_{\rm gas, disk}$
 to obtain the instantaneous SFR, $\psi$. We do this using an adaptive integrator that adopts 
a 15 point Gauss-Kronrod rule (due to its speed), available in the GSL {\tt C++} libraries. We enforce a $1$\% accuracy.

\shark\ has several implementations to calculate $f_{\rm mol}$, which are described below.

\begin{itemize} 
\item The {\tt BR06} model. \citet{Blitz06} found that the H$_2$ to HI ratio, $R_{\rm mol}\equiv \Sigma_{\rm H_2}/\Sigma_{\rm HI}$, 
correlates with the local hydrostatic pressure as 

\begin{equation}
R_{\rm mol} =  \left( \frac{P}{P_0} \right)^{\alpha_P}, 
\label{eq:at_mol}
\end{equation}

\noindent where $P_0$ and 
$\alpha_P$ are parameters measured in observations and have values 
$P_0/\kappa_{\rm B} = 1,500-40,000\,\rm cm^{-3}\,\rm K$ and $\alpha_P \approx 0.7-1$
\citep{Blitz06,Leroy08,Leroy13}. 
 We calculate the hydrostatic pressure from the
surface densities of gas and stars following \citet{Elmegreen89}, 

\begin{equation}
P = \frac{\pi}{2} \,G\,\Sigma_{\rm gas}(\Sigma_{\rm gas} + \frac{\sigma_{\rm gas}}{\sigma_{\star}}\,\Sigma_{\star}),
\label{eq:press}
\end{equation}

\noindent where $\Sigma_{\rm gas}$ and $\Sigma_{\star}$ are the 
total gas (atomic plus molecular) and stellar surface densities, respectively, 
and $\sigma_{\rm gas}$ and $\sigma_{\star}$ are the gas and stellar velocity dispersions.
The stellar surface density is assumed to follow an exponential profile with 
a half-mass stellar radius of $r_{\star,\rm disk}$.
We adopt $\sigma_{\rm gas}=10\,\rm km\,s^{-1}$ \citep{Leroy08}
 and calculate $\sigma_{\star}=\sqrt{\pi\,G\,h_{\star}\,\Sigma_{\star}}$.
Note that $\sigma_{\rm gas}$ is treated as a free parameter in the sense that 
the user can set it to different values, though we recommend 
to adopt values that do not deviate much from the observational measurement.
Here, $h_{\star}$ is the stellar scale height, 
and we adopt the observed relation $h_{\star}=r_{\star}/7.3$ \citep{Kregel02}, with  
 $r_{\star}$ being the half-stellar mass radius (see $\S$~\ref{sizes}).

\item The {\tt GD14} model. \citet{Gnedin14} presented the results of cosmological hydrodynamical 
simulations that include the formation of H$_2$. These simulations also included gravity,
hydrodynamics, non-equilibrium chemistry combined with equilibrium cooling rates for metals, and a
3-dimensional, on the fly, treatment of radiative transfer, using an Adaptive Mesh Refinement (AMR)
code. Compared to earlier implementations \citep{Gnedin11}, \citet{Gnedin14} paid special attention to 
the effect of line overlap in the Lyman and Werner bands in H$_2$ shielding. \citet{Gnedin14} presented 
a model for $R_{\rm mol}$ that describes the simulation results well. This model depends on the dust-to-gas ratio, $D_{\rm MW}$,  
and the local radiation field, $U_{\rm MW}$, with respect to that of the solar neighbourhood (i.e. they therefore are dimensionless parameters). 
We estimate these two parameters as $D_{\rm MW} = Z_{\rm gas}/Z_{\odot}$ and $U_{\rm MW} = \Sigma_{\rm gas}/\Sigma_{\rm MW}$, 
where $Z_{\rm gas}$ is the metallicity of the ISM. We adopt $Z_{\odot}=0.134$ \citep{Asplund09} and $\Sigma_{\rm MW}=2.5\,\rm M_{\odot}\,yr^{-1}$ \citep{Bonatto11}. 

The approximation we use for $U_{\rm MW}$ is based on the argument of \citet{Wolfire03} that pressure balance between the warm and the cold
neutral media is achieved only if the density is larger than a minimum density, which is proportional to $U_{\rm MW}$. Thus, 
if we assume that pressure equilibrium between the warm/cold media is a requirement for the formation of the ISM, we can then assume 
that $U_{\rm MW} \propto \rho_{\rm gas}$, with $\rho_{\rm gas}$ being the gas density. Since galaxies show a close to constant 
$\sigma_{\rm gas}$, we can assume that the gas scale height is close to constant, which allow us to replace $\rho_{\rm gas}$ by $\Sigma_{\rm gas}$ above. 

Based on $D_{\rm MW}$ and $U_{\rm MW}$ we calculate $R_{\rm mol}$  following  \citet{Gnedin14},

\begin{equation}
R_{\rm mol} = \left( \frac{\Sigma_{\rm gas}}{\Sigma_{\rm R=1}}\right)^{\alpha_{\rm GD}}, 
\end{equation}

\noindent where

\begin{equation}
\alpha_{\rm GD} = 0.5 + \frac{1}{1 + \sqrt{U_{\rm MW} D^2_{\rm MW}/600}},
\end{equation}

\begin{equation}
\Sigma_{\rm R=1} = \frac{50\,\rm M_{\odot}\,pc^{-2}}{g}\,\frac{\sqrt{0.01 + U_{\rm MW}}}{1 + 0.69\sqrt{0.01+U_{\rm MW}}},
\end{equation}

\noindent and 

\begin{equation}
g = \sqrt{D^2_{\rm MW} + D^2_{\star}}.
\end{equation}

\noindent Here, $D_{\star}\approx 0.17$ for scales $>500\,\rm pc$.

\item The {\tt KMT09} model. \citet{Krumholz09}, hereafter KMT09, calculated $\nu_{\rm SF}$ and $f_{\rm mol}$ in Eq.~\ref{SFLaw}  for a spherical cloud with SF
regulated by supersonic turbulence. KMT09 assume that $f_{\rm mol}$ is determined by the balance between the photodissociation of H$_2$ 
molecules by the interstellar far-UV radiation and the formation of molecules on the surface of dust grains, and 
calculated it theoretically to be a function of the total gas surface density of the cloud and of the gas metallicity (see Eq.~$2$ 
in KMT09). The gas surface density of the cloud is related to the disk gas surface density via a cumpling factor, $f_{\rm c}$. 
The latter is argued to be $f_{\rm c}\approx 5$ when averaging over $1$~kpc region in local galaxy disks. 
KMT09 estimated $\nu_{\rm SF}$  from the theoretical model of turbulent fragmentation of \citet{Krumholz05}. In this model, 
$\nu_{\rm SF}$ depends on the cloud surface density, which in spiral galaxies is assumed to be constant, with an observed value of 
$\Sigma_{0} \approx 85\,\rm M_{\odot}\,\rm pc^{-2}$. In starbursts (SBs), however, the ambient pressure is expected to increase significantly, which is accompanied by 
gas surface densities that can become larger than $\Sigma_{0}$. KMT09 argue that clouds will therefore have a density $\Sigma_{\rm cl}={\rm max}[\Sigma_{0}, \Sigma_{\rm gas}]$, which 
leads to $\nu_{\rm SF}$ to be described as

\begin{eqnarray}
\nu_{\rm SF} = \left\{ \begin{array}{rl}
  \nu^0_{\rm SF}\,\left(\frac{\Sigma_{\rm gas}}{\Sigma_{0}}\right)^{-0.33}, &\mbox{$\Sigma_{\rm gas} < \Sigma_{0},$}\\
  \nu^0_{\rm SF}\,\left(\frac{\Sigma_{\rm gas}}{\Sigma_{0}}\right)^{0.33}, &\mbox{ $\Sigma_{\rm gas} \ge \Sigma_{0}.$}
       \end{array} \right.
\label{KMT09_nusf}
\end{eqnarray}

\item The {\tt K13} model. \citet{Krumholz13} developed a theoretical model for the transition
from HI-to-H$_2$ that depends on the total column density of neutral
hydrogen, the gas metallicity and the interstellar radiation field. 
A key property in the \citet{Krumholz13} model is the density of the
cold neutral medium (CNM).  At densities $n_{\rm H}\gtrsim 0.5\,\rm
cm^{-3}$, the transition from HI to H$_2$ is mainly determined by the
minimum density that the CNM must have to ensure pressure balance with
the warm neutral medium (WNM, which is HI dominated). The assumption
is that the CNM is supported by turbulence, while the WNM is thermally
supported (see also \citealt{Wolfire03}). At $n_{\rm H}\lesssim
0.5\,\rm cm^{-3}$ the transition from HI to H$_2$ is mainly determined
by the hydrostatic pressure, which has three components: the
self-gravity of the WNM ($\propto \Sigma^2_{\rm HI}$), the gravity
between the CNM and WNM ($\propto \Sigma_{\rm HI}\Sigma_{\rm H_2}$),
and the gravity between the WNM and the stellar plus DM
component ($\propto \Sigma_{\rm HI}\Sigma_{\rm sd}$, where
$\Sigma_{\rm sd}$ is the surface density of stars plus DM). Note that the exact value of $n_{\rm H}$ at which the
transition between these two regimes takes place is a strong function
of gas metallicity. For this model, we adopt the same dust-to-gas mass ratio and 
local radiation field as in the {\tt GD14} model. We then define two densities, one that 
corresponds to the CNM density in the regime of two-phase equilibrium, $n_{\rm CNM,2p}$, and 
the CNM density set by hydrostatic balance, $n_{\rm CNM,hydro}$. The former (latter) is expected to dominate 
at high (low) gas surface densities. These densities are defined as:

\begin{equation}
n_{\rm CNM,2p} \approx 23\,U_{\rm MW}\,\left(\frac{1+3.1\,D^{0.365}_{\rm MW}}{4.1}\right)^{-1}\, {\rm cm}^{-3},
\label{ncnm_k13}
\end{equation}

\noindent and 

\begin{equation}
n_{\rm CNM,hydro}=\frac{P_{\rm th}}{1.1\,k_{\rm B}\,T_{\rm CNM,max}},
\end{equation}

\noindent where $T_{\rm CNM,max}$ is the maximum temperature at which the CNM can exist ($\approx 243$~K; \citealt{Wolfire03}), 
and 

\begin{equation}
P_{\rm th}\approx \frac{\pi\,G\,\Sigma^2_{\rm gas}}{4\,\alpha}\left[1+\sqrt{1+\frac{32\,\zeta_{\rm d}\,\alpha\,f_{\rm w}\,\sigma^2_{\rm gas}\, \rho_{\rm sd}}{\pi\,G\,\Sigma^2_{\rm gas}}}\right].
\end{equation}

\noindent Here $\alpha\approx 5$ represents how much of the midplane pressure support comes from turbulence, magnetic fields and cosmic rays,
compared to the thermal pressure \citep{Ostriker10}, $\zeta_{\rm d}\approx 0.33$ is a numerical factor that depends on
the shape of the gas surface isodensity contour,  
$f_{\rm w}=0.5$ is the ratio between the mass-weighted mean square thermal velocity dispersion and the square of the sound speed of
the warm gas (the value adopted here originally comes from \citealt{Ostriker10}) and $f_{\rm c}$ is the clumping factor (as in KMT09).
The value of the gas density in the CNM
is then taken to be $n_{\rm CNM}={\rm max}(n_{\rm CNM,2p},n_{\rm CNM,hydro})$.

K13 defines a dimensionless radiation field parameter:

\begin{equation}
\chi=7.2\,U_{\rm MW} \left(\frac{n_{\rm CNM}}{10\,{\rm cm}^{-3}}\right)^{-1},
\end{equation}

\noindent and writes $f_{\rm mol}$ as 

\begin{eqnarray}
f_{\rm H_2}= \left\{ \begin{array}{rl}
 1-{0.75\,s}/({1+0.25\,s}), &\mbox{ $s<2$,} \\
  0, &\mbox{ $s \ge 2,$}
       \end{array} \right.
\label{H2FracK13}
\end{eqnarray}

\noindent where

\begin{eqnarray}
s&\approx& \frac{{\rm ln}(1+0.6\,\chi+0.01\,\chi^2)}{0.6\,\tau_{\rm c}},\\
\tau_{\rm c}&=& 0.066\,f_{\rm c}\,D_{\rm MW}\,\left(\frac{\Sigma_{\rm gas}}{\rm M_{\odot} {\rm pc}^{-2}}\right).
\label{sdefinition}
\end{eqnarray}

\noindent We use Eq.~\ref{KMT09_nusf} to estimate $\nu_{\rm SF}$ for this model.
\end{itemize}

\subsubsection{Star formation in bulges}\label{sec:sfb}

SBs, which can be triggered by either galaxy mergers or disk instabilities, build up the central bulge in \shark. 
Thus, when we refer to SBs, we mean star formation taking place in the central bulge.

There is strong evidence that SBs follow a similar relation to normal star-forming galaxies 
studied in \citet{Leroy13} but with a timescale significantly shorter \citep{Daddi10b,Genzel15,Tacconi17}. 
We then adopt the same calculation of $R_{\rm mol}$, $\Sigma_{\rm gas}$ and $P$ above for bulges, replacing the disk properties 
with the bulge's. The only important difference is that we apply a boost factor to the star formation efficiency $\nu_{\rm SF,burst} = \eta_{burst} \nu_{\rm SF}$, 
with $\eta_{burst}$ taking values in the range $\approx 1-10$, according to observations \citep{Daddi10b,Scoville16,Tacconi17}. 

We implicitly assume that the gas in the bulge also settles in an exponential disk with scale length $r_{\rm bulge}/1.67$.   
To avoid calculating SFRs for very small quantities of gas left in the bulges, we decide to transfer the bulge gas 
to disk if it drops below $\rm min_{\rm gas,bulge}$. We find that for the 
resolution of the L210N1504, our default option, values $<10^{5.5}\rm \, M_{\odot}$ gives the same results. 
Our default option is $\rm min_{\rm gas,bulge} = 10^5\rm \, M_{\odot}$.

\subsubsection{Stellar feedback}\label{sec:stellarfeed}

\shark\ separates stellar feedback into two main components: the outflow rate of the gas that escapes from the galaxy, 
$\dot{m}_{\rm outflow}$, and the ejection rate of the gas that escapes from the halo, $\dot{m}_{\rm ejected}$. 
We implement different descriptions of SNe feedback, but 
$\dot{m}_{\rm outflow}$ and $\dot{m}_{\rm ejected}$ are related in the same way in all the model variants.

We can describe $\dot{m}_{\rm outflow}=\psi\,{\rm f}(z,V_{\rm circ})$, where $\psi$ is the instantaneous SFR, $z$ is the redshift and $V_{\rm circ}$ 
is the maximum circular velocity of the galaxy. The ejection rate of the halo should be $>0$ 
only in the case where the injected total energy of the outflow is larger than the {specific} binding energy 
of the halo. \citet{Muratov15} used the FIRE simulation suite to estimate several properties of the stellar driven outflows, 
including the terminal wind velocity, $V_{\rm w}$. \citet{Muratov15} found that 

\begin{equation}
\frac{V_{\rm w}}{\rm km\,s^{-1}}=1.9\,\left(\frac{V_{\rm circ}}{\rm km\,s^{-1}}\right)^{1.1}.
\end{equation}

\noindent We use this terminal velocity to compute the excess energy that will be used to eject gas out of the halo as 
\begin{equation}
{\rm E}_{\rm excess} = \epsilon_{\rm halo}\,\frac{V^2_{\rm w}}{2} \psi\,{\rm f}(z,V_{\rm circ}).
\label{eq:epshalo}
\end{equation}

\noindent Here $\epsilon_{\rm halo}$ is a free parameter. The net ejection rate is therefore calculated as,

\begin{equation}
\dot{m}_{\rm ejected} = \frac{{\rm E}_{\rm excess}}{V^2_{\rm circ}/2} - \dot{m}_{\rm outflow}.
\end{equation}

\noindent If $\dot{m}_{\rm ejected} < 0$ no ejection from the halo takes place and we limit 
$\dot{m}_{\rm outflow} = {\rm E}_{\rm excess}/(V^2_{\rm circ}/2)$.

As discussed above we implemented several models for ${\rm f}(z,V_{\rm circ})$, 
described below.

\begin{itemize}
\item The {\tt Lacey16} model. \citet{Bower12} presented a version of GALFORM that distinguishes 
the components $\dot{m}_{\rm outflow}$ and $\dot{m}_{\rm ejected}$, describing the function $f$ in a very simple 
fashion as,

\begin{equation}
{\rm f} = \left(\frac{V_{\rm circ}}{v_{\rm hot}}\right)^{\beta},
\label{SNGal}
\end{equation} 

\noindent with $\beta < 0$. Here there is no redshift dependence. In the \citet{Bower12} model 
$\beta = -3.2$ and $v_{\rm hot}=350\,\rm km\,s^{-1}$. Note that in the standard GALFORM implementation of \citet{Lacey15}, 
$\dot{m}_{\rm ejected}\equiv \dot{m}_{\rm outflow}$, and thus, in \shark\ we also enforce that equivalency.

\item The {\tt Guo11} model. \citet{Guo11} described SNe feedback as 

\begin{equation}
{\rm f} = \epsilon_{\rm disk}\, \left[0.5 +  \left(\frac{V_{\rm circ}}{v_{\rm hot}}\right)^{\beta}\right].
\label{SNLGAL}
\end{equation}

\noindent \citet{Guo11} adopted $v_{\rm hot}=70\,\rm km\,s^{-1}$, $\beta = -3.5$ and $\epsilon_{\rm disk}=6.5$, all of which 
were adjusted to fit the stellar mass function.
\item The {\tt Muratov15} model. \citet{Muratov15} presented a detailed analysis of the stellar driven outflows 
produced in the FIRE simulation suite by tracking explicitly the SPH particles and using 
kinematics to distinguish between outflowing and inflowing gas. \citet{Muratov15} found that the outflow rates 
relative to $\psi$ (also termed `mass loading') evolved significantly with redshift. They provide a best fit to the scaling 
between $\dot{m}_{\rm outflow}$ and $\psi$, z, $V_{\rm circ}$ as

\begin{equation}
{\rm f} = \epsilon_{\rm disk}\,(1+z)^{\rm z_{\rm P}}\,\left(\frac{V_{\rm circ}}{v_{\rm hot}}\right)^{\beta},
\label{SNFIRE}
\end{equation}

\noindent with $\epsilon_{\rm disk} = 2.9$, $\rm z_{\rm P} = 1.3$, $v_{\rm hot}=60\,\rm km\,s^{-1}$ and $\beta = -3.2$ 
if $V_{\rm circ}< v_{\rm hot}$ and $\beta=-1$ if $V_{\rm circ}> v_{\rm hot}$.
The redshift scaling in FIRE implies $L^{*}$ galaxies have mass loadings of $\approx 10$ at $z=6$ and $<1$ at $z=0$. 

\item The {\tt Lagos13} model. \citet{Lagos13} presented a detailed modelling of the expansion of SNe driven bubbles in 
a two-phase ISM. The authors followed the evolution of these bubbles from the early epoch of adiabatic expansion, to the momentum driven expansion 
until either confinement in the disk or break-out from the disk. They used this model to estimate $\dot{m}_{\rm outflow}$ and find 

\begin{eqnarray}
{\rm f} &=& \epsilon_{\rm disk}\, \left(\frac{V_{\rm circ}}{v^{\prime}_{\rm hot}}\right)^{\beta},\label{SNLAG13}\\
v^{\prime}_{\rm hot}  &=& v_{\rm hot} \, (1+z)^{\rm z_{\rm P}},\label{vel_power}
\end{eqnarray}

\noindent \citet{Lagos13} found values of $v_{\rm hot} = 425\,\rm km\,s^{-1}$, $\beta = 2.7$, $\rm z_{\rm P} = -0.2$, $\epsilon_{\rm disk}=1$. Note that \citet{Lagos13} found that the mass 
loading decreases with increasing redshift in tension with the 
findings of \citet{Muratov15}. This discrepancy is not necessarily due to the Lagos et al. model being a simpler description of the stellar feedback process, which 
is evidenced by previous hydrodynamical simulations (e.g. \citealt{Creasey12}; \citealt{Hopkins12}) finding results similar to those in 
\citet{Lagos13}. This is clearly a controversial topic and thus justifies our decision of implementing several different models of stellar 
feedback and leaving the parameters to vary. Note that assuming $\rm z_{\rm P} > 0$ in Eq.~\ref{vel_power} mimics the effect reported in \citet{Muratov15}. 
In fact, in our default \shark\ model, we adopt the {\tt Lagos13} model but with $\rm z_{\rm P} > 0$.
\end{itemize}

We also allow for two variants of the {\tt Lacey16} and {\tt Lagos13} models. In the case of the former, we implement a redshift dependence with the same form as 
in Eq.~\ref{vel_power}. We refer to this variant as {\tt Lacey16RedDep}. For {\tt Lagos13} we also apply a variant that reproduces the break 
of the {\tt Muratov15} mass loading function, implemented such that at $V_{\rm circ} > v_{\rm hot}$, $\beta = -1$. We refer to this model as 
{\tt Lagos13Trunc}. Note that {\tt Lacey16RedDep} is different from {\tt Lagos13}  in that the former assumes $\dot{m}_{\rm ejected}\equiv \dot{m}_{\rm outflow}$, though 
$ \dot{m}_{\rm outflow}$ would have the same functional form.

\subsubsection{Reincorporation of ejected gas}\label{sec:reinco}

The gas expelled from the halos of galaxies by stellar feedback is assumed to be
reincorporated in a timescale that is mass dependent. We follow the method developed by \citet{Henriques13}
 and describe the reincorporation rate as  

\begin{eqnarray}
\dot{m}_{\rm reinc} = \frac{m_{\rm ejected}}{\tau_{\rm reinc}\,\left(M_{\rm halo}/M_{\rm norm}\right)^{\gamma}}.
\label{Eqreinc}
\end{eqnarray}

\noindent Here, $m_{\rm ejected}$ is the reservoir of ejected mass, $\tau_{\rm reinc}$, $M_{\rm norm}$ and 
$\gamma$ are free parameters. \citet{Henriques13} found that in their model the best values for these parameters to fit the 
SMF of galaxies at $z=0$ and $z=2$ simultaneously,   
were $18\,\rm Gyr$, $10^{10}\,\rm M_{\odot}$ and $-1$, respectively. In \shark, a value $\tau_{\rm reinc} = 0$ is interpreted as the user adopting  
instantaneous reincorporation.

\subsubsection{Recycled fraction and yield}\label{Sec:Randp}

For chemical enrichment in \shark, we
 adopt the instantaneous recycling approximations
for the metals in the ISM. This implies that the metallicity
of the ISM gas mass instantaneously absorbs the fraction of recycled mass and newly synthesised metals
 in recently formed stars, neglecting the time delay for the ejection of gas and metals from stars.

The recycled mass injected back to the ISM by newly born stars
is calculated from the initial mass function (IMF) as,

\begin{equation}
R=\int_{m_{\rm min}}^{m_{\rm max}}\, (m-m_{\rm rem})\phi(m)\, {\rm d} m, 
\label{Eq:ejec}
\end{equation}

\noindent where $m_{\rm rem}$ is the remnant mass and the IMF is defined
as $\phi(m)\propto dN(m)/dm$. Similarly, we define the yield as

\begin{equation}
p =\int_{m_{\rm min}}^{m_{\rm max}}\, m_{\rm i}(m)\phi(m) {\rm d} m, 
\label{Eq:yield}
\end{equation}

\noindent where $m_{\rm i}(m)$ is the mass of newly synthesised metals ejected by stars of initial mass $m$.
 The minimum and maximum mass in the integrations are taken to be $m_{\rm min}=1\, M_{\odot}$ and
$m_{\rm max}=120\, M_{\odot}$. Stars with masses
$m<1\, M_{\odot}$ have lifetimes longer than the age of the Universe, and therefore they do not contribute
to the recycled fraction and yield. We use the stellar population model of \citet{Conroy09} to calculate 
$m_{\rm rem}$ and $m_{\rm i}(m)$ in Eqs.~\ref{Eq:ejec}~and~\ref{Eq:yield}, respectively.

In \shark\ we assume the stellar IMF is assumed to be universal and take the shape of a \citet{Chabrier03} IMF. 
This is a widely adopted IMF in observations and so it facilitates comparisons. 
Under this assumption we obtain $p=0.029$ and $R=0.46$. {Note that these values are subject to the assumed 
models for the remnant mass and yields, and therefore are left as `free' parameter. We, however, advice to apply only 
small perturbations to the values suggested here, unless well informed.}

\subsubsection{Galaxy mergers}\label{sec:mergers}

When DM halos merge, we assume that the galaxy hosted by the main progenitor halo (see $\S$~\ref{sec:evolvemergertree} for details) becomes
the central galaxy, while all the other galaxies become satellites orbiting
the central galaxy. These orbits gradually decay towards the centre due to energy and angular
momentum losses driven by dynamical friction with the halo material, including other satellites.
We distinguish between two types of satellite galaxies as described in $\S$~\ref{sec:evolvemergertree}. 
We calculate a dynamical friction timescale for orphan satellites only, and merge those with the central 
once that clock goes to zero.

Depending on the amount of gas and baryonic mass
 involved in the galaxy merger, a SB can be triggered. The time for the satellite to hit the central galaxy
is called the orbital timescale, $\tau_{\rm merge}$, which is calculated following \citet{Lacey93} as

\begin{equation}
\tau_{\rm merge}=f_{\rm df}\, \Theta_{\rm orbit}\, \tau_{\rm dyn}\, \left[\frac{0.3722}{{\rm ln}(\Lambda_{\rm Coulomb})}\right]\, \frac{M}{M_{\rm sat}}.
\label{eqtaum}
\end{equation}

\noindent Here, $f_{\rm df}$ is a dimensionless adjustable
parameter which is $f_{\rm df}\le 1$, $\Theta_{\rm orbit}$ is a function of the orbital parameters,
$\tau_{\rm dyn}\equiv R_{\rm v}/V_{\rm v}$ is the dynamical timescale of the halo,
${\rm ln}(\Lambda_{\rm Coulomb})={\rm ln}(M/M_{\rm sat})$ is the Coulomb logarithm,
$M$ is the halo mass of the central
galaxy and $M_{\rm sat}$ is the mass of the satellite, including the mass of the DM halo in which the galaxy
was formed. Note that the parameter $f_{\rm df}$ provides the flexibility to choose 
to merge galaxies right after the subhalos disappear from the catalogs ($f_{\rm df} = 0$).
In addition, this parameter may be $>0$ but $<1$ if the subhalos tend to disappear deep into the potential well when their number of particles 
drop below the threshold imposed by the subhalo finder (which is the ideal case). In this case, the dynamical friction timescales 
should be a lot shorter than Eq.~\ref{eqtaum} with $f_{\rm df}=1$, as that was originally calculated for subhalos at the virial radius. 
\citet{Simha17} argued that the true dynamical friction timescale for orphan satellites would be smaller than Eq.~\ref{eqtaum} with $f_{\rm df}=1$ 
if satellites are modelled as we do in \shark\ (i.e. by not allowing type 1 satellites to merge). Thus, 
we leave $f_{\rm df}$ to vary freely, though we recommend to adopt values between $0$ and $1$.

The orbital function, $\Theta_{\rm orbit}$ is defined as 

\begin{equation}
\Theta_{\rm orbit}=\left[\frac{J}{J_{\rm c}(E)}\right]^{0.78}\, \left[\frac{r_{\rm c}(E)}{R_{\rm v}}\right]^{2},
\label{Eq:merger:Lacey}
\end{equation}

\noindent where $J$ is the initial angular momentum and $E$ is the energy of the satellite's orbit, and
$J_{\rm c}(E)$ and $r_{\rm c}(E)$ are the angular momentum and radius of a circular orbit with the same energy as
that of the satellite, respectively. Thus, the circularity of the orbit corresponds to $J/J_{\rm c}(E)$. 
The dependence of $\Theta_{\rm orbit}$ on $J$ in
Eq~\ref{Eq:merger:Lacey} is a fit to numerical simulations \citep{Lacey93}.
The function
$\Theta_{\rm orbit}$ is well described by a log normal distribution with
median value $\langle{\rm log}_{10} \Theta_{\rm orbit} \rangle=-0.14$ and
dispersion $\langle({\rm log}_{10} \Theta_{\rm orbit}-\langle{\rm log}_{10} \Theta_{\rm orbit} \rangle)^2 \rangle^{1/2}=0.26$, and its
value is not correlated with satellite galaxy properties. Therefore, for each satellite, the value of $\Theta_{\rm orbit}$
is randomly chosen from the above distribution. 

If $\tau_{\rm merge}<t-t_{\rm orphan}$ for a satellite galaxy, with $t_{\rm orphan}$ being the time the galaxy became 
orphan, we proceed to merge it with the central galaxy at a time $t$.
If the total mass of gas plus stars of the primary (largest) and secondary galaxies involved in a merger are
$M_{\rm p}=M_{\rm cold,p}+M_{\star,\rm p}$ and
$M_{\rm s}=M_{\rm cold,s}+M_{\star,\rm s}$, the outcome of the galaxy merger depends on the galaxy mass ratio,
$M_{\rm s}/M_{\rm p}$,  and
the fraction of gas in the primary galaxy, $M_{\rm cold,p}/M_{\rm p}$ as:

\begin{itemize}
\item $M_{\rm s}/M_{\rm p}>f_{\rm ellip}$ drives a major merger. In this case all the stars
present are rearranged into an spheroid. In addition, any cold gas in the merging system is assumed to undergo
a SB and the stars formed are added to the spheroid component. We adopt $f_{\rm ellip}=0.3$, which is within the range found in simulations (e.g. \citealt{Baugh96}).
\item $f_{\rm burst}<M_{\rm s}/M_{\rm p}\le f_{\rm ellip}$ drives minor mergers. In this case all the stars
in the secondary galaxy are accreted onto the primary galaxy spheroid, leaving intact the stellar disk
of the primary. In minor mergers the triggering of a SB depends on the cold gas content of the
primary galaxy. If the minor merger has $M_{\rm cold,p}/M_{\rm p}>f_{\rm gas,burst}$, a SB
is driven. The perturbations introduced by the secondary galaxy suffice to drive all the cold gas from
both galaxies to the new spheroid, where it produces a SB. If $M_{\rm cold,p}/M_{\rm p}<f_{\rm gas,burst}$, 
the gas mass of the secondary is accreted by the disk of the primary.
\item $M_{\rm s}/M_{\rm p}\le f_{\rm burst}$ results in the primary disk being unperturbed. As before, the stars accreted
from the secondary galaxy are added to the spheroid, but the overall gas component (from both galaxies) stays in the disk,
along with the stellar disk of the primary. 
\end{itemize}

In the time between satellites becoming orphans and merging onto the central galaxy, we have no self-consistent information 
on their orbits. This is a problem if we want to study clustering and if we want to build lightcones from the \shark\ outputs. 
In order to mitigate this issue, we position orphan satellites randomly in a 3D~NFW halo with the properties of the host halo 
where the orphan galaxy lives. We do this following the analytic quantile function for an NFW profile 
described in \citet{Robotham18}. For velocities, we assign them by using the virial
theorem in an NFW halo and assuming isotropic velocities. 

\subsubsection{Disk instabilities}\label{sec:diskins}

If the disk becomes sufficiently massive that its self-gravity is dominant, then it is unstable to small perturbations
by minor satellites or DM substructures. The criterion for instability was described in \citet{Ostriker73} and \citet{Efstathiou82} as, 

\begin{equation}
\epsilon=\frac{V_{\rm circ}}{\sqrt{1.68\,G\, M_{\rm disk}/r_{\rm disk}}}.
\label{DisKins}
\end{equation}

\noindent Here, $V_{\rm circ}$ is the maximum circular velocity,
$r_{\rm disk}$ is the half-mass disk radius and
$M_{\rm disk}$ is the disk mass (gas plus stars). The numerical factor $1.68$ converts the disk half-mass radius into a scalelength, assuming an exponential profiles. 
If $\epsilon<\epsilon_{\rm disk}$ the disk is considered to be unstable. 
In \shark, gas and stellar disks can have different sizes, and thus to evaluate Eq.~\ref{DisKins} we compute a mass-weighted $r_{\rm disk}$ between the two disk 
components. 
Following \citet{Lacey15}, we assume that in the case of unstable disks, stars and gas in the disk are accreted onto the spheroid and the gas inflow drives a SB.
Several detailed hydrodynamical simulations have been performed to study the effect of these ``violent disk instabilities''
 and have shown that they can form galaxies with steep stellar profiles, similar to early-type galaxies (e.g. \citealt{Ceverino15}, \citealt{Zolotov15}). 
Large, cosmological hydrodynamical simulations show that this path of formation is present in their compact elliptical galaxies, but it is not dominant compared 
to galaxy mergers (e.g. \citealt{Wellons15}; \citealt{Clauwens18}; \citealt{Lagos18}). 

Simple theoretical arguments indicate that $\epsilon_{\rm disk}$ should be of the order of unity \citep{Efstathiou82}. However, because the process of bar creation and thickening of the disk  
can be a very complex phenomenon \citep{Bournaud11}, we treat $\epsilon_{\rm disk}$ as a free parameter in \shark\ rather than forcing it to be $\equiv 1$. 
In addition, $\epsilon_{\rm disk}$ is not expected to be the same for stars and gas \citep{Romeo11}.

\subsubsection{Photoionisation feedback}\label{sec:reio}

At very early epochs in the Universe, right after the epoch of cosmological recombination, the background light 
consists of the black body radiation from the CMB. At this stage the universe remains neutral until the first generation 
of stars, galaxies and quasars start emitting photons and ionising the medium around them. Eventually, the ionised 
pockets grow and merge. This corresponds to the reionization epoch of the Universe (e.g. \citealt{Barkana01}). 
The large ionising radiation density significantly affects small halos, 
 maintaining the baryons at temperatures hotter than the virial temperature, and thus suppressing               
cooling. In \shark, we implement two models  for photo-ionisation feedback. 
The first one assumes that no gas is allowed to cool in haloes with a
circular velocity below $V_{\rm crit}$ at redshifts below $z_{\rm
reion}$ \citep{Benson03}. We adopt $V_{\rm crit}=30\,\rm km \,s^{-1}$ and $z_{\rm reion}=10$ following \citet{Okamoto08}.
This is the model adopted in the {GALFORM} semi-analytic model \citep{Lacey15}, and as such we term it 
the {\tt Lacey16} model for reionisation.

A second, more sophisticated model, follows the results of the one-dimensional collapse simulations 
of \citet{Sobacchi13}, which suggest a threshold velocity parameter that is redshift dependant.
\citet{Sobacchi13} provide a parametric form for the halos that are affected by photo-ionisation 
that is redshift dependant in terms of halo mass. \citet{Kim15} 
adapted the Sobacchi \& Mesinger parametric form to depend instead on the halo's $V_{\rm circ}$ 
 by using the spherical collapse model of \citet{Cole96}, which predicts $M_{\rm halo}\propto V^3_{\rm circ}$. 
 Thus, halos with circular velocities below $v_{\rm thres}(z)$ are not allowed to cool down 
 their halo gas, with $v_{\rm thres}(z)$ being:

\begin{equation}
v_{\rm thresh}(z) = v_{\rm cut}\,(1+z)^{\alpha_{\rm v}}\,\left[1 - \left(\frac{1+z}{1+z_{\rm cut}} \right)^2\right]^{2.5/3}.
\label{reioeq}
\end{equation}
 
\noindent Here, $v_{\rm cut}$, $z_{\rm cut}$ and $\alpha_{\rm v}$ are free parameters that are constrained by the
 \citet{Sobacchi13} simulation. In this model $z_{\rm cut}$ corresponds to the redshift of UV background exposure of galaxies, 
 which, as \citet{Kim15}, we fix to a single value for simplicity. In principle we leave $v_{\rm cut}$, $z_{\rm cut}$ and $\alpha_{\rm v}$ 
 to vary freely but suggest the user to adopt the
 values in \citet{Sobacchi13}, $v_{\rm cut}\approx 30\rm \, km\,s^{-1}$, $z_{\rm cut}\approx 10$ and  $\alpha_{\rm v}=-0.2$.
 Note that \citet{Kim15}, using this model in the GALFORM semi-analytic model, adopted 
  $v_{\rm cut}\approx 50\rm \, km\,s^{-1}$ and $\alpha_{\rm v}=-0.8$. We termed this model the {\tt Sobacchi13} model.

\subsubsection{Black hole growth and AGN feedback}\label{sec:agnfeedback}

In \shark, DM halos more massive than $m_{\rm halo,seed}$ are seeded with supermassive black holes (SMBHs) of mass $m_{\rm seed}$. 
These two mass scales are treated as free parameters. 

SMBHs can then grow via three channels: (i) BH-BH mergers, (ii) accretion during SBs and (iii) accretion in the hot-halo regime. 
Chanel (i) happens when there are galaxy mergers and both galaxies host a SMBH. In that case, the resulting SMBH is simply the addition of the 
two SMBH masses. Chanel (ii) can happen both during galaxy mergers and during violent disk instabilities. In that case BHs grow following the phenomenological 
description of \citet{Kauffmann00}, and increase their mass by 

\begin{equation}
\delta\,m_{\rm BH,sb} = f_{\rm smbh}\,\frac{m_{\rm gas}}{1+(v_{\rm smbh}/V_{\rm vir})^2},
\label{bhgrow_sbs}
\end{equation}

\noindent where $m_{\rm gas}$ and $V_{\rm vir}$ are the cold gas mass reservoir of the starburst and the virial velocity, respectively.
$f_{\rm smbh}$ and $v_{\rm smbh}$ are free parameters. The former parameter is the main responsible for controlling the normalization of the 
BH-bulge mass relation (see $\S$~\ref{sec:BH}). The dependence on $V_{\rm vir}$ indicates that the rate of accretion is regulated by the binding energy of the system. 
If the binding energy is small, less gas makes onto the central region of the galaxy where the SMBH resides. We can estimate a typical SMBH 
accretion rate during SBs from Eq~\ref{bhgrow_sbs} and assuming that a typical accretion timescale is of the order of the 
bulge dynamical timescale, $\tau_{\rm acc,sb}=e_{\rm sb}\,r_{\rm bulge}/v_{\rm bulge}$, where $e_{\rm sb}$ is an e-folding parameter of the order of 
unity. The accretion rate during SBs is therefore,

\begin{equation}
\dot{m}_{\rm BH,sb} = \frac{\delta,m_{\rm BH,sb}}{\tau_{\rm acc,sb}}.
\end{equation}

For the BH growth in the hot halo regime, also termed `radio-mode accretion' by \citet{Croton06}, we implement two models, the {\tt Croton16} \citep{Croton16} and {\tt Bower06} \citep{Bower06} models. 
Below we describe these two models.

\begin{itemize}
\item The {\tt Croton16} model. Here, we assume a Bondi-Hoyle \citep{Bondi52} like 
accretion mode,

\begin{equation}
\dot{m}_{\rm BH,hh} = 2.5\,\pi\,G^2\,\frac{m^2_{\rm BH}\,\rho_{0}}{c^3_{\rm s}},
\end{equation}

\noindent where $c_{\rm s}$ and $\rho_{0}$ are the sound speed and average density of the hot gas 
in the halo that will rain down to the SMBH. 
We approximate $c_{\rm s}\approx V_{\rm vir}$. 
For $\rho_0$, we follow \citet{Croton06} and calculate it from equating the sound travel time 
across a shell of diameter twice the Bondi radius to the local cooling time. This is also termed 
``maximal cooling flow'' by \citet{Nulsen00}. This leads to 

\begin{equation}
\dot{m}_{\rm BH,hh} = \kappa_{\rm R}\,\frac{15}{16} \pi\,G\,\mu\,m_{\rm p}\,\frac{\kappa_{\rm B}\,T_{\rm vir}}{\Lambda}\,m_{\rm BH}.
\label{Eq.RM}
\end{equation}

\noindent $\kappa_{\rm R}$ is a free parameter that was introduced by \citet{Croton06} to counteract the approximations 
used to derive the accretion rate. $\kappa_{\rm B}$ and $\Lambda$ are the Boltzmann's constant and 
the cooling function that depends on $T_{\rm vir}$ and the hot gas metallicity.
With this accretion rate we can estimate a BH luminosity as $L_{\rm BH} = \eta\,\dot{m}_{\rm BH,hh}\,c^2$, 
where $\eta$ is the luminosity efficiency, which strictly depends on the BH spin \citep{Lagos09b}, but here is assumed to be $=0.1$ (approximately 
corresponding to a spin of $0.1$). $c$ is the speed of light.

We use  $L_{\rm BH}$ to estimate how much heating the BH provides and adjust the cooling rate in response to this source 
of energy. The heating rate is calculated as 

\begin{equation}
\dot{m}_{\rm heat} = \frac{L_{\rm BH}}{0.5\,V^2_{\rm vir}}.
\end{equation}

\noindent Based on $\dot{m}_{\rm heat}$ we then calculate the radius within which the energy injected by the AGN equals that of the 
energy of the halo gas internal to that radius that would be lost if the gas were to cool \citep{Croton16}. This heating radius, $r_{\rm heat}$
is estimated as:

\begin{equation}
r_{\rm heat} = \frac{\dot{m}_{\rm heat}}{\dot{m}_{\rm cool}}\,r_{\rm cool}.
\end{equation}

\noindent We modify the cooling rate in response to this heating source as 

\begin{equation}
\dot{m}^{\prime}_{\rm cool} = \left(1-\frac{r_{\rm heat}}{r_{\rm cool}}\right)\,\dot{m}_{\rm cool}.
\end{equation}

\noindent If $r_{\rm heat}/r_{\rm cool} > \alpha_{\rm cool}$ then the cooling flow is completely shut down, i.e. $\dot{m}^{\prime}_{\rm cool} = 0$. 
Here, $\alpha_{\rm cool}\sim 1$ is an adjustable parameter close to unity. Note that values $>1$ would give $\dot{m}^{\prime}_{\rm cool}<0$, which 
in the code we set to $0$, giving the same results as 
$\alpha_{\rm cool}=1$. Thus, in this model, it only makes sense to adopt values $\lesssim 1$.  
If the formation of a hot corona was perfectly modelled, then it would only make sense to adopt 
$\alpha_{\rm cool}=1$. However, the several simplifications made in SAMs regarding the halo gas density and how metal enrichment happens warrants 
some flexibility to be allowed in the exact transition between the rapid cooling and hot halo regimes.

In this model, the heating radius is forced to only move outwards. This is due to the heating due to radio jets retain the memory 
of past heating episodes. 

\item The {\tt Bower06} model. Here, AGN feedback is assumed to be effective only in halos undergoing
quasi-hydrostatic cooling. In this situation,
mechanical energy input by the AGN is expected to stabilise the flow and regulate the rate at which the gas cools.
 Whether or not a halo is undergoing quasi-hydrostatic cooling depends on the cooling and free-fall times:
the halo is in this regime if $t_{\rm cool}(r_{\rm cool})>\alpha^{-1}_{\rm cool}\, t_{\rm ff}(r_{\rm cool})$, where
$t_{\rm ff}$ is the free fall time at $r_{\rm cool}$, and $\alpha_{\rm cool}\sim 1$ is an adjustable parameter close to unity.
During radiative cooling, BHs have a growth rate given by
$\dot{m}_{\rm BH}=L_{\rm cool}/0.2\, c^2$, where $L_{\rm cool}$ is the cooling luminosity and $c$ is the speed of light.

AGNs are assumed to be able to quench gas cooling only if the available AGN power is comparable to the cooling luminosity,
$L_{\rm cool}<\epsilon_{\rm SMBH}\, L_{\rm Edd}$, where $L_{\rm Edd}$ is the Eddington luminosity
the central black hole and $\epsilon_{\rm SMBH}\sim 1$ is a free parameter. Note that because this model explicitly compares the 
dynamical and cooling timescales, 
it is not compatible with the {\tt Croton06} cooling model (as by definition $t_{\rm cool}(r_{\rm cool})/t_{\rm ff}(r_{\rm cool}) = 1$), 
and it can only use the {\tt Benson10} cooling model.  
\end{itemize}

\subsubsection{Environmental effects}\label{sec:env}

In \shark\ we have two models for the treatment of the halo gas in the case of satellite galaxies.
In the model of `instantaneous ram pressure stripping' or `strangulation' \citep{Lagos14}, 
we assume that as soon as they become satellites, 
their halo gas is instantaneously stripped and transferred to the hot gas of the central. 
Thus, gas can only accrete onto the {\em central} galaxy in a
halo, and not onto any {\em satellite} galaxies. 
However, the cold gas 
in the disks of galaxies is not stripped. 
{If the satellite galaxy continues to form stars, the ejected gas due to outflows is transferred to the 
halo gas of the central at the beginning of the next snapshot and before cooling rates are calculated.}
Another option is to allow satellite galaxies to retain their hot halo and continue to use it up 
until it exhausts. This model corresponds to the configuration parameter {\tt stripping} set to false. 
In future versions, we will be implementing 
more sophisticated environmental models.

\subsubsection{Disk and bulge sizes}\label{sizes}
\begin{table*}
\setlength\tabcolsep{2pt}
\centering\footnotesize
\caption{\shark\ models and parameters. Here we show the names these variables have in the configuration file, the associated name of the variables in the equations 
presented in $\S$~\ref{ModelDescription} and the physical processes in which they appear. We show the values chosen for our 
default \shark\ model in parenthesis in the middle column.}
\begin{tabular}{@{\extracolsep{\fill}}l|cc|p{0.45\textwidth}}
\hline
\hline
    Parameter & suggested value range & variable/equation\\
\hline
    halo properties and angular momentum & & \\
\hline
	{\tt halo$_{-}$profile} & {\tt nfw} & Eq.~\ref{EqNFW} \\
	{\tt lambda$_{-}$random} & $\,\,\,0$ (Eq.~\ref{jhalo}) or $1$ (random distribution)\, ($1$)\\
	{\tt size$_{-}$model} & {\tt Mo98} & Size calculation\\
\hline
    gas cooling & & \\
\hline
	{\tt lambdamodel} & {\tt cloudy} or {\tt sutherland} ({\tt cloudy})& $\Lambda$ in Eq.~\ref{tcool}  \\ 
	{\tt model} & {\tt Croton06} or {\tt Benson10} ({\tt Croton06}) & Described in $\S$~\ref{sec:cooling} \\
\hline
    gas accretion &  & \\
\hline
	{\tt pre$_{-}$enrich$_{-}$z} & $>0-10^{-5}$ ($10^{-7}$) & $Z_{\rm min}$ in $\S$~\ref{DMhalosAcc}\\
\hline
    chemical enrichment &  & \\
\hline
    {\tt recycle} & $0.4588$ for a Chabrier IMF & $R$ in Eq.~\ref{Eq:ejec} \\
    {\tt yield}  & $0.02908$ for a Chabrier IMF & $p$ in Eq.~\ref{Eq:yield}\\
    {\tt zsun} & $0.018$ & adopted solar metallicity \\
\hline
    stellar feedback  & &  \\
\hline
        {\tt model} & {\tt Muratov15}, {\tt Lagos13}, {\tt Lagos13Trunc}, {\tt Lacey16}, & $\S$~\ref{sec:stellarfeed}\\
                & {\tt Lacey16RedDep} or {\tt Guo11} ({\tt Lagos13}) & \\
	{\tt v$_{-}$sn} & $50-500\rm \, km\, s^{-1}$ ($110\rm \, km\, s^{-1}$)& $v_{\rm hot}$ in Eqs.~\ref{SNGal}-\ref{SNLAG13} \\
	{\tt beta$_{-}$disk} & $0.5-5$ ($4.5$) & $\beta$ in Eqs.~\ref{SNGal}-\ref{SNLAG13}\\
	{\tt redshift$_{-}$power} & $-0.5$ to $1.5$ ($0.12$) & $\rm z_{\rm P}$ in Eqs.~\ref{SNFIRE} and \ref{vel_power}\\
	{\tt eps$_{-}$halo} & $0.1-10$ ($2$) &  $\epsilon_{\rm halo}$ in Eq.~\ref{eq:epshalo}\\ 
	{\tt eps$_{-}$disk} & $1-10$ ($1$) &  $\epsilon_{\rm disk}$ in Eq.~\ref{SNLGAL}\\
\hline
    star formation & &  \\
\hline
	{\tt model} & {\tt BR06}, {\tt GD14}, {\tt KMT09} or {\tt K13} ({\tt BR06}) & in $\S$~\ref{sec:sf}\\
	{\tt nu$_{-}$sf} & $0.25-1.25\,\rm Gyr^{-1}$ ($1\,\rm Gyr^{-1}$)& $\nu_{\rm SF}$ in Eq.~\ref{SFLaw}\\
	{\tt boost$_{-}$starburst} & $1-10$ ($10$) & $\eta_{burst}$ in $\S$~\ref{sec:sfb}\\
	{\tt sigma$_{-}$hi$_{-}$crit} & $0.01-0.1\,\rm M_{\odot}\,pc^{-2}$ ($0.1\,\rm M_{\odot}\,pc^{-2}$) & $\Sigma_{\rm thresh}$ in $\S$~\ref{sec:sf}\\
	{\tt po} & $10,000-45,000\,\rm K\,cm^{-3}$ ($34,673,\rm K\,cm^{-3}$) & $P_0$ in Eq.~\ref{eq:at_mol}; only relevant for {\tt BR06}\\
	{\tt beta$_{-}$press} & $0.7-1$ ($0.92$) & $\alpha_{\rm P}$ in Eq.~\ref{eq:at_mol}; only relevant for {\tt BR06}\\
	{\tt gas$_{-}$velocity$_{-}$dispersion} & $7-10\,\rm km\,s^{-1}$ ($10\,\rm km\,s^{-1}$) & $\sigma_{\rm gas}$ in Eq.~\ref{eq:press};  only relevant for {\tt BR06} and {\tt K13}\\
	{\tt clump$_{-}$factor$_{-}$kmt09} & $1-10$ ($5$) & only relevant for {\tt KMT09} and {\tt K13}\\
\hline
   reincorporation & & \\
\hline
	{\tt tau$_{-}$reinc} & $1-30\,\rm Gyr$ ($25\,\rm Gyr$) & $\tau_{\rm reinc}$ in Eq.~\ref{Eqreinc}\\ 
	{\tt mhalo$_{-}$norm} & $10^9-10^{11}\,\rm M_{\odot}$ ($10^{10}\,\rm M_{\odot}$) &  $M_{\rm norm}$ in Eq.~\ref{Eqreinc}\\
	{\tt halo$_{-}$mass$_{-}$power} & $-2$ to $0$ ($-1$) &  $\gamma$ in Eq.~\ref{Eqreinc}\\
\hline
   reionisation & & \\
\hline
	{\tt model} & {\tt Lacey16} or {\tt Sobacchi13} ({\tt Sobacchi13}) & in $\S$~\ref{sec:reio}\\
	{\tt zcut} & $7-11$ ($10$) & in $\S$~\ref{sec:reio}\\
	{\tt vcut} & $20-50\rm \,km\,s^{-1}$ ($35\rm \,km\,s^{-1}$) & in $\S$~\ref{sec:reio}\\
	{\tt alpha$_{-}$v} & $-1$ to $0$ ($-0.2$) & only relevant for {\tt Sobacchi13} model, Eq.~\ref{reioeq}\\
\hline
   AGN feedback \& BH growth & & \\
\hline
	{\tt model} & {\tt Bower06} or {\tt Croton16} ({\tt Croton16}) & AGN feedback model $\S$~\ref{sec:agnfeedback}\\
	{\tt mseed} & $0-10^5\,\rm M_{\odot}/h$ ($10^4\,\rm M_{\odot}/h$) & $m_{\rm seed}$ in $\S$~\ref{sec:agnfeedback}\\
	{\tt mhalo$_{-}$seed} & $0-10^{11}\,\rm M_{\odot}/h$ ($10^{10}\,\rm M_{\odot}/h$) & $m_{\rm halo,seed}$ in $\S$~\ref{sec:agnfeedback}\\
	{\tt f$_{-}$smbh} & $10^{-5}-10^{-2}$ ($8\times 10^{-3}$) & $f_{\rm smbh}$ in Eq.~\ref{bhgrow_sbs}\\
	{\tt v$_{-}$smbh} & $100-1000\,\rm km\,s^{-1}$ ($400\,\rm km\,s^{-1}$) & $v_{\rm smbh}$ in Eq.~\ref{bhgrow_sbs}\\
        {\tt tau$_{-}$fold} & $0.5-10$ ($1$) & $e_{\rm sb}$ in $\S$~\ref{sec:agnfeedback}\\
	{\tt alpha$_{-}$cool} & $0.3-3$ ($0.5$) & used in both {\tt Bower06} and {\tt Croton16}; $\S$~\ref{sec:agnfeedback}\\
	{\tt accretion$_{-}$eff$_{-}$cooling} & $0.07-0.4$ ($0.1$) & $\eta$ in $\S$~\ref{sec:agnfeedback}; only relevant for {\tt Croton16} \\
	{\tt kappa$_{-}$agn} & $10^{-5}-10$ ({\bf $3\times 10^{-3}$}) & $\kappa_{\rm r}$ in Eq.~\ref{Eq.RM}; only relevant for {\tt Croton16}\\
	{\tt f$_{-}$edd} & $0.0001-0.1$ ($0.01$) & $\S$~\ref{sec:agnfeedback}; only relevant for {\tt Bower06}\\
\hline
\end{tabular}
\label{tab:parameters}
\end{table*}

\begin{table*}
\setlength\tabcolsep{2pt}
\centering\footnotesize
\caption{Continuation of Table~\ref{tab:parameters}.}
\begin{tabular}{@{\extracolsep{\fill}}l|cc|p{0.45\textwidth}}
\hline
\hline
    Parameter & suggested value range & variable/equation\\
\hline
	galaxy mergers and bulge size& & \\
\hline
   {\tt major$_{-}$merger$_{-}$ratio} & $0.2-0.4$ ($0.3$) & $f_{\rm ellip}$ in $\S$~\ref{sec:mergers}\\ 
   {\tt minor$_{-}$merger$_{-}$burst$_{-}$ratio} & $0.05-0.2$ ($0.1$) & $f_{\rm burst}$ in $\S$~\ref{sec:mergers}\\
   {\tt gas$_{-}$fraction$_{-}$burst$_{-}$ratio} & $0-1$ ($0.3$) & $f_{\rm gas, burst}$ in $\S$~\ref{sec:mergers}\\
   {\tt f$_{-}$orbit} & $0.5-2$ ($1$) & $f_{\rm orbit}$ in Eq.~\ref{size:merger}\\
   {\tt cgal} & $0.45-0.5$ ($0.49$) & $c_{\rm gal}$ in Eq.~\ref{size:merger}\\ 
   {\tt tau$_{-}$delay} & $0-1$ ($0.1$) & $f_{\rm df}$ in Eq.~\ref{size:merger}\\
   {\tt fgas$_{-}$dissipation} & {\bf $0-1.5$ ($1$)} & $R_{\rm 0}$ in Eq.~\ref{size:merger2}; set to $=0$ if no dissipation is considered.\\
   {\tt merger$_{-}$ratio$_{-}$dissipation} & $0-0.3$ ($0.3$) & $m_{\rm r,diss}$ in $\S$~\ref{sec:mergers}\\
\hline
   disk instabilities and bulge size & & \\
\hline
   {\tt stable} & $0-4$ ({\bf $0.8$}) & $\epsilon_{\rm disk}$ in Eq.~\ref{sec:diskins}\\
   {\tt fint} & $1-3$ ($2$) & $f_{\rm int}$ in Eq.~\ref{sizebulge_di}\\
\hline
   environment & & \\
\hline
   {\tt stripping} & {\tt true} or {\tt false} ({\tt true}) & $\S$~\ref{sec:env}\\
\hline
\end{tabular}
\label{tab:parameters2}
\end{table*}

To estimate the disk scale radii, $r_{\rm s}$, 
we follow the exchange of specific angular momentum between the cooling gas, gas and stellar disk. 
For the cooling gas we assume it has the same specific angular momentum of the DM halo,

\begin{eqnarray}
	j_{\rm cool} = \frac{J_{\rm h}}{M_{\rm halo}},
\end{eqnarray}

\noindent where $J_{\rm h}$ is calculated as in Eq.\ref{jhalo}. $j_{\rm cool}$ 
is then input in the set of ODEs that control the exchange of angular momentum 
(Eqs.~\ref{eqn:sff:j}-\ref{eqn:sff:jf}). The gaseous and stellar disks also 
 exchange angular momentum at a rate $\dot{J}_{\rm g,s}$. 
In its simplest form, 

\begin{equation}
	\dot{J}_{\rm g,s} = \psi \,j_{\rm gas}, 
        \label{AMexchange1}
\end{equation}

\noindent where $\psi$ is the instantaneous SFR and  
$j_{\rm gas}$ is the specific angular momentum of the gaseous disk.
This may, however, be changed for more sophisticated models, for example 
by considering that gas that forms stars tend to be the low specific angular momentum 
gas \citep{Mitchell18}. In future work, we explore this natural extension 
for \shark. In our standard model we adopt Eq.~\ref{AMexchange1}.
%
%
%

The half-mass gas and stellar disk sizes are then calculated as 
$r_{\rm gas} = f_{\rm norm}\, j_{\rm gas} / V_{\rm circ}$ and 
$r_{\star} = f_{\rm norm}\, j_{\star} / V_{\rm circ}$. Here, we set 
$f_{\rm norm} = 0.677$, following the relation between 
$r\,v_{\rm circ}$ and $j$ that \citet{Swinbank17} reported for the 
{\sc EAGLE} simulations. Note that the value of $f_{\rm norm}$ is slightly smaller than 
the idealized value ($0.835$) adopted by \citet{Guo11} and \citet{Zoldan18}.




For the case of SBs (driven by mergers and disk instabilities), angular momentum 
is not a well defined quantity, and thus we do not follow the explicit exchange of angular momentum between 
gas and stars as we do for disks, but assume that they are always well mixed if a SB is triggered. 
We calculate a pseudo specific angular momentum for bulges following \citet{Cole00}, in the form 
$j_{\rm B} = r_{\rm B} \,v(r_{\rm B})$, where $r_{\rm B}$ is the half-mass radius of the bulge (described below) and 
$v(r_{\rm B})$ is the circular velocity at $r_{\rm B}$. 

In the case of galaxy major mergers, the resulting radius of the bulge is calculated from the virial theorem as in \citet{Cole00},

\begin{eqnarray}
 \frac{(M_{\rm s}+M_{\rm p})^2}{r_{\rm new}}&=&\frac{M_{\rm s}}{r_{\rm s}}+\frac{M_{\rm p}}{r_{\rm p}}+\frac{f_{\rm orbit}}{c_{\rm gal}}\, \frac{M_{\rm s}\, M_{\rm p}}{r_{s}+r_{\rm p}},
\label{size:merger}
\end{eqnarray}

\noindent where $c_{\rm gal}$ and $f_{\rm orbit}$ are estimated from the binding energy of each of the galaxies and
the mutual orbital energy, respectively, $M_{\rm s}$ and $M_{\rm p}$ are the secondary and primary galaxy masses, respectively, 
and $r_{\rm s}$ and $r_{\rm p}$ are the half-baryon mass radii of the secondary and primary galaxies, respectively. 
In the case of the secondary, because they have had their host subhalo stripped, we only consider the baryon mass, while in the 
case of the primary we also include the DM mass that is enclosed within $r_{\rm p}$. The latter is done because 
during a merger the DM the inner parts of galaxies is expected to have similar dynamics than the stars. With this in consideration 
we define $M_{\rm p} = M_{\rm p,bar} + 2\,M_{\rm halo}(r_{\rm p})$, in which the factor $2$ implicitly assumes 
that the DM has the same spatial distribution as the baryons within $r_{\rm p}$. 
A value of $c_{\rm gal}=0.5$ is adopted, which is valid for both the
exponential and the $r^{1/4}$ profiles (i.e. $c_{\rm gal}$ is very weakly dependent on the density profile), 
and $f_{\rm orbit}=1$, which corresponds to
the orbital energy of two point masses moving in a circular orbit with separation $r_{\rm p}+r_{\rm s}$.

For minor mergers we replace $M_{\rm p}$ for the mass of the central galaxy that will end up in the bulge following the merger, 
and $r_{\rm p}$ for an effective half-mass radius calculated from mass weighting the sizes of all the baryon components 
of the central that will end up in the bulge. The latter means that in the cases of minor mergers that trigger a SB, 
$M_{\rm p}$ will include the bulge mass and disk gas mass, and $r_{\rm p}$ is an effective half-mass radius 
including bulge and the gas disk. 

In \shark\ we also include the merger dissipation model suggested by \citet{Hopkins09}. \citet{Hopkins09} presented a 
suite of binary merger simulations with mass ratios above $1:6$, adopting different initial gas fractions. Hopkins et al. 
found that the sizes of the merger remnants were smaller than Eq.~\ref{size:merger} due to dissipation effects that 
are increasingly more important in gas rich mergers. More recent cosmological hydrodynamical simulations show 
this effect very clearly, as gas very efficiently infalls to the galaxy centre in gas rich mergers \citep{Lagos17}. 
\citet{Hopkins09} suggest to shrink the sizes of the merger remnants following

\begin{eqnarray}
	r^{\prime}_{\rm new} = \frac{r_{\rm new}}{1+\frac{R_{\rm gas}}{R_{\rm 0}}},
\label{size:merger2}
\end{eqnarray}

\noindent where $r_{\rm new}$ is the radius calculated assuming no dissipation (Eq.~\ref{size:merger}), 
$R_{\rm gas}=M_{\rm cold}/M_{\star}$, 
$M_{\rm cold}$ and $M_{\star}$ are the total ISM and stellar mass of the resulting merger remnant, and $R_{\rm 0}\approx 0.3$ as shown 
in \citet{Hopkins09}. If the user sets $R_{\rm 0} \equiv 0$ we assume no dissipation takes place. 
Note that because the simulation experiments of  \citet{Hopkins09} were focused on major mergers, we include an additional parameter, 
$m_{\rm r,diss}$, which is the mass ratio of the merger above which we trigger the dissipation calculation. 

In the case of disk instabilities, we follow a similar procedure as for galaxy mergers but using as input system the galaxy disk and 
bulge of the galaxy before the disk instability, with masses and radii of $M_{\rm disk}$, $M_{\rm bulge}$, $r_{\rm disk}$ and $r_{\rm bulge}$, respectively. 
Note that masses here include both stars and gas, and radii are calculated as the mass weighted average stellar plus gas radii.
The resulting galaxy is a new spheroid containing all the mass of the disk plus bulge.

\begin{eqnarray}
 \frac{(M_{\rm disk}+M_{\rm bulge})^2}{r_{\rm new}}&=&c_{\rm disk}\,\frac{M_{\rm disk}}{r_{\rm disk}}+c_{\rm bulge}\,\frac{M_{\rm bulge}}{r_{\rm bulge}}+\nonumber\\
&&f_{\rm int}\, \frac{M_{\rm disk}\, M_{\rm bulge}}{r_{\rm disk}+r_{\rm bulge}}.
\label{sizebulge_di}
\end{eqnarray}
 
\noindent Here, $c_{\rm disk}$ and $c_{\rm bulge}$ have the same meaning as $c_{\rm gal}$. 
The last term represents the gravitational interaction energy of the disk and bulge. According to \citet{Lacey15}, 
$f_{\rm int}\approx 2$ is a good approximation for a large range of $r_{\rm disk}$ and $r_{\rm bulge}$.

\subsubsection{Evolving galaxies: the interplay between physical processes}\label{sec:odes}

The SF activity in \shark\ is regulated by three channels:
(i) accretion of gas which cools from the hot gas halo onto the disk,
(ii) SF from the cold gas and, (iii) reheating
and ejection of gas due to stellar feedback. These channels modify the
mass and metallicity of each of the baryonic components: stellar
mass, $M_{\star}$, cold gas mass, $M_{\rm cold}$, hot halo gas mass,
$M_{\rm hot}$, the ejected gas reservoir, $M_{\rm ejec}$, and their respective masses in metals, $M_{\star}^Z$,
$M_{\rm cold}^Z$, $M_{\rm hot}^Z$, $M_{\rm ejec}^Z$. The system of equations
relating these quantities is:

\begin{eqnarray}
\dot  M_{\star{\hphantom{col}}}     &=&  (1-R) \psi  \label{eqn:sff} \\
\dot  M_{\rm cold}^{\hphantom{Z}}  &=& \crate - (1-R+\beta) \psi \\
\dot  M_{\rm cold,halo{\hphantom{l}}}   &=& - \crate \\
\dot  M_{\rm hot,halo{\hphantom{l}}}   &=& \dot{m}_{\rm outflow} - \dot{m}_{\rm ejected} \\
\dot  M_{\rm ejec{\hphantom{l}}}   &=& \dot{m}_{\rm ejected} \\
\dot  M_{\star{\hphantom{olj}}}^Z   &=& (1-R)  Z_{\rm cold} \psi \\
\dot  M_{\rm cold}^Z &=& \crate Z_{\rm cold,halo} + \nonumber\\
		   &&(p - (1+\beta-R)Z_{\rm cold}) \psi \\
\dot  M_{\rm cold,halo{\hphantom{l}}}^Z &=& -\crate Z_{\rm cold,halo}\\
\dot  M_{\rm hot,halo{\hphantom{l}}}^Z &=& (\dot{m}_{\rm outflow} - \dot{m}_{\rm ejected})\,Z_{\rm cold}\\
\dot  M_{\rm ejec{\hphantom{l}}}^Z &=& Z_{\rm cold}\,\dot{m}_{\rm ejected}.\label{eqn:sfflast}
\label{eqn:sflc}
\end{eqnarray}

\noindent where

\begin{eqnarray}
\beta\equiv \frac{\dot{m}_{\rm outflow}}{\psi},\,Z_{\rm cold}\equiv \frac{M_{\rm
cold}^Z}{M_{\rm cold}^{\hphantom{Z}}},\,Z_{\rm cold,halo}\equiv \frac{M^Z_{\rm cold,halo}}{M_{\rm cold,halo}}
\end{eqnarray}

\noindent are the mass loading, the metallicity of the cold
gas and the metallicity of the cold (ISM) gas in the halo (the one that is actively cooling), respectively.
In the set of Eqs.~\ref{eqn:sff}-\ref{eqn:sfflast}, $\psi$ denotes the instantaneous SFR,
$\crate$ the cooling rate, $p$ denotes the yield
(the fraction of mass converted
into stars that is returned to the ISM in the form of metals) and $R$ is
the fraction of mass recycled to the ISM (in the form
of stellar winds and SN explosions).
The expressions for $\dot M_{\rm cold,halo{\hphantom{l}}}$ and $\dot M_{\rm hot,halo{\hphantom{l}}}$ 
{assume that the cold halo gas is not affected by the outflowing gas from the galaxy, until the cooling rate is calculated again.}

Simultaneously to the mass and metal exchange, in the case of star formation in disks, 
we solve for the angular momentum exchange between 
these components:

\begin{eqnarray}
	\dot  J_{\star{\hphantom{col}}}     &=&  (1-R)\, \dot{J}_{\rm g,s}  \label{eqn:sff:j} \\
	\dot  J_{\rm cold}^{\hphantom{Z}}  &=& \crate\, j_{\rm cool}- (1-R+\beta)\, \dot{J}_{\rm g,s} \\
	\dot  J_{\rm cold,halo{\hphantom{l}}}   &=& - \crate\, j_{\rm cool}\\
	\dot  J_{\rm hot,halo{\hphantom{l}}}   &=& \dot{m}_{\rm outflow}\,  j_{\rm out} - \dot{m}_{\rm ejected}\, j_{\rm out}\\
	\dot  J_{\rm ejec{\hphantom{l}}}   &=& \dot{m}_{\rm ejected}\, j_{\rm out}. \label{eqn:sff:jf}
\label{eqn:sflc}
\end{eqnarray}

\noindent Here, $J\equiv |\vec{J}|$, and $\dot{J}_{\rm g,s}$ is as described in $\S$~\ref{sizes}.
In the case of the hot halo and ejected gas mass components, the angular momentum growth depends on the 
specific angular momentum of the outflowing gas. This in principle allows for 
outflows to affect the angular momentum of the disk in a differential form, which would be the case 
of the outflow rate being an explicit function of radius (as it has been proposed by detailed stellar feedback models, 
e.g. \citealt{Creasey12,Hopkins12,Lagos13}). In \shark, for {\sc v1.1}, we assume the simplest solution, which is 
$j_{\rm out} = \dot{J}_{\rm g,s}/\psi$, but the current code design allows the user to extend 
the model to assume different angular momentum loading functions. 
Eqs.~\ref{eqn:sff:j}~to~\ref{eqn:sff:jf} are not solved for starbursts, which can be triggered by galaxy mergers and disk instabilities. 
This is because in these cases and as described in $\S$~\ref{sizes}, angular momentum is not a well defined quantity, and in addition, during galaxy mergers, gas dissipation can significantly modify 
the sizes of galaxies, resulting in large losses of specific angular momentum.

\begin{figure*}
\begin{center}
\includegraphics[trim=2mm 13mm 5mm 25mm, clip,width=0.4\textwidth]{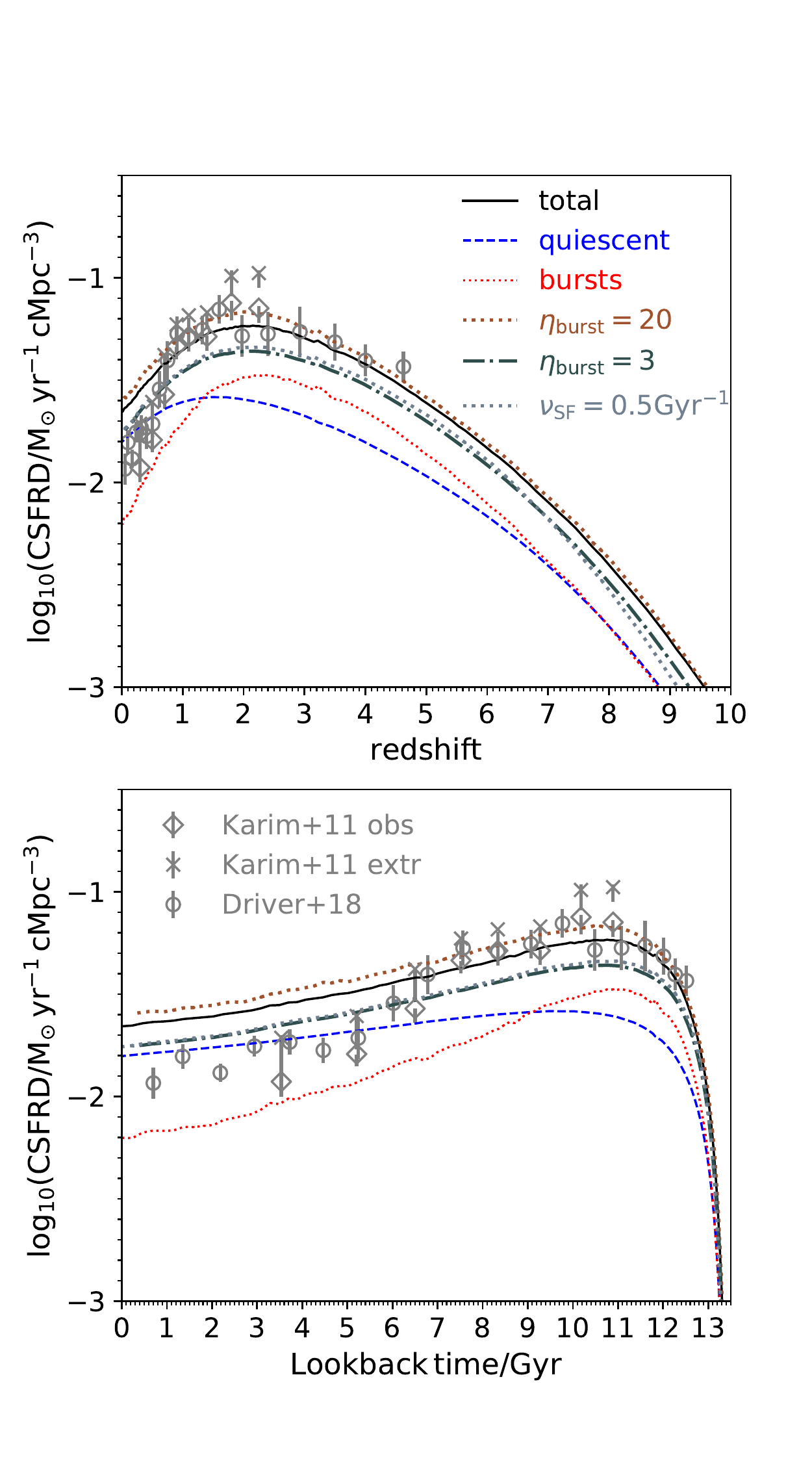}
\includegraphics[trim=2mm 13mm 5mm 25mm, clip,width=0.4\textwidth]{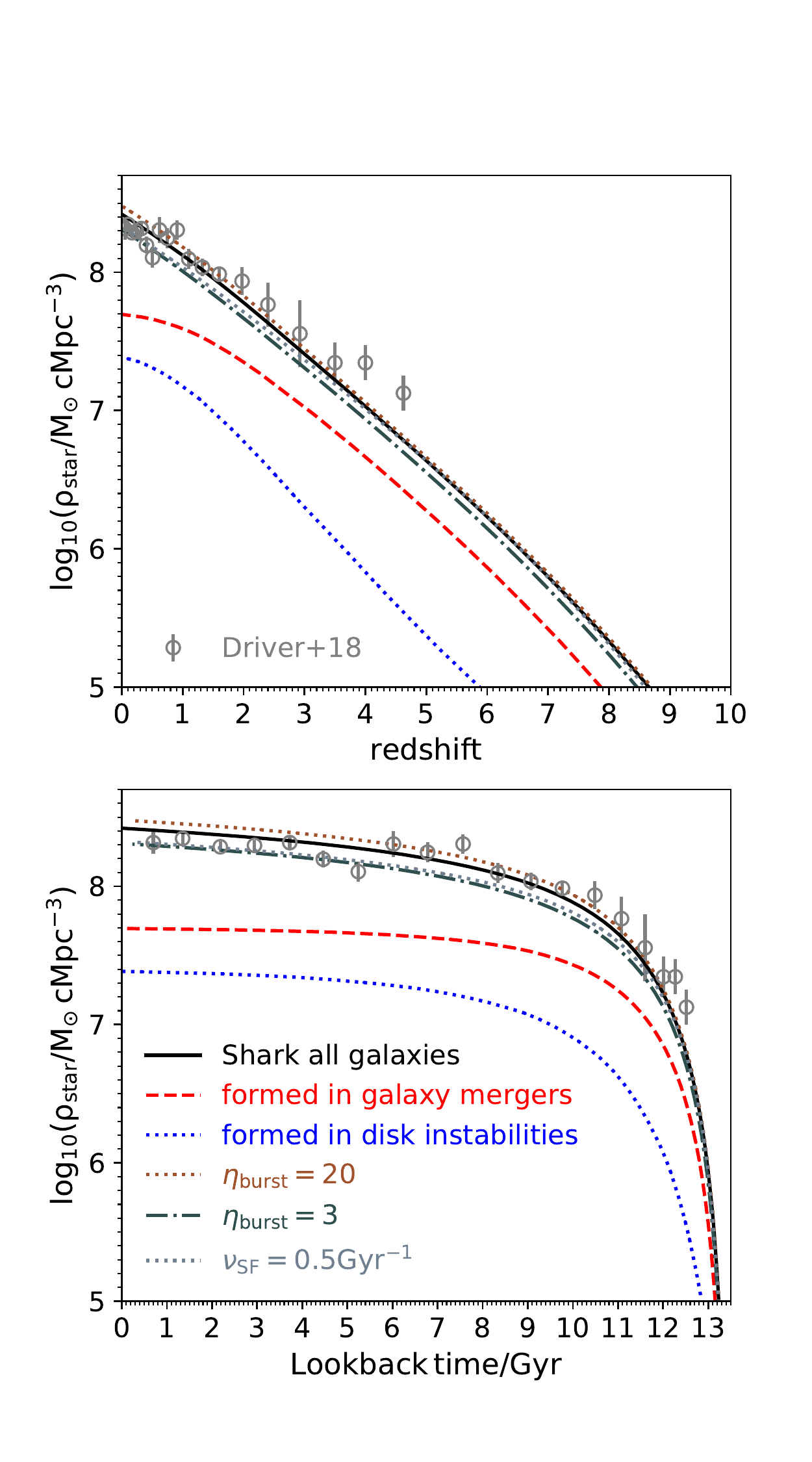}
\caption{Cosmic SFR density (left panel) and stellar mass density (right panel) evolution for our default \shark\ model. Solid lines show all galaxies, while the dotted and dashed line show in the left panel show 
the contribution from the SBs and quiescent modes of star formation, and the dotted and dashed line in the right panel show the contribution 
of all stars formed during SBs driven by galaxy mergers and disk instabilities, respectively. 
The latter can de driven by both disk instabilities and mergers. Observations from \citet{Karim11} and 
\citet{Driver17} are also shown. We show three model variants adopting a lower/higher SF efficiency in SBs ($\eta_{\rm burst}=3$ and $\eta_{\rm burst}=20$, respectively; see $\S$~\ref{sec:sfb}) 
and in quiescent SF ($\nu_{\rm SF}=0.5\,\rm Gyr^{-1}$; see Eq.~\ref{SFLaw}), as labelled.}
\label{GlobalsSFR}
\end{center}
\end{figure*}

We solve these equations numerically using the Runge-Kutta Cash-Karp with adaptive stepsizes 
of the {\tt C++} GSL library. The accuracy to which the equations are solved 
are a parameter the user inputs into \shark\ (see Table~\ref{tab:execparameters} for the default value and name of this 
variable in the code).

The Eqs.~\ref{eqn:sff}-\ref{eqn:sfflast} are the same in both star formation modes, quiescent (i.e. star formation in disks) and SB modes. 
The only difference is that during SBs $\crate\equiv 0$. Ideally we would like to solve for the quiescent and SB
 modes simultaneously, and \shark\ will be progressing towards that more realistic representation of how 
 star formation, outflows and inflows take place in galaxies. However, in the implementation of {\sc v1.1}, 
we first solve for quiescent star formation and then for SBs.

\section{Basic Results and Performance}\label{basicresults}

In this section we present some basic results of \shark, focusing on some traditional tests, such as the SMF, and 
overall growth of galaxy stellar mass and SFR, but also on the gas content of galaxies and the universe. We also show 
scaling relations that relate galaxies' masses, sizes, and metallicities in different components. For this section, we use our 
default model (see values adopted in parenthesis in Tables~\ref{tab:parameters}-\ref{tab:parameters2}) using as backbone the L210N1536 simulation (unless 
otherwise stated), but also show model variations to aid our discussion and to show the reader some key systematic uncertainties 
in the model. All the observations shown throughout this paper have been scaled to our adopted cosmology and 
IMF when necessary. 

The primary constraints we use to tune the free parameters are the $z=0,\,1,\,2$ SMFs, the $z=0$ 
the black hole-bulge mass relation and 
the mass-size relations. These are observations that we tried to fit as best as we could, 
although based on a visual inspection approach. Any other observables shown here are 
therefore ``predictions'' of the model (i.e. results that we 
did not fit for). 

\subsection{Baryon budget and its evolution}

\begin{figure*}
\begin{center}
\includegraphics[trim=3.5mm 5mm 10mm 2mm, clip,width=0.33\textwidth]{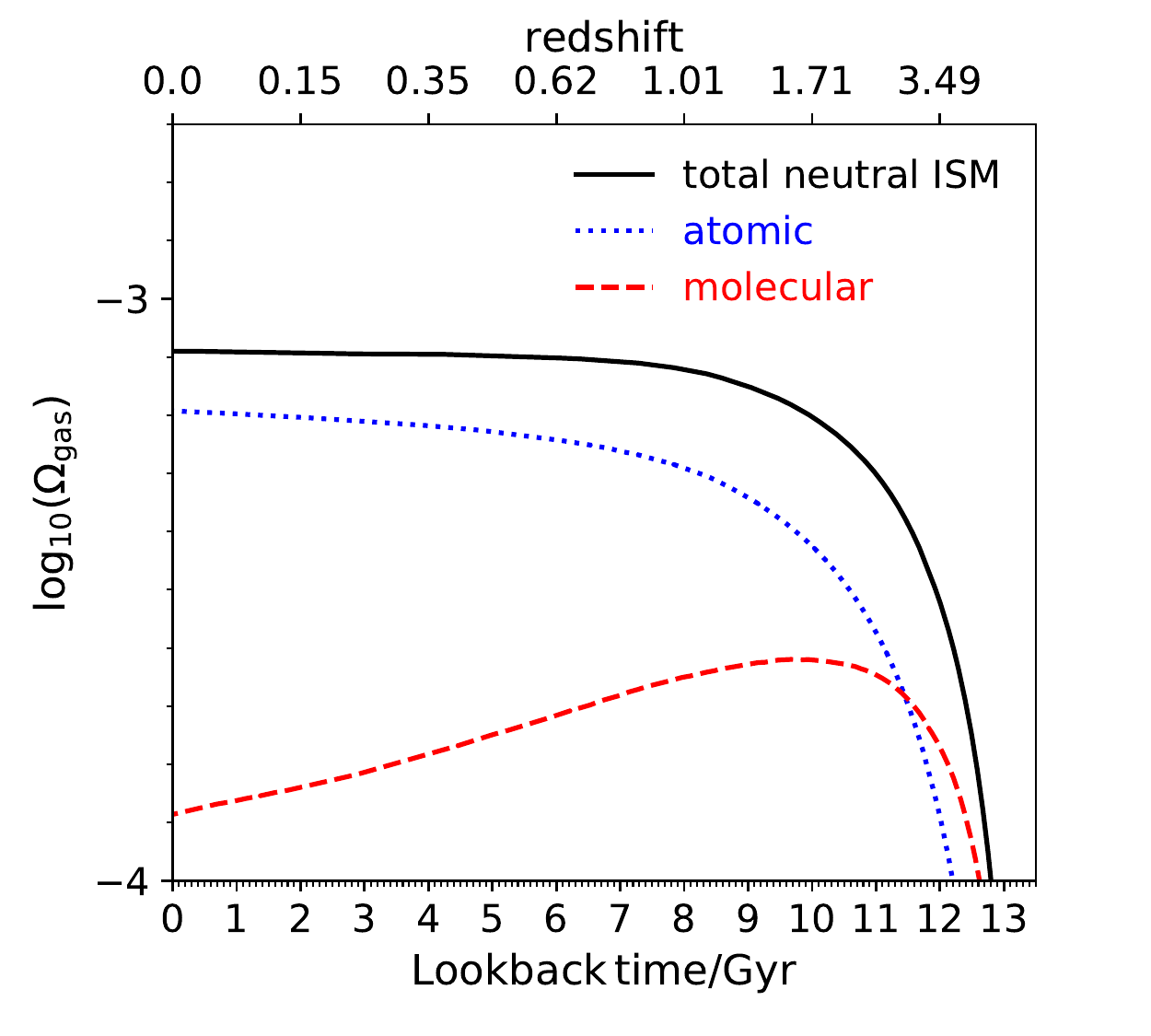}
\includegraphics[trim=3.5mm 5mm 10mm 2mm, clip,width=0.33\textwidth]{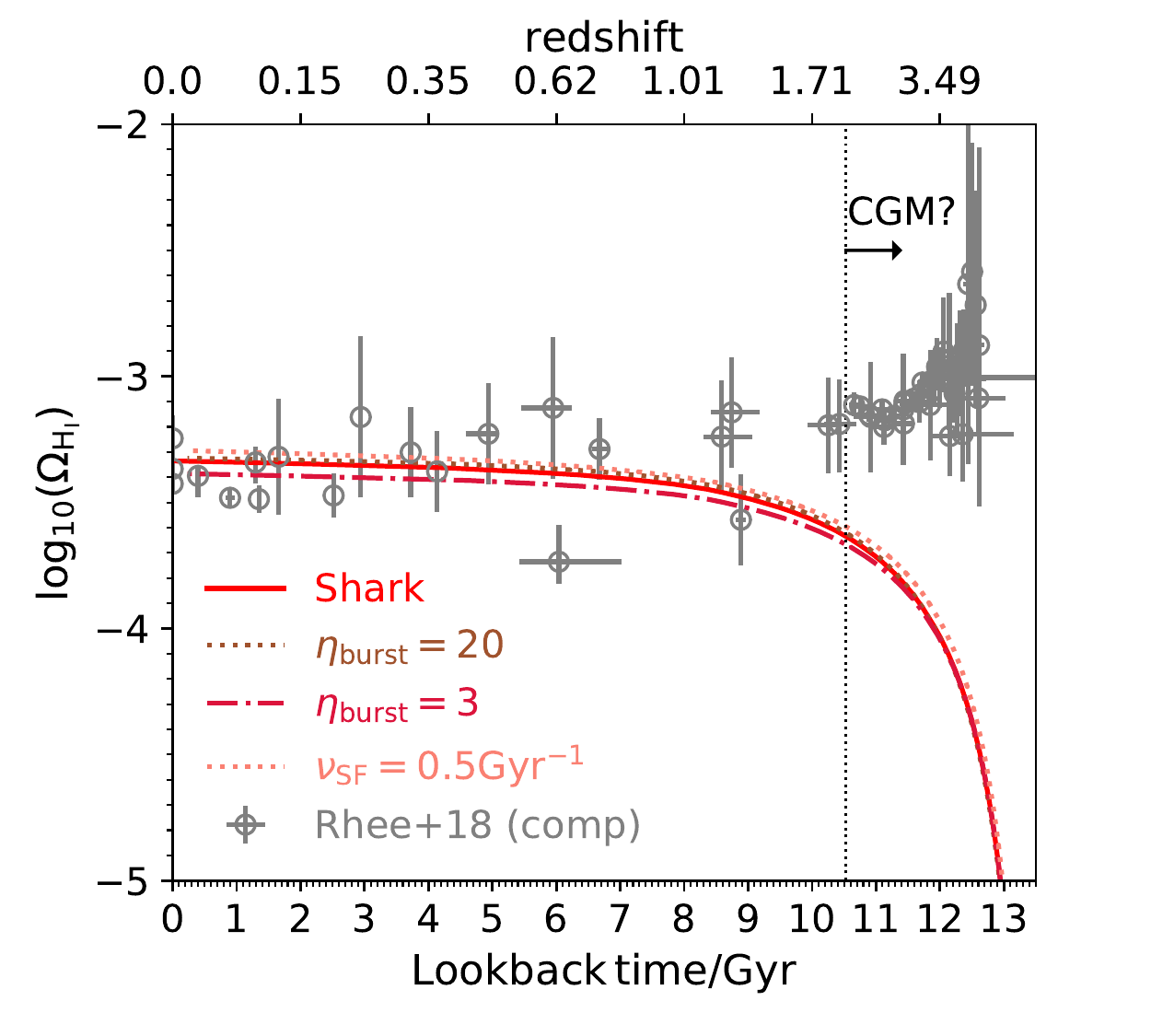}
\includegraphics[trim=3.5mm 5mm 10mm 2mm, clip,width=0.33\textwidth]{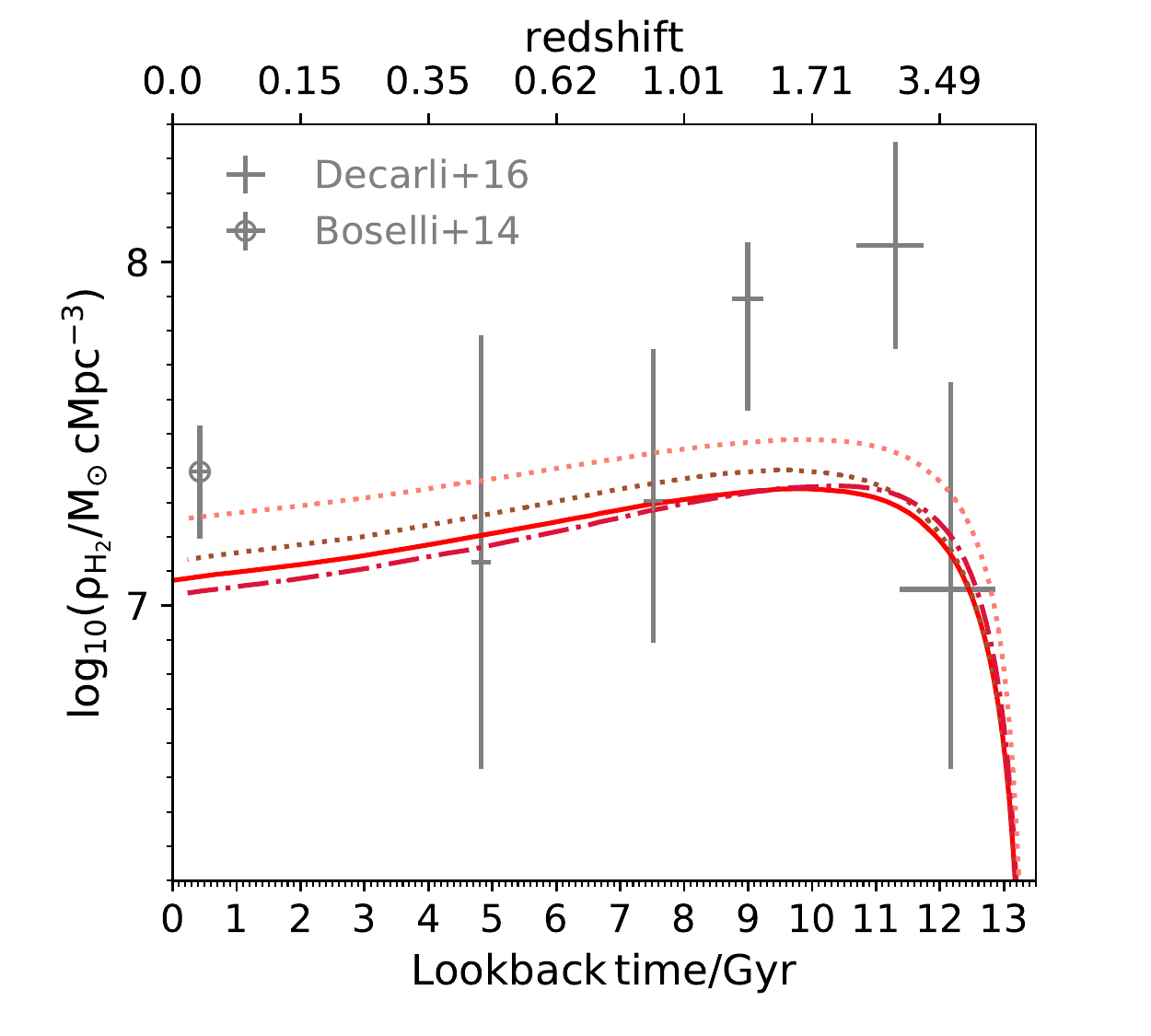}
\caption{{\it Left panel:} Evolution of $\Omega_{\rm gas}\equiv\rho_{\rm gas}/\rho_{\rm crit}$, with $\rho_{\rm gas}$ and 
$\rho_{\rm crit}$ being the gas and  critical densities, respectively. 
We show this for the total neutral gas in the ISM, HI and H$_2$, as labelled in our {default \shark\ model}.
{\it Middle panel:} Evolution of $\Omega_{\rm HI}$ compared to the observational compilation of \citet{Rhee18}. 
The vertical dotted lines denotes the approximate redshift at which the hydrodynamical simulations of \citet{VanDeVoort12}  
predict the transition from the HI being dominated by the ISM to the CGM of galaxies takes place. This transition is relevant as the \shark\ line 
here includes only HI in the ISM while the observations are not biased to detect ISM.
We also show the two model variations shown in Fig.~\ref{GlobalsSFR}.
{\it Right panel:} Evolution of $\rho_{\rm H2}$ compared to observations of \citet{Boselli14b} and \citet{Decarli16}. In this case 
we show density rather than $\Omega$ as it is the most common way observers express the density of H$_2$. 
We also show in the middle and right panels the same model variants as in Fig.~\ref{GlobalsSFR}.}
\label{GlobalsGas}
\end{center}
\end{figure*}

\begin{figure}
\begin{center}
\includegraphics[trim=4.5mm 42.5mm 8mm 34mm, clip,width=0.45\textwidth]{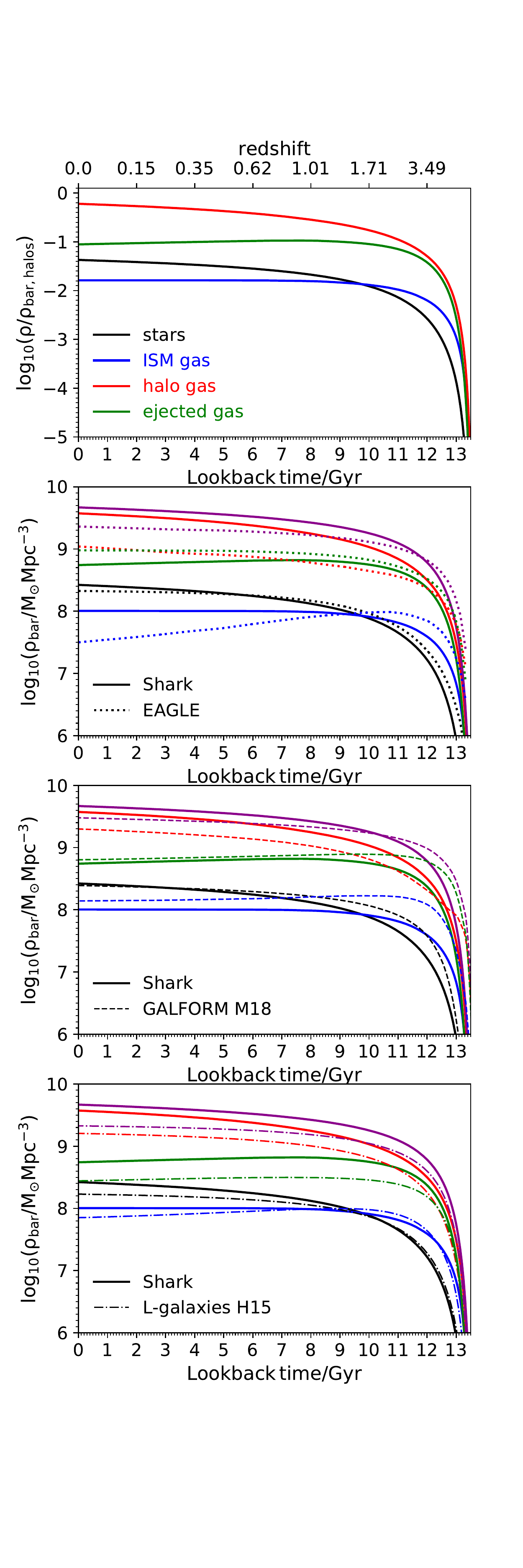}
\caption{{\it Top panel:} The evolution of the contribution of several baryon reservoirs (see label) to the total baryon content of the universe locked up in halos (or that 
was at some point in halos) in our default \shark\ model. {\it Second to bottom panels:} Comparison of the density evolution of the baryon components in \shark\ with those of the 
EAGLE hydrodynamical simulations as analysed by \citet{Mitchell18} (second panel), the GALFORM model in the variant of \citet{Mitchell18} (third panel) 
and the L-galaxies model in the variant of \citet{Henriques15} (bottom panel). Here, we also show the total baryon budget (the sum of the components shown) 
as magenta lines.}
\label{Globalsall}
\end{center}
\end{figure}

Fig.~\ref{GlobalsSFR} show the evolution of the cosmic SFR and stellar mass densities, both in linear redshift (top) and lookback time (bottom). The former 
is useful to see in more detail how the model performs at high redshift, while the opposite is true for the latter. We compare with the observations 
of \citet{Karim11}, who used radio continuum detections and stacking, and from \citet{Driver17}, who presented a combined analysis of the surveys GAMA, G23-COSMOS and 3D-HST. 

\begin{figure*}
\begin{center}
\includegraphics[trim=5mm 18mm 8mm 32mm, clip,width=0.85\textwidth]{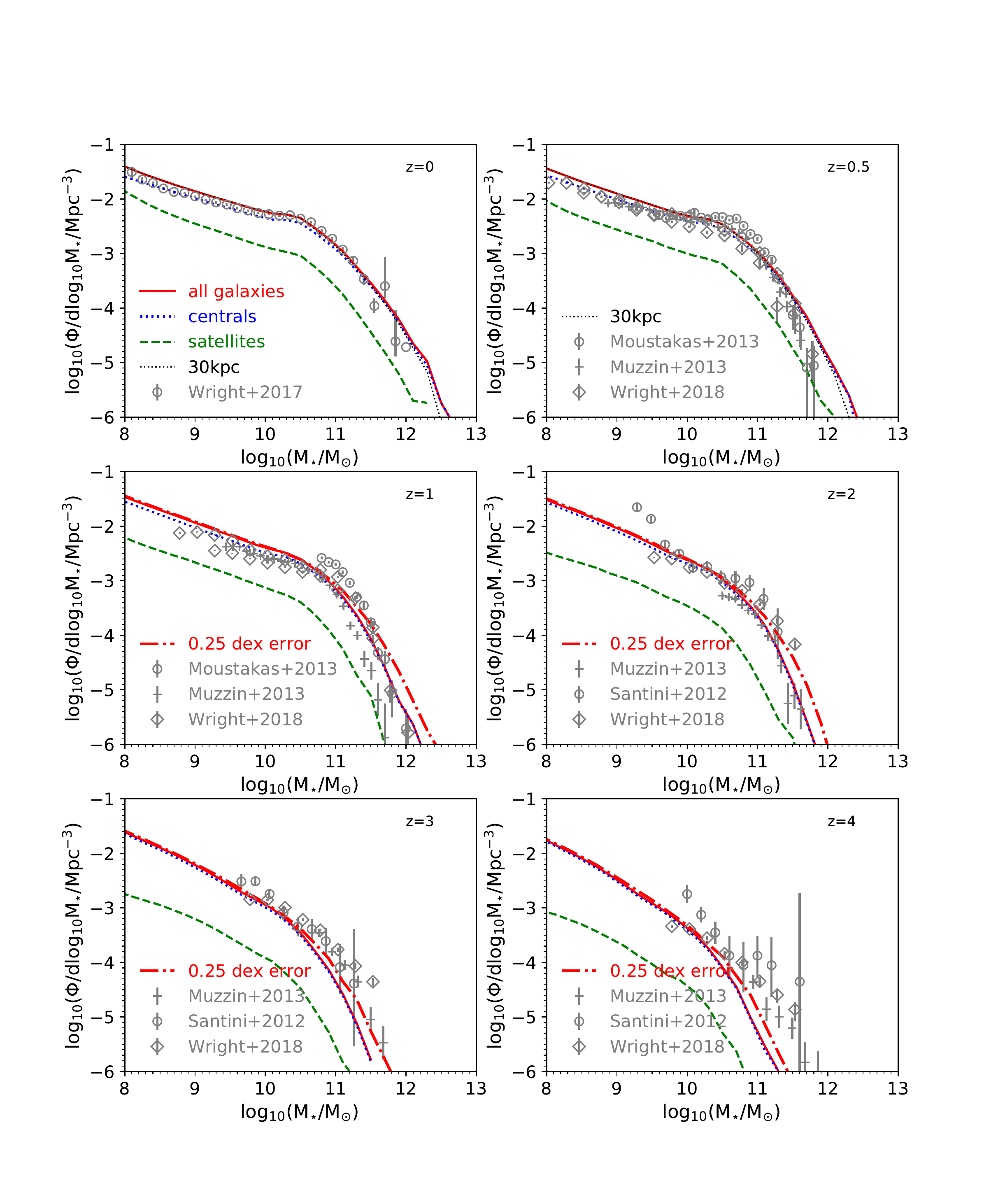}
\caption{Galaxy SMF at $z=0$, $z=0.5$, $z=1$, $z=2$, $z=3$ and $z=4$, as labelled, for our default \shark\ model. 
Solid lines show all galaxies, while dotted and dashed lines show central and satellite galaxies, respectively.
We also show observations from \citet{Wright17}, Wright et al. (submitted), \citet{Muzzin13}, 
\citet{Moustakas13}, \citet{Santini12}, as labelled. For reference, the dot-dashed line in the middle and bottom panels
show the SMF if we assume stellar masses have a Gaussian uncertainty of width $0.25$~dex. 
We also show at $z=0$ and $z=0.5$ the SMF if we use the mass enclosed in a $30$~physical kpc aperture rather than the total stellar mass 
(black dotted line in the top panels).}
\label{SMF}
\end{center}
\end{figure*}

Regarding the cosmic SFR density, 
\shark\ agrees nicely with the observations at $z\gtrsim 1$, while producing slightly too much SFR at $z\lesssim 1$ by up to $\approx 0.15$~dex.
We split the contribution from star formation in disks and SBs (blue and red lines, respectively) and find that overall disks dominate the cosmic SFR at $z\lesssim 1.7$ and $z\gtrsim 7$, with 
SBs making a significant contribution at $1.7\lesssim z \lesssim 5$ and becoming negligible at lower redshifts. Thus, the reason for the overproduction of stars 
at $z\lesssim 1$ is due to SF in disks. In principle we could increase the strength of AGN feedback to suppress SF in disks by suppressing 
cooling flows. However, we find that there is a tension at $z<1$ with the high mass end of the SMF (Fig.~\ref{SMF}), such that 
a more effective AGN feedback would move the massive end towards lower masses, undershooting the observational SMF. 
Interestingly, 
\shark\ reproduces quite well the stellar mass density evolution (right panels in Fig.~\ref{GlobalsSFR}), 
with some minor tension arising at $z\gtrsim 4$. 
This shows that the tension seen in the cosmic SFR vanishes when looking at the stellar mass density. There has been a long standing 
tension between these two measurements (see discussion in \citealt{Driver17}) and thus, we decide to not 
tune for the cosmic SFR. 

The right panels of Fig.~\ref{GlobalsSFR} show the contribution to the stellar mass density from SBs triggered 
by galaxy mergers and disk instabilities. Galaxy mergers are the main driver of stellar mass growth due to SBs. 
Disk instabilities start contributing more significantly at $z\lesssim 1$ and by $z=0$ they contribute about $35$\% of the total 
stellar mass ever formed in SBs. 
We find that the contribution from SBs have an important effect on the overall cosmic SFR density. 
As an example of that, we show in Fig.~\ref{GlobalsSFR} the predictions of a model in which the SF law for SBs is assumed to have a lower/higher  
normalization relative to SF in disks ($\eta_{\rm burst}=3$ and $\eta_{\rm burst}=20$; see $\S$~\ref{sec:sfb}). 
We find that decreasing $\eta_{\rm burst}$, decreases the cosmic SFR density throughout time by $\approx 0.1-0.15$~dex, but increasing it has less of an effect.
In \shark, we find that the effect of $\eta_{\rm burst}$ saturates at values $\gtrsim 7$, which is due to the gas depletion timescale being always comfortably shorter than 
the Hubble time in those cases. This is no longer true when $\eta_{\rm burst}\lesssim 7$ and when the SF efficiency is considerably lower than the $1\,\rm Gyr^{-1}$ adopted 
in our default model. The latter can be seen from the variation with $nu_{\rm SF}=0.5\,\rm Gyr^{-1}$ in Fig.~\ref{GlobalsSFR}. 
A higher $\eta_{\rm burst}$ in the case of $nu_{\rm SF}=0.5\,\rm Gyr^{-1}$, allows us to recover the same overall 
cosmic SFR density evolution as our default model.

Fig.~\ref{GlobalsGas} shows the evolution of the total ISM mass in galaxies, and the contributions from atomic and molecular gas (left panel) in 
our default \shark\ model.
Molecular gas dominates the ISM budget of galaxies at $z\gtrsim 2.5$, while HI dominates at lower redshifts. The H$_2$ density peaks at $z \approx 1.7$, which is lower 
than the peak of the cosmic SFR density ($z\approx 2$). This is due to the different effect SBs have on the cosmic SFR and H$_2$ densities. 
SBs contribute significantly to the SFR density, helping to shift the SFR density peak to higher redshifts, while their contribution to the H$_2$ density is minor. 
The latter is due to the short H$_2$ depletion times adopted for SBs.
 This is similar to what was seen in the {\sc Eagle} hydrodynamical simulations \citep{Lagos15}. The offset between the peaks of the 
cosmic SFR and $\Omega_{\rm mol}$ is exacerbated for larger values of $\eta_{\rm burst}$ (the parameter controlling the normalisation of the SF law of bursts relative to star formation 
in disks; see $\S$~\ref{sec:sfb}), while becoming smaller if $\eta_{burst}\rightarrow 1$. This is seen as a shift in the cosmic SFR density peak in that model (see dot-dashed line in 
Fig.~\ref{GlobalsSFR}).

The middle and right panels of Fig.~\ref{GlobalsGas} compare our predicted HI and H$_2$ densities (only hydrogen; i.e. removing the contribution from Helium) 
with the observations
of \citet{Boselli14b} and \citet{Decarli16}, in the case of H$_2$, and \citet{Rhee18}, for HI. 
We see that \shark\ reproduces the HI observations at $z\lesssim 1.5$ very well, while deviating significantly at higher redshifts. The latter is 
not necessarily a failure of the model, as here we only show HI in the ISM of galaxies, while observations are not biased to detecting ISM gas. 
\citet{VanDeVoort12}, using cosmological hydrodynamical simulations of galaxy formation, showed that at $z\gtrsim 2$ the neutral gas abundance of the universe 
from $z=3.5$ to $z=0$, 
is expected to be dominated by circumgalactic gas rather than the ISM of galaxies. 
Since \shark\ does not account for the fraction of HI in the halo gas, 
we do not necessarily expect to be able to reproduce the observations at $z \gtrsim 1.5$. The model variants adopting a different efficiency in SF in disks or SBs has little 
effect on the HI density evolution despite having an effect on the evolution of the cosmic SFR and H$_2$ densities (see discussion below). 

In the case of H$_2$, we find that \shark\ predicts an H$_2$ density that agrees well with observations, except at the peak, where our model produces too little H$_2$ abundance. 
The peak of the H$_2$ evolution is very sensitive to how star formation is modelled in \shark. In the model variant in which SBs are less efficient ($\eta_{\rm burst}\approx 3$) than 
in our default model, 
the H$_2$ density peak is $\approx 0.08$~dex lower than in our default \shark\ model, while increasing 
$\eta_{\rm burst}$ has little effect.
Adopting $\nu_{\rm SF}=0.5\,\rm Gyr^{-1}$ has a much more important effect on the H$_2$ density evolution, increasing it by $\approx0.2$~dex. 

Fig.~\ref{Globalsall} shows the breakdown of the $4$ key baryon components in \shark, the total stellar mass, 
ISM mass, halo gas (gas inside the halos and outside galaxies) and ejected gas (gas outside halos). 
Note that these baryons are those currently, or that were in the past, in halos. 
The fraction of baryons outside halos can be as large as $\approx 60$\% according to observations 
\citep{Shull12,Driver17}, while the fraction of DM that is unbound to halos 
with masses $\gtrsim 10^{10}\,\rm M_{\odot}$ can be up to $50$\% \citep{Shattow15}, accounting for most of that gas (if we assume the universal 
baryon fraction).

In \shark, the total baryon budget of galaxies is dominated by ISM gas at lookback times $\gtrsim 9$~Gyr ($z\approx 1.3$), while stellar 
mass becomes dominant at later times. However, galaxies never dominate the baryon content of the universe. Halo gas is overall the 
most dominant baryon component, with the ejected halo gas contributing similarly to the halo gas at lookback times 
$\gtrsim 12$ ($z\gtrsim 3.5$). The latter is due to the high specific star formation rates of high redshift galaxies 
and the redshift dependence of the outflow rate. The latter translates into more powerful outflows at high redshift (as $\rm z_{\rm P} > 0$ in Eq.~\ref{vel_power}). 
The latter is adopted following the FIRE results of \citet{Muratov15}. 

\citet{Mitchell18} studied the growth of the same baryon components of Fig.~\ref{Globalsall} for the GALFORM semi-analytic model and the EAGLE
hydrodynamical simulations. In the second to fourth panels of Fig.~\ref{Globalsall} we compare \shark\ with EAGLE, GALFORM and the L-galaxies 
SAMs, respectively. 
\shark\ behaves similarly to L-galaxies in the sense that halo gas always dominate the baryon budget. 
GALFORM has the ejected gas component dominating 
at early times (lookback time $\gtrsim 9.5$), but the halo gas dominates over most of the history of the universe. 
EAGLE displays a different behavior to \shark, GALFORM and L-galaxies as the 
majority of baryons throughout the history of the universe are locked up in the `ejected' component, except at lookback times $<4.5$~Gyr where
the halo gas starts to dominate. Note that 
\citet{Mitchell18} calculated this ejected mass as the difference between the baryons accounted for inside the halos and the universal baryon fraction, 
thus it is not strictly the same as the ejected component in SAMs. The comparison is, however, a fair one, as all the SAMs by construction 
have a total baryon fraction (including the ejected component) per halo equal or close to the universal baryon fraction. 
Thus, the most striking difference here is that overall EAGLE has much less baryons inside 
halos than the three SAMs. This can happen either because the gas accretion onto halos is less efficient than the DM accretion
or because feedback is very effective at ejecting gas from halos. In the former case, feedback can still play a key role as
outflows can interact with the gas that is 
outside halos, preventing it from ever inflowing onto halos. 
Note that the overall efficiency of outflowing gas in \shark, GALFORM and EAGLE are similar at lookback times $\lesssim 9$~Gyr and $>7$~Gyr, respectively, 
 as the ejected gas baryon component 
is similar between the models (green lines in the second and third panels of Fig.~\ref{Globalsall}), 
while the L-galaxies one is lower by a factor of $\approx 2.5-3$ compared to \shark\ and GALFORM, and $\approx 4$ compared to EAGLE. 
Another important effect is the exact definition of halo mass, as the amount of baryons scales with it. Different definitions, 
FOF mass, $M_{\rm crit, 200}$, $M_{\rm mean,200}$, etc., can differ by factors of up to a few \citep{Jiang14}. 
The \shark\ DM mass function agrees very well with the mean $200$ density mass function of 
\citet{Sheth01}, produced using the \citet{Planck15} cosmology and {\sc HMF calc} \citep{Murray13}. 
These masses, however, are generally larger than the Dhalo masses used in GALFORM and the {\tt subfind} masses of EAGLE and 
L-galaxies. The latter leads to the total baryon density of \shark\ being slightly larger than the other models (dark magenta lines in 
Fig.~\ref{Globalsall}).

In summary, the main difference between the SAMs and EAGLE 
is that in the latter the amount of baryons inside halos is much smaller. 
This is an important physical effect that is currently not included in any SAM 
to our knowledge, but could relax significantly the AGN feedback efficiency required in 
SAMs to prevent the large amounts of halo gas from cooling and 
forming stars. {This is because in EAGLE, the baryon fractions inside halos is less than the universal one even for massive halos 
($M_{\rm halos}\sim10^{14}\rm \,M_{\odot}$; see Fig.~$8$ in \citealt{Mitchell18}), with deviations being larger at increasing redshift; while in 
SAMs, the baryon fraction inside halos becomes significantly lower than the universal one only at $M_{\rm halo}\lesssim 10^{11.5}\rm\, M_{\odot}$.}
Remarkably, the stellar mass density evolution of \shark\ and EAGLE are within $15$\% of each other, 
while with GALFORM and L-galaxies differences increase to up to $40$\% and $25$\%, respectively. This is crucial evidence 
showing that effectively the same stellar mass density can be obtained for very different physical reasons. 

\subsection{Stellar masses and their scaling with halo mass}

\begin{figure}
\begin{center}
\includegraphics[trim=2mm 10mm 2mm 22mm, clip,width=0.45\textwidth]{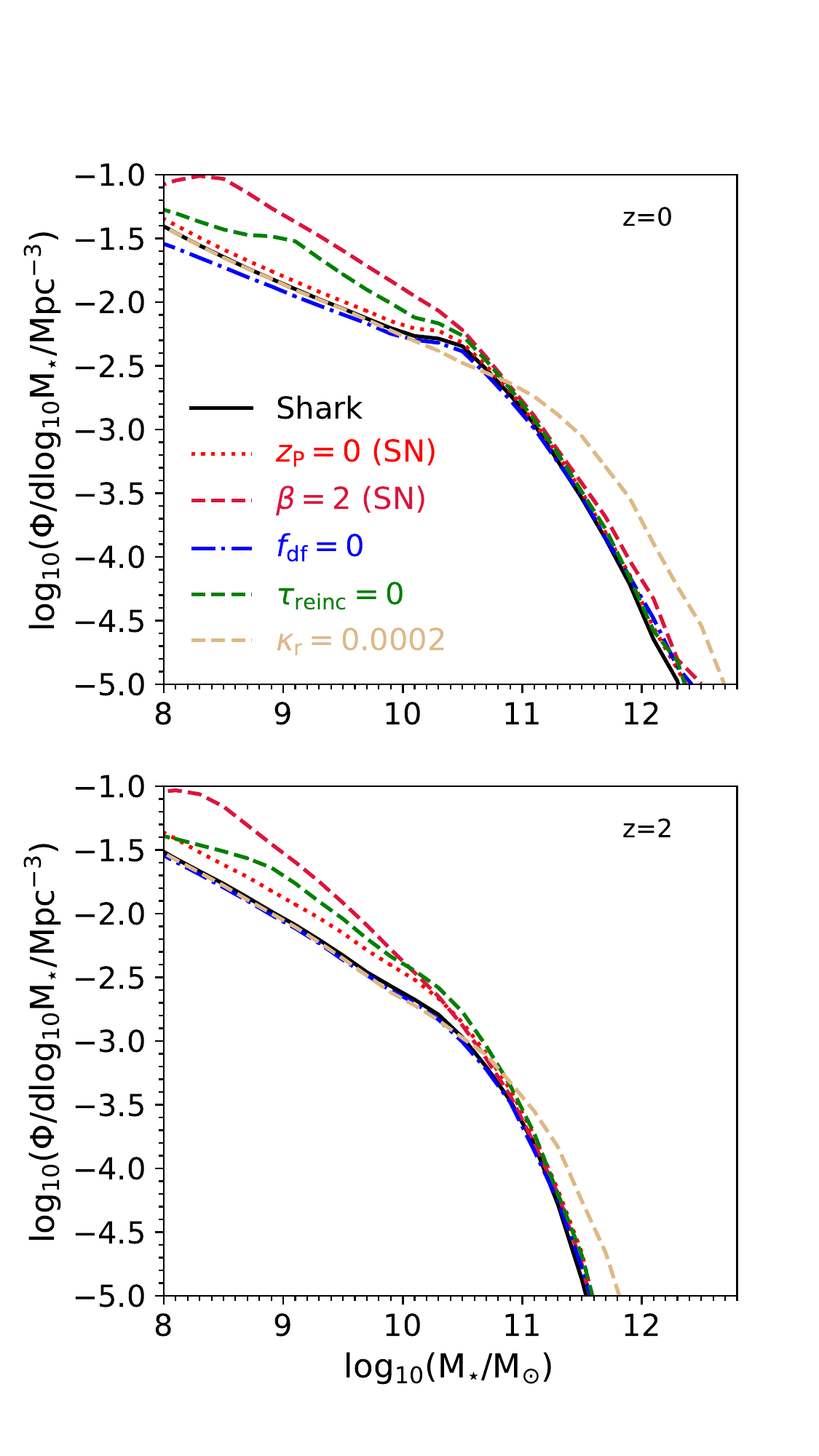}
\caption{Galaxy SMF at $z=0$ and $z=2$, as labelled, for our default \shark\ model and 5 variations 
perturbing the stellar feedback parameter (red lines), dynamical friction timescale (blue line),  
reincorporation timescale of the ejected gas (green line) and the AGN feedback efficiency (tan line). See text for details.}
\label{SMF2}
\end{center}
\end{figure}

\begin{figure*}
\begin{center}
\includegraphics[trim=5mm 18mm 8mm 32mm, clip,width=0.85\textwidth]{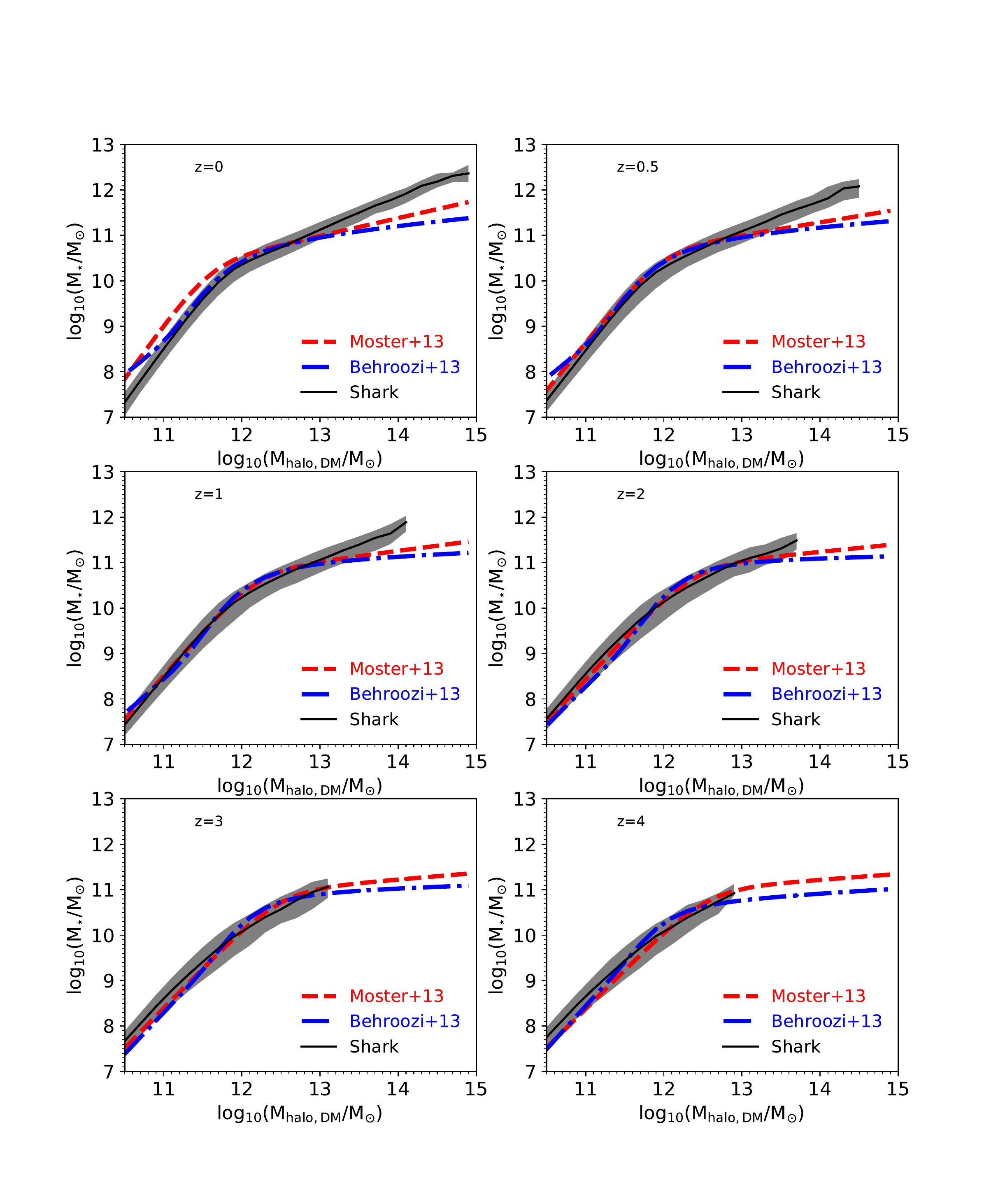}
\caption{Stellar mass as a function of halo mass for centrals galaxies in our default \shark\ model 
at $z=0$, $z=0.5$, $z=1$, $z=2$, $z=3$ and $z=4$, as labelled. Solid lines with shaded regions show the median and the 
$16^{\rm th}-84^{\rm th}$ percentile ranges. Here, we show all bins with $\ge 10$ objects. 
We also show the empiral results of \citet{Moster13} and \citet{Behroozi13}.}
\label{SMHM}
\end{center}
\end{figure*}

SMFs and the stellar-halo mass relation have become a key test for models and are now 
the usual observational choice for the tuning of free parameters in semi-analytic models 
\citep{Guo11,Henriques13,Knebe17,Cora18} and hydrodynamical simulations \citep{Schaye14,WangL15,Pillepich18}.  
We therefore also study these relations to understand the behaviour of \shark.

Fig.~\ref{SMF} shows the SMF of \shark\ at several redshifts, from $z=0$ to $z=4$. 
This is shown for the entire galaxy population and for centrals and satellites separately. Here, satellites
include galaxies in satellite subhalos and orphan galaxies, as described in $\S$~\ref{sec:evolvemergertree}. We also 
show a compilation of observations covering the same redshift range. 
We are able to reproduce very well the SMF up to $z=3$, while at higher redshifts the model 
struggles to reproduce the high-mass end, even in the case a small Gaussian uncertainty of 
$0.25$~dex is included. The latter is a typical error at high redshift \citep{Marchesini09,Mitchell13}, and is enough to alleviate the small tension at the massive end 
at $z=2$ and $z=3$. 
At the high-mass end and at $z<1$, \shark\ produces slightly more mass than observed at fixed number density.
This may not be very well converged at the resolution of the L210N1536, as our higher resolution SURFS run L40N512 
produces slightly less massive galaxies at fixed number density, in better agreement with 
the observations (see discussion in Appendix~\ref{ConvTests}).
In addition, ({see discussion in \citealt{Bernardi13}}) 
the exact high-mass end of the SMF is a fairly contested topic in observations, as the exact contribution from the intra-cluster light 
is uncertain. This is why several cosmological hydrodynamical simulations tend to show the stellar mass contained 
in an spherical aperture \citep{Schaye14,Pillepich18}.
We show the effect of using a fixed aperture of $30$ physical kpc in the SMF at $z=0$ and $z=0.5$. We calculate the stellar mass enclosed 
by assuming an exponential and a Plummer profiles for the stellar disk and bulge, respectively, with half-stellar mass radii 
calculated as in $\S$~\ref{sizes} (see $\S$~\ref{sec:sizes:comp} for a comparison with observations). The number density at the very high-mass 
end decreases, albeit only a small amount.
Fig.~\ref{SMF} also shows that central galaxies always dominate at the high-mass end, while satellite galaxies have a negligible contribution 
at high redshift, but become increasingly important towards low redshift at the low-mass end.

The normalization and flatness of the SMF at the low-mass end throughout redshift has been challenging to reproduce 
in semi-analytic models \citep{Henriques13}, but as Fig.~\ref{SMF} shows, \shark\ does not struggle with this. 
This is the result of a combination of effects: dynamical friction modelling, mass loading of the stellar feedback model depending very strongly 
on $V_{\rm circ}$, a slow reincorporation timescale and the weak redshift dependence of the stellar feedback efficiency.
At the massive end, the main physical process controlling the steep decline in number density is 
 AGN feedback in high mass halos.
This is shown in Fig.~\ref{SMF2}, in which we show model variations adopting a dynamical friction efficiency $f_{\rm df}=0$ (i.e. satellites merge onto the central 
as soon as they become orphans; blue dotted line), a weaker scaling of the mass loading on $V_{\rm circ}$ ($\beta=2$; see Eqs.~\ref{SNGal}-\ref{SNLAG13}; red, dashed line), 
removing the redshift dependence of the stellar feedback efficiency ($\rm z_{\rm P}=0$; see Eqs.~\ref{SNFIRE} and \ref{vel_power}; red, dotted line), 
assuming instantaneous reincorporation of the ejected gas ($\tau_{\rm reinc}=0$; see Eq.~\ref{Eqreinc}; green, dashed line), and 
adopting an AGN feedback efficiency $10$ times lower ($\kappa_{\rm r} = 0.0002$; see Eq.~\ref{Eq.RM}; tan, dashed line).

Merging galaxies instantaneously after they become orphans ($f_{\rm df}=0$) 
has the effect of producing a flatter low-mass end of the SMF and an 
overall lower abundance of galaxies below the knee, which is 
more clearly seen at $z=0$. In the case of stellar feedback, removing the redshift dependence of the mass loading ($\rm z_{\rm P}=0$) has little effect by 
$z=0$, but at $z=2$ the model produces a higher abundance of galaxies below the knee of the SMF, in tension with the observations. 
A weaker scaling of the mass loading on $V_{\rm circ}$ ($\beta=2$) has a dramatic effect on the abundance of low mass galaxies at both 
redshifts, though not affecting the high-mass end. This shows that a strong dependence on $V_{\rm circ}$ is required in order to reproduce 
the flat low-mass end. Finally, the reincorporation timescale has also a clear effect on the low-mass end, which is more obvious at 
$z=2$. \citet{Henriques13} suggested that long reincorporation timescales are required to fit the low-mass end of the SMF at 
high redshift. Although this is seen in \shark, it is important to stress that this is not a unique solution, as Fig.~\ref{SMF2} shows 
the complex dependence that the low-mass end of the SMF has on several, different physical processes. 
At the massive end, reducing the AGN feedback efficiency by a factor of $10$ produces a shallower decrease in the number density beyond the knee of the SMF. 
The effect is, however, non-linear as the number density of galaxies with $M_{\rm star}\approx 10^{12}\,\rm M_{\odot}$ increases by only $\approx 2$. 

Although our primary constraint to tune the free parameters is the SMF, we did not use the 
stellar-halo mass relation in that process. Thus, we can use it to study whether \shark\ places the right amount of stellar 
mass in different halo masses. This is an important test as it can be viewed as a halo star formation efficiency.
Fig~\ref{SMHM} shows the stellar-halo mass relation at several redshifts from $z=0$ to $z=4$, compared to the empirical estimates of \citet{Moster13} and \citet{Behroozi13}. 
The agreement between \shark\ and \citet{Behroozi13} is good over the entire redshift range {at halo masses $\lesssim 10^{13}\,\rm M_{\odot}$}, while at $z=0$ 
we find that the stellar mass is lower at fixed halo mass compared to \citet{Moster13}. 
The differences between \citet{Moster13} and \citet{Behroozi13} can be considered part of the systematic uncertainties 
of the measurement.
{At high halo masses, $M_{\rm halo}\gtrsim 10^{13}\,\rm M_{\odot}$, we find that \shark\ has a slow that} is slightly too steep 
compared to \citet{Moster13} and \citet{Behroozi13}, which means that by $M_{\rm halo}\approx 10^{15}\,\rm M_{\odot}$, 
central galaxies in \shark\ {are $\approx 1.4-8.5$ (i.e. the $16^{\rm th}-84^{\rm th}$ percentile range) times too massive compared to \citet{Moster13}}.
{This is not entirely surprising, as both the \citet{Moster13} and \citet{Behroozi13} empirical stellar-halo mass relations were calibrated using the 
stellar mass function of \citet{Li09}, which has a high-mass end that is steeper than the \citet{Wright17} stellar mass function. The latter is the one 
we use as calibration reference.}

\citet{Guo15} presented a comparison between EAGLE \citep{Schaye14} and two widely known semi-analytic models 
L-galaxies \citep{Guo11} and GALFORM \citep{Gonzalez-Perez13}. One of the key comparisons was the stellar-halo mass relation 
and they found that the three models agreed relatively well but they displayed very different scatter, with EAGLE having the 
tightest relation ($\approx 0.4$~dex). An interesting feature of \shark\ and that is evident in Fig~\ref{SMHM} is that 
it predicts a very tight relation with a $1\sigma$ scatter of $\approx 0.35-0.5$~dex depending on the halo mass. This value 
is much smaller than the semi-analytic models L-galaxies and GALFORM, and similar to EAGLE,  {though still larger than 
the empirical estimates of \citet{Moster13} and \citet{Behroozi13}.}
\citet{Mitchell16} showed that 
adopting a scaling with the halo velocity, as adopted in \shark, or the galaxy velocity for stellar feedback has an important 
effect on the scatter. In the future we will investigate 
the effect different physical processes have on the scatter of this relation in \shark.

\subsection{Gas mass functions and scaling relations}

\begin{figure}
\begin{center}
\includegraphics[trim=3mm 3mm 5mm 12mm, clip,width=0.45\textwidth]{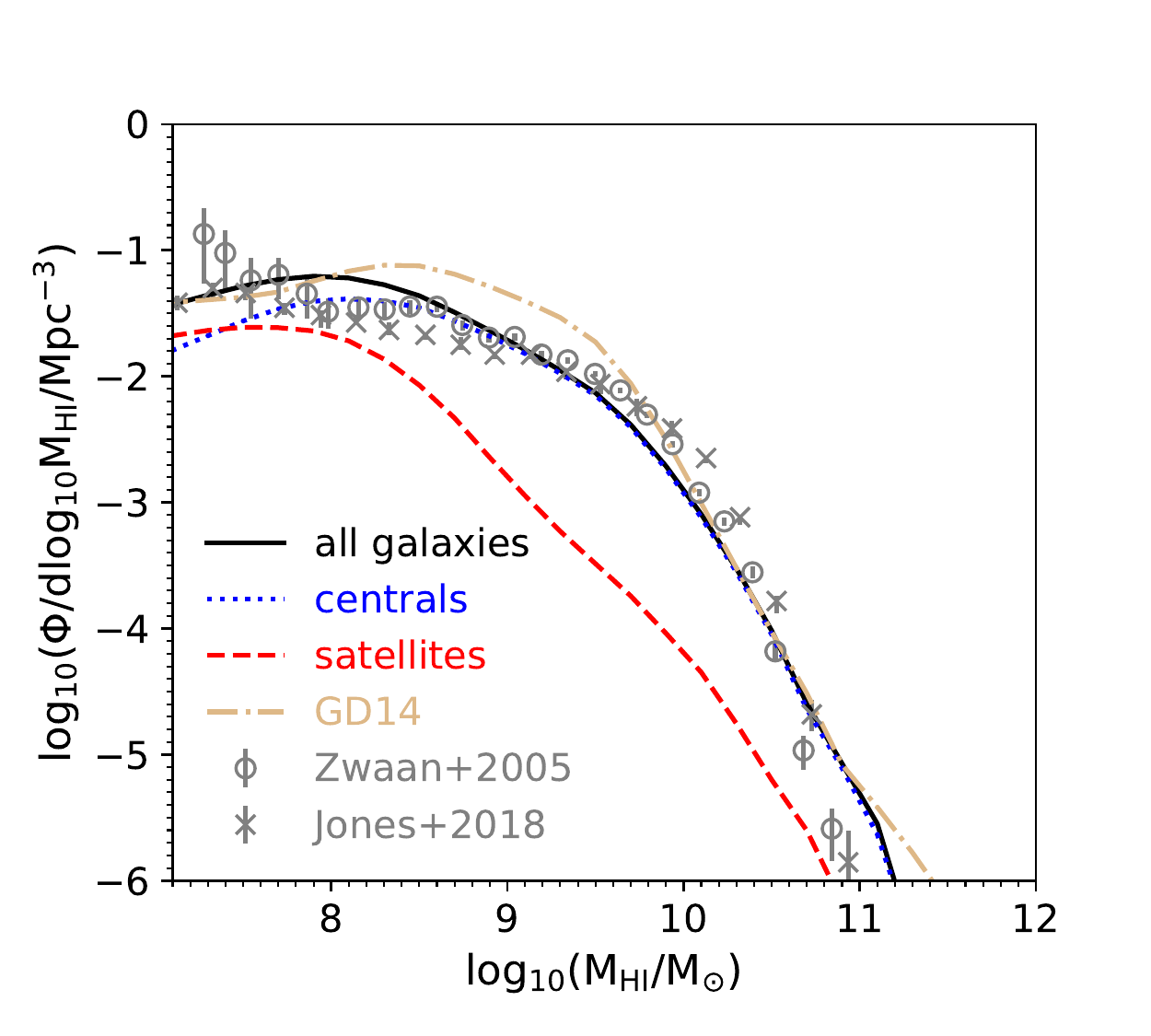}
\includegraphics[trim=3mm 3mm 5mm 12mm, clip,width=0.45\textwidth]{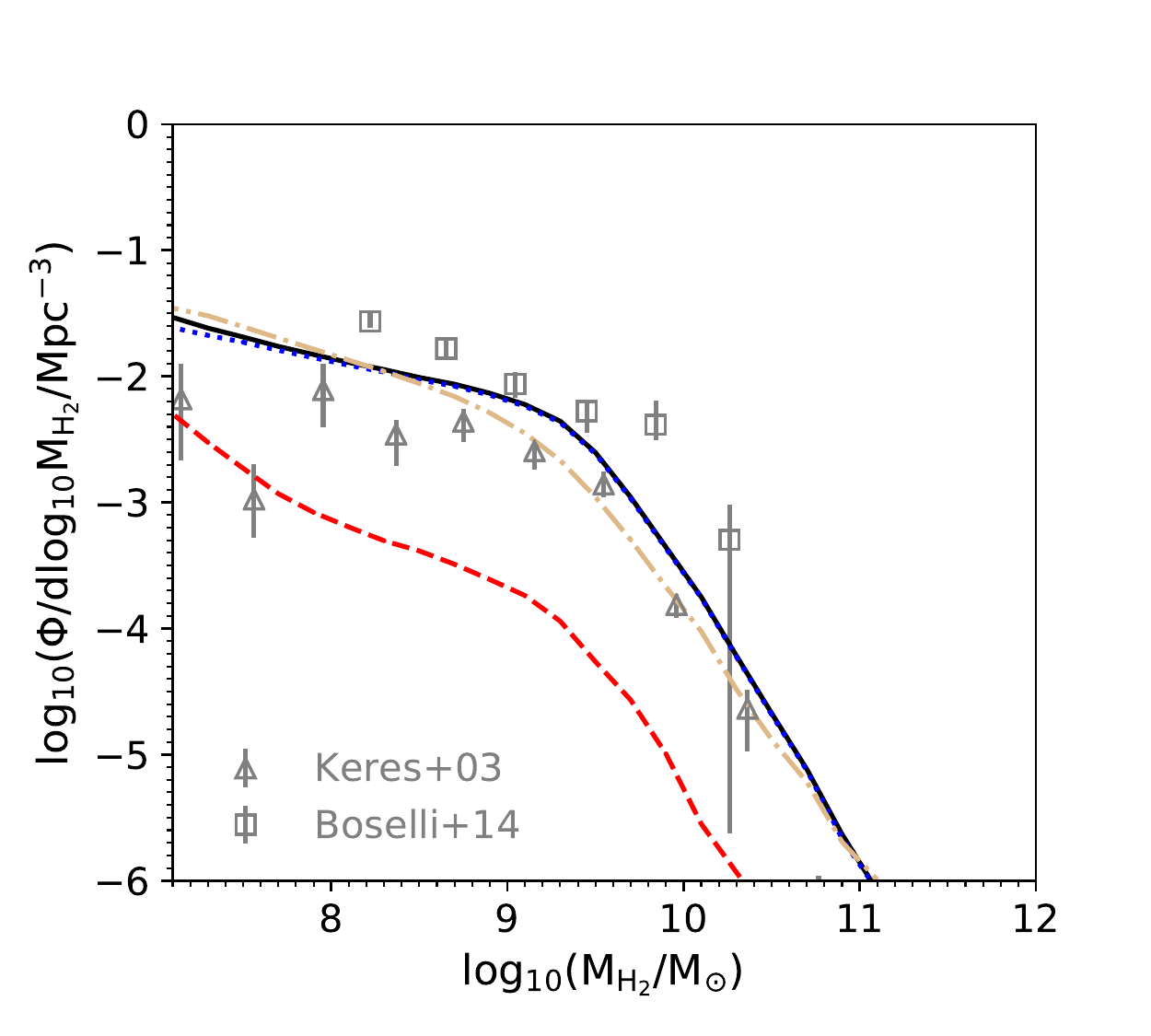}
\caption{Atomic (top) and molecular (bottom) hydrogen MFs at $z=0$, for our default \shark\ model.
Solid lines show all galaxies, while dotted and dashed lines show central and satellite galaxies, respectively. 
We show observations of the HI MF from \citet{Zwaan05} and \citet{Jones18}, and 
from \citet{Keres03} and \citet{Boselli14b} for the H$_2$ MF. We also show as dot-dashed line a variant of the model adopting the 
GD14 star formation law (rather than the BR06; see $\S$~\ref{sec:sf}).}
\label{GasMF}
\end{center}
\end{figure}

\citet{Lagos10,Lagos11} showed that the HI and H$_2$ mass functions (MFs) and their scaling with stellar mass 
are a key test to the star formation modelling in simulations. The stellar mass and SFR of galaxies are only mildly 
affected by the modelling of star formation, because inflows and outflows tend to quickly self-regulate and 
erase the effect a higher/lower efficiency star formation model could have. However, 
the gas properties of galaxies are very sensitive to this choice. We did not use these MFs in the process 
of parameter tuning, and thus comparing the HI and H$_2$ abundance of galaxies with observations represents 
a test of how well the model does.

Fig.~\ref{GasMF} shows the $z=0$ HI and H$_2$ MFs of our default \shark\ model. 
We show all the galaxies, and the contribution from centrals and satellites. 
We show the observed HI MF of \citet{Zwaan05} and \citet{Jones18}, which were obtained 
using the HI Parkes All-Sky Survey (HIPASS; \citealt{Meyer04}) and the Arecibo Legacy Fast ALFA Survey (ALFALFA; \citealt{Haynes18}), 
respectively. In the case of the H$_2$ MF, we show the inferences of \citet{Keres03} and \citet{Boselli14b}. 
These observations correspond to CO$\rm (1-0)$, which we convert to H$_2$ by adopting a Milky-Way like $\rm CO(1-0)$-H$_2$
conversion factor \citep{Bolatto13}, $N_{\rm H_2}/\rm cm^{-2}=2\times10^{-20}\, I_{\rm CO}/\rm K\,km\,s^{-1}$.
Note that these observations provide an indirect measurement of the MF as they are not blind surveys, 
but correspond to the follow-up galaxy samples selected from their $60\,\rm \mu m$ in the case of \citet{Keres03} and 
from the Herschel Reference Survey in the case of \citet{Boselli14b}.

\begin{figure*}
\begin{center}
\includegraphics[trim=20mm 15mm 20mm 30mm, clip,width=0.9\textwidth]{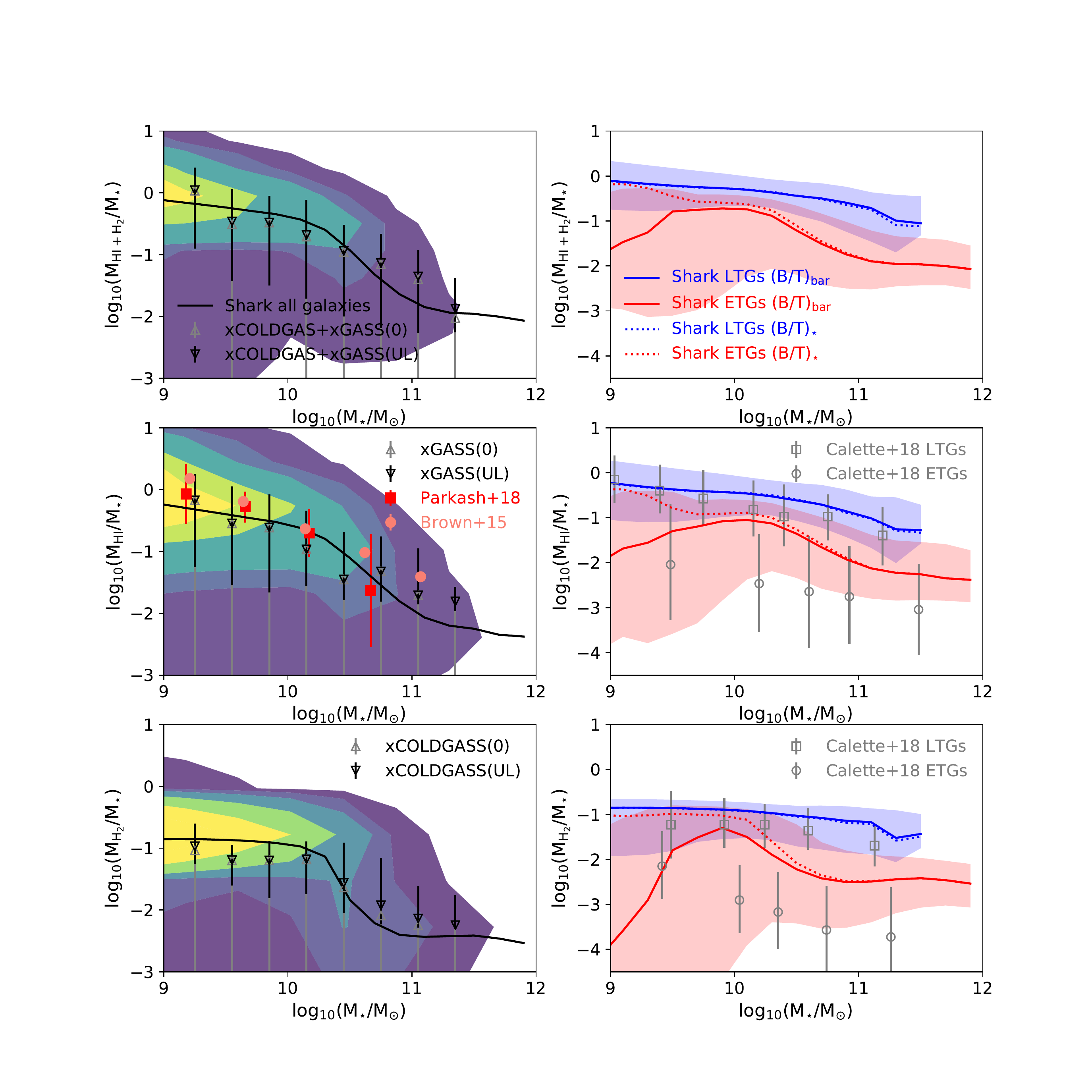}
\caption{Gas scaling relations at $z=0$ 
for our default \shark\ model: total neutral gas (atomic plus molecular; top panels), HI (middle panels) and molecular gas (bottom panels) 
fractions as a function of stellar mass. Left panels show all the galaxies, while the right panels show 
 the population split into ETGs and LTGs, as labelled. We show two definitions of 
ETGs/LTGs, one based on the stellar plus gas mass fraction contributed by the bulge, $(\rm B/T)_{\rm bar}$, 
and another one based on stellar mass alone, $(\rm B/T)_{\star}$. Ratios $\ge 0.5$ and $<0.5$ are considered 
ETGs and LTGs, respectively.
Lines show the median of the populations, while the contours in the left panel 
show percentile ranges from $99^{\rm th}$ to $10^{\rm th}$, and the shaded regions on the right panels show 
the $16^{\rm th}-84^{\rm th}$
percentile ranges. 
The latter are shown only for the $(\rm B/T)_{\rm bar}$ classification, for clarity.
Medians are shown for all bins with $\ge 10$ objects.
Symbols with errorbars show the 
median and the $1\sigma$ scatter of the observations of xGASS \citep{Catinella18}, the stellar mass-selected sample of \citet{Parkash18}, HI stacking of \citet{Brown15} and 
xCOLDGASS \citep{Saintonge17} (left panels), and the observationally derived gas fraction scaling relations for early- and late-type galaxies 
from \citet{Calette18} (right panels). Downwards triangles show the observations if non-detections are set to their upper limit {(referred to as `UL')}, while 
upward triangles show the results of non-detections are set to zero {(referred to as `0')}.}
\label{Scalings2}
\end{center}
\end{figure*}

\shark\ predicts an HI MF in reasonable agreement with the observations, except for the slight overproduction 
of galaxies with $M_{\rm HI} \gtrsim 10^{11}\,\rm M_{\odot}$. 
This may not be entirely surprising as for \shark\ we show the total HI mass in the ISM of galaxies, while observations 
are subject to some level of HI self-absorption which may lie at the $10-20$\% level (going up to $40$\% in dense regions; \citealt{Braun09}).
The decrease in number density at $M_{\rm HI} \lesssim  10^{8}\,\rm M_{\odot}$
 is due to a combination of the resolution of the L210N1536 simulation, which does not allow us to have a complete sample of the halos that would host these galaxies, 
 $M_{\rm halo}\lesssim 10^{10}\,\rm M_{\odot}$, and the photo-ionisation modelling (see convergence tests in Appendix~\ref{ConvTests}).
As expected, the HI MF is dominated over the entire mass range by central galaxies, with satellites 
becoming important only at $M_{\rm HI} \lesssim 10^{8}\,\rm M_{\odot}$. The latter is due to the environmental effects included in \shark, which 
assume instantaneous stripping of the halo gas of galaxies as soon as they become satellites and photo-ionisation feedback affecting low-mass central galaxies. 
In the absence of gas accretion, satellites can 
quickly use up their gas by continuing star formation and the driving of outflows. The presence of outflows means that the ISM gas consumption can happen faster than 
the instantaneous ISM depletion timescale. For low-mass centrals, the \citet{Sobacchi13} model applied in \shark, prevents gas cooling 
from taking place in these galaxies and thus the HI is not replenish.  

The H$_2$ MF of \shark\ is also in reasonable agreement with the observational inferences shown in the bottom 
panel of Fig.~\ref{GasMF}. We do, however, warn that the systematic uncertainties, mostly due to the 
$\rm CO(1-0)$-H$_2$ conversion factor, and the fact that we lack a CO blind survey, are large \citep{Obreschkow09d}. 
An interesting feature is that the low-mass end of the H$_2$ MF in \shark\ is flatter than the HI MF. 
This means that low HI mass galaxies have a larger contribution to the cosmic HI density that the contribution of their low H$_2$ mass 
counterparts to the cosmic H$_2$ density.
We also show in Fig.~\ref{GasMF} a variation of our default \shark\ model, adopting the GD14 rather than the BR06 star formation law (see $\S$~\ref{sec:sf} for details). 
Interestingly, the H$_2$ MF is only slightly affected by this choice, while the HI MF changes dramatically at $M_{\rm HI}\lesssim 10^{10.2}\rm \, M_{\odot}$. 
This is due to galaxy self-regulation, in which the dense gas mass adapts to give the same SFR and stellar mass growth, while the HI is in principle free to change 
regardless of self-regulation (see discussion in \citealt{Lagos14b}). 
This is a good example of the effect of self-regulation on the galaxy's ISM content.

The left panels of Fig.~\ref{Scalings2} show the neutral hydrogen (HI plus H$_2$), HI and H$_2$ gas to stellar mass ratios as a function of stellar mass at $z=0$ for our default \shark\ model.
We compare with the observations of \citet{Saintonge17}, for H$_2$, and \citet{Brown15}, \citet{Parkash18} and \citet{Catinella18}, for HI. 
\citet{Catinella18} and \citet{Parkash18} studied individual galaxies with $M_{\star}>10^9\,\rm M_{\odot}$, and thus are better suited at measuring the scatter of the gas fraction scaling relations
than \citet{Brown15}, who performed HI spectral stacking. {In the latter, a higher gas fraction is expected, as it is effectively a mean HI-to-stellar mass ratio in real space 
rather than the median on the logarithmic space.}
We find that \shark\ predicts gas fractions in reasonable agreement with 
the observations. {\shark\ may be producing slightly too much neutral gas at fixed stellar mass, but still comfortably within the scatter and uncertainties in the 
observations.}
In the future we will compare the HI-to-stellar mass ratio distributions in different stellar mass bins, as this may be able to offer new constraints to models \citep{Lemonias13}.

In the right panels of Fig.~\ref{Scalings2} we show the same gas scaling relations but separating late- (LTGs) and early-type (ETGs) galaxies. We define these in two 
ways, based on the bulge-to-total stellar plus gas mass ratio, $(\rm B/T)_{\rm bar}$, and on the bulge-to-total stellar mass ratio, $(\rm B/T)_{\star}$, where 
ratios above (below) $0.5$ correspond to ETGs (LTGs). LTGs are characterised by much higher gas fractions at fixed stellar mass than ETGs, with those differences being larger for H$_2$ than for HI. 
The latter is due to the star formation law assumed for star formation in bulges being dependent on the surface density of the gas. As the latter decreases, 
the conversion of HI into H$_2$ becomes less efficient (due to lower pressure), and thus less stars are formed, depleting the HI gas in longer timescales. Thus, HI can be preserved for longer 
compared to H$_2$. 
We also show observational inferences of \citet{Calette18} for the gas fractions for LTGs and ETGs, who analysed in a self-consistent way a large compilation of HI and H$_2$ observations in the local Universe. 
We find that the observations suggest that the difference between LTGs and ETGs is larger than that obtained in \shark at $M_{\star}\gtrsim 10^{10}\rm\,M_{\odot}$, regardless of the 
way we select ETGs/LTGs in the model,  
which may be connected with the fact that we do not yet include 
in \shark\ environmental processes that strip the ISM of galaxies. 
Those will be included in future versions of \shark, and this issue will be explored in detail.
At $M_{\star}\lesssim 10^{10}\rm\,M_{\odot}$, we see a reversal of the relation in the case of the $(\rm B/T)_{\rm bar}$, due to the effect of gas-rich disks being assigned to the LTG population. 
Those are classify at ETGs if we instead use the $(\rm B/T)_{\star}$, 
as their stellar mass is dominated by the bulge, but their baryon mass (gas plus stars) is dominated by the disk, due to the presence of very gas-rich disks.

\subsection{Galaxy sizes and morphology}\label{sec:sizes:comp}

\begin{figure}
\begin{center}
\includegraphics[trim=4mm 10mm 10mm 27mm, clip,width=0.45\textwidth]{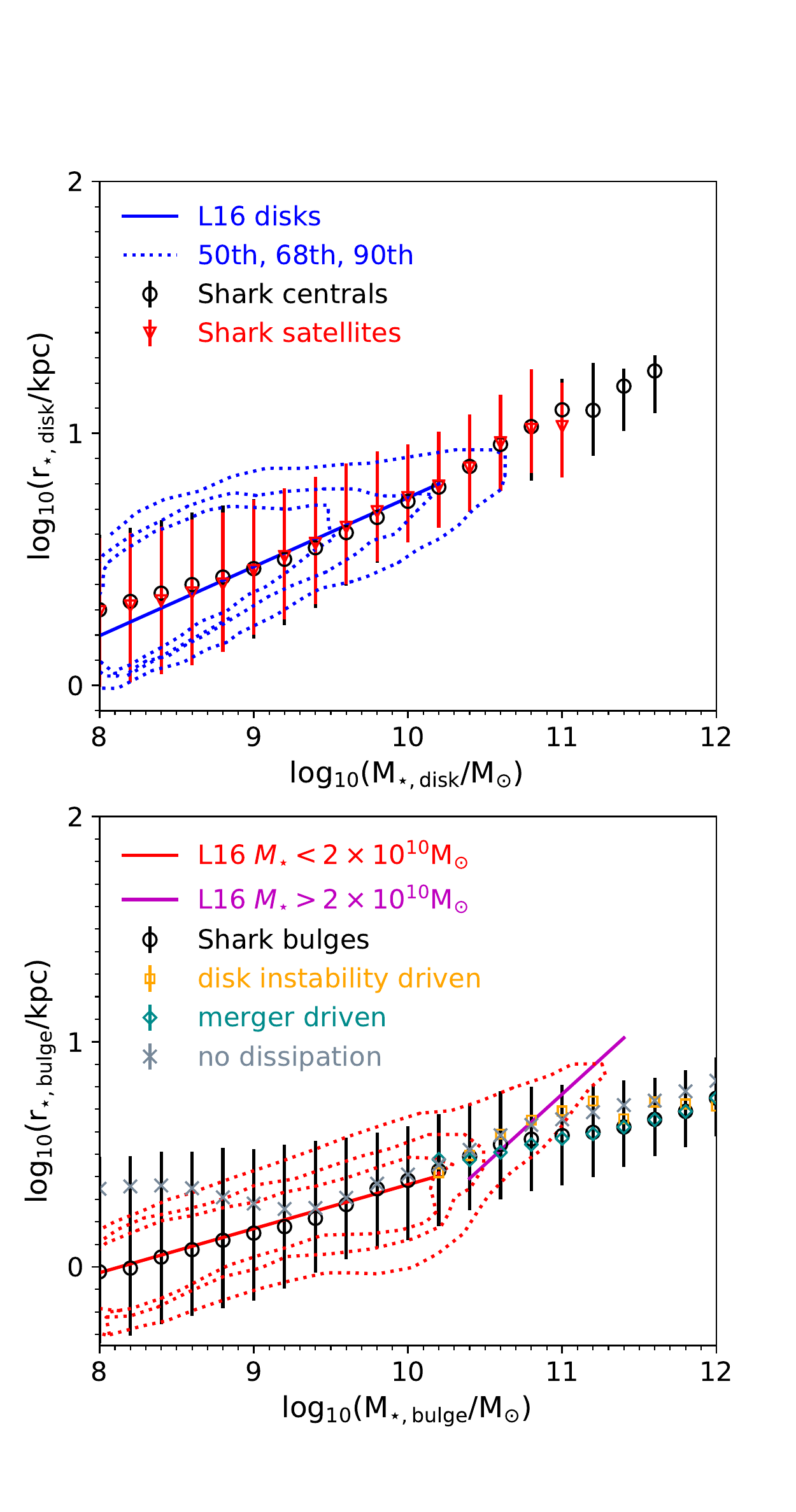}
\caption{Size-stellar mass relation for disks (top panel) and bulges (bottom panel) at $z=0$. Note that here 
we plot disk (bulge) half-stellar mass radii vs. disk (bulge) stellar mass. 
Here, we show all bins with $\ge 10$ objects.
The top panel shows only 
LTGs in \shark\ ($\rm (B/T)_{\star}<0.5$), separating into centrals and satellites. 
The bottom panel shows \shark\ ETGs ($\rm (B/T)_{\star}\ge 0.5$).
Symbols with errorbars show the medians and $16^{\rm th}-84^{\rm th}$ 
percentile ranges, respectively.
In the bottom panel we also show the relation for bulges with stellar masses $>10^{10}\,\rm M_{\odot}$ that mostly grew by galaxy mergers (diamonds) or 
by disk instabilities (squares). We also show a model variant that assumes no gas dissipation during mergers (see Eq.~\ref{size:merger2}; crosses).
Solid and dotted lines show the best fit and {the $50^{\rm th}$, $68^{\rm th}$ and $90^{\rm th}$ percentile regions of} the GAMA observations of \citet{Lange16}.} 
\label{Scalings}
\end{center}
\end{figure}

Fig.~\ref{Scalings} shows the disk size-disk mass relation, and the equivalent for bulges. We compare with the observations of 
\citet{Lange16}, who performed profile light fitting to decompose their galaxies into disks and bulges. 
They then measured the sizes and the stellar masses  
of both components. 
Galaxy disks agree very nicely with the observations of \citet{Lange16}. Central galaxies in \shark\ tend to have more extended disks than 
satellites, due to the latter forming at earlier times, where DM halos had lower specific angular momentum. 
The region with $M_{\rm disk}\gtrsim 10^{11}\,\rm M_{\odot}$ is scarcely populated in observations. This is also the case 
in \shark\, as we find that of the $764,135$ galaxies with $M_{\rm disk}>10^8\,\rm M_{\odot}$ that have $\rm (B/T)_{\star}<0.5$, only 
$50$ have $M_{\rm disk}>10^{11}\,\rm M_{\odot}$.

\shark\ bulges have sizes that are {good} agreement with the observations, though they tend to be slightly too large 
at around bulge masses of $10^{10}\,\rm M_{\odot}$. We find that this is mostly due to the effect of galaxy mergers.
This can be seen from the squares and diamonds in the lower panel of Fig.~\ref{Scalings}, which show the sizes of 
bulges with stellar masses $>10^{10}\,\rm M_{\odot}$
that grew mostly due to disk instabilities and galaxy mergers, respectively. 
We only show massive bulges in these two cases as at lower masses bulges mostly form and 
grow due to galaxy mergers. 
Bulges formed via disk instabilities tend to be 
$\approx 0.15$~dex {\bf larger} than those produced by galaxy mergers 
around a bulge mass of {\bf $10^{11}\,\rm M_{\odot}$}. The latter is mostly due to the effect of gas dissipation during mergers 
that is included in our default \shark\ model becoming negligible 
 around that stellar mass (see description in $\S$~\ref{sec:mergers}). If we do not include the effects of dissipation, bulges 
become unrealistically large (see crosses in Fig.~\ref{Scalings}) below a bulge mass of $10^{10}\,\rm M_{\odot}$. 
Both mass ends are affected by gas-rich mergers, 
at later (early) times in the case of the low-(high-)mass end. Thus, we find that dissipation is a key process 
that has to be considered in order to reproduce realistic bulge sizes. This agrees with the conclusion of 
\citet{Zoldan18} using the GAEA SAM. 
Reproducing the bulge sizes of galaxies has been a long standing challenge for SAMs (e.g. \citealt{Lacey15}; \citealt{Zoldan18}), and thus 
we consider the agreement with the bulge sizes obtained by \shark\ an important success.

\begin{figure}
\begin{center}
\includegraphics[trim=5mm 3mm 10mm 12mm, clip,width=0.45\textwidth]{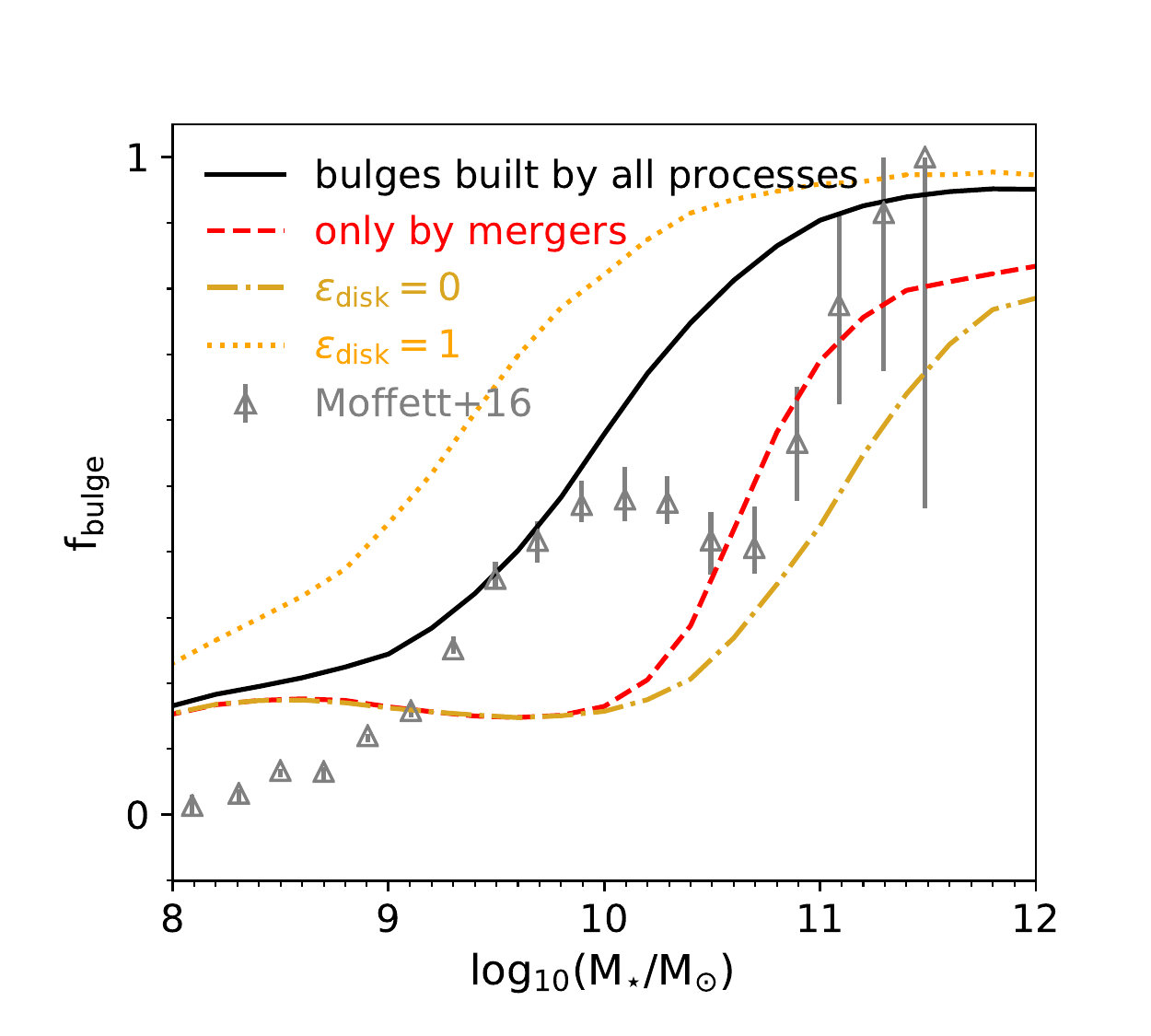}
\caption{The total fraction of stellar mass contributed by bulges as a function of stellar mass at $z=0$ in 
our default \shark\ model. 
Symbols with errorbars show the median and $1\sigma$ uncertainty in the observations 
of \citet{Moffett16}. We also show two model variants changing the threshold $\epsilon_{\rm disk}$ below which galaxies are considered unstable (see Eq.~\ref{sec:diskins}), as labelled.}
\label{Bulgefraction}
\end{center}
\end{figure}

Another important test for galaxy formation models, is whether they place the right amount of stellar mass into disks and bulges. 
\citet{Moffett16} measured the SMF separating galaxies into different morphological types and 
also into disks/bulges. With this, they derived the fractional contribution from bulges/disks to the total stellar mass in bins of stellar mass. 
This is quite a difficult measurement to do in observations as light profile fitting is required, which can be robustly done in very disk- and bulge-dominated galaxies, 
but when both components contribute similarly, the measurement is less robust \citep{Robotham17}. 
In Fig.~\ref{Bulgefraction}, we compare \shark\ with these measurements for two measurements of bulge mass\footnote{Bulges here include the elliptical galaxy population and hence 
$f_{\rm bulge}\rightarrow 1$ as $M_{\star}\rightarrow 10^{12}\,\rm M_{\odot}$.}. The first one is considering all the mass in the central 
concentration (regardless of whether it was formed due to mergers or disk instabilities; solid line), and the second one assumes that the bulge mass formed through disk instabilities 
(either through the starburst triggered by the gas being fueled to the centre or the stars that are transferred from the disk to the bulge) 
is part of the disk (dashed line). The bulge mass formed via mergers include both the stars formed via merger-driven starbursts and stars that were accreted by the bulge 
as a result of a merger (but that were formed in the disk of the primary and/or in the secondary galaxies).
The latter is done as pseudobulges in \citet{Moffett16} were added up to the disk rather than the bulge, and those are thought to form 
through secular processes taking place in the disks of galaxies \citep{Kormendy04}. The effect of assigning the bulge mass formed via disk instabilities to the disk has the effect 
of shifting the transition from disk- to bulge-dominated stellar budget to higher stellar masses, much closer to the \citet{Moffett16} observations.
In \shark, we find that the formation of elliptical galaxies (i.e. spheroid dominated, massive galaxies) is dominated by galaxy mergers rather than 
disk instabilities, as disk instabilities only increase the bulge contribution by $\approx 10$\% at $10^{11}\,\rm M_{\odot}$. 
This is qualitatively similar to the finding in large cosmological hydrodynamical simulations that elliptical galaxies form primarily via galaxy mergers \citep{Wellons16,Clauwens18,Lagos18}.

The transition from disk- to bulge-dominated stellar budget is very sensitive to the value of the threshold $\epsilon_{\rm disk}$ below which galaxies are considered unstable 
(see Eq.~\ref{sec:diskins}). This is seen in the model variants shown in Fig.~\ref{Bulgefraction}, adopting $\epsilon_{\rm disk}=0$ (dot-dashed line) and 
{\bf $\epsilon_{\rm disk}=1$} (dotted line). When no disk instabilities take place, the transition to bulge-dominated stellar budget moves by $\approx 0.5$~dex to higher stellar masses, 
while adopting a much larger $\epsilon_{\rm disk}$ moves the transition by $\approx -0.5$~dex. We find that values of $\epsilon_{\rm disk} \lesssim 0.4$, lead to the bulge growth to be dominated by 
galaxy mergers, while the opposite is true for larger values of $\epsilon_{\rm disk}$. 
We find that in \shark\ $f_{\rm bulge}$ tends to {\bf $0.15$} as stellar mass decreases. This is due to gas-rich mergers taking 
place in dwarf galaxies. We find that centrals tend to have a lower incidence of bulge-dominated galaxies than 
satellites at $M_{\star}\lesssim 10^{10.8}\,\rm M_{\odot}$, at which point the trend reverses (not shown here). 
 
\subsection{The BH population}\label{sec:BH}

\begin{figure}
\begin{center}
\includegraphics[trim=5mm 3mm 10mm 12mm, clip,width=0.45\textwidth]{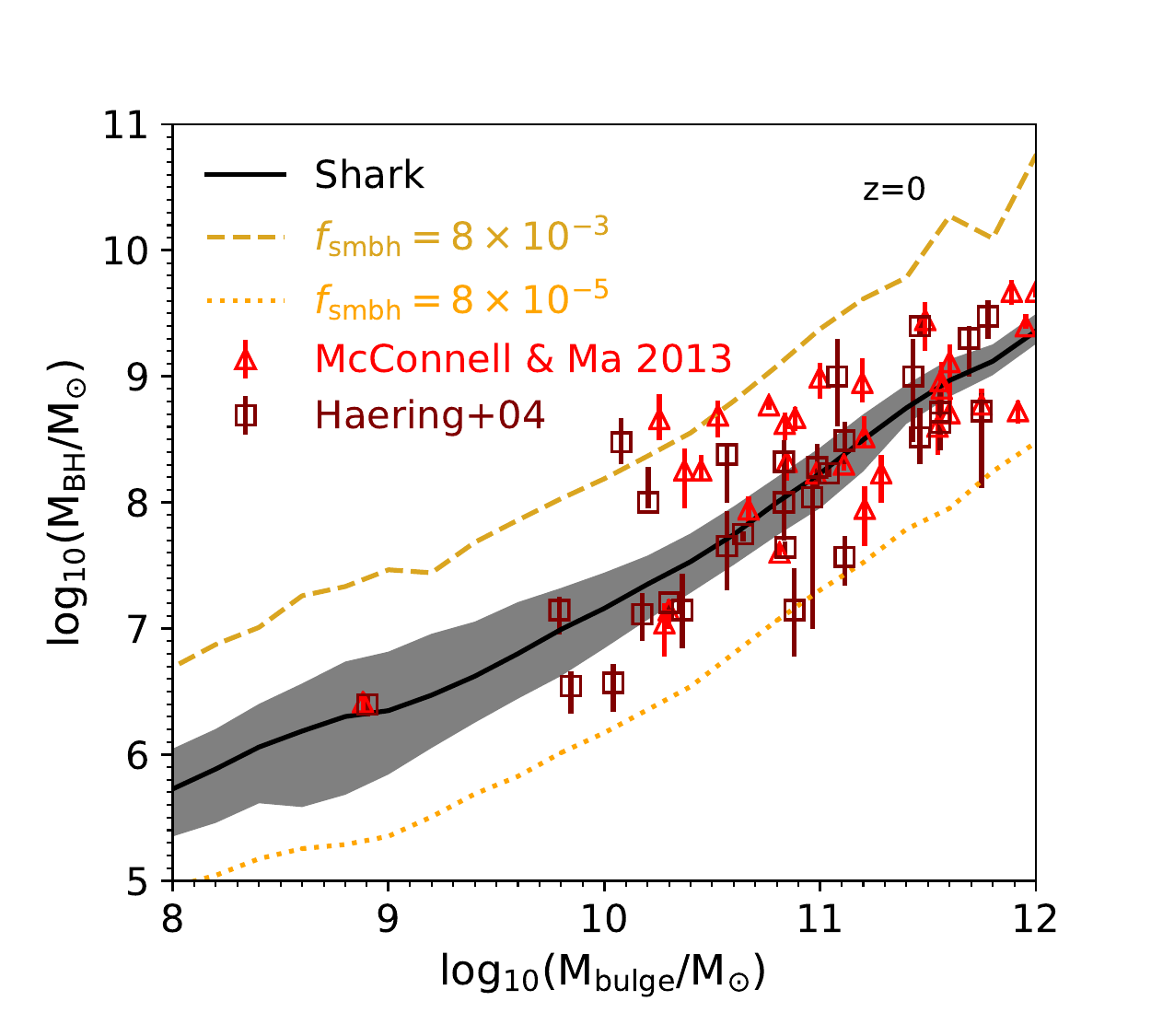}
\caption{The BH-bulge mass relation at $z=0$. The solid line and shaded region correspond to the median and the $16^{\rm th}-84^{\rm th}$ 
percentile ranges for our default \shark\ model. 
Here, we show all bins with $\ge 10$ objects.
Triangles with errorbars show the local observations of \citet{Haering04} and \citet{McConnell13}, as labelled.
We also show the medians of two model variations that change the efficiency of the gas inflow towards the SMBH during starbursts by $\times 10$ ($f_{\rm smbh}=8\times 10^{-3}$) and $:10$ ($f_{\rm smbh}=8\times 10^{-5}$), 
as labelled.}
\label{ScalingsBH}
\end{center}
\end{figure}

Like most SAMs and hydrodynamical simulations, \shark\ uses the BH-bulge mass relation to tune its free parameters (e.g. \citealt{Schaye14}; \citealt{Lacey15}). 
In \shark, this relation is pretty much controlled by a single parameter, and that is the efficiency at which gas flows towards the central 
BH during starbursts ($f_{\rm smbh}$ in Eq.~\ref{bhgrow_sbs}). We show in Fig.~\ref{ScalingsBH} the BH-bulge mass relation at $z=0$ for our default \shark\ model, and two variants 
adopting a $f_{\rm smbh}$ $10$ times higher/lower than our optimal value. These changes almost linearly translate into the same change in the normalisation of the BH-bulge relation (see 
dashed and dotted lines in Fig.~\ref{ScalingsBH}). We also show in Fig.~\ref{ScalingsBH} the observations of \citet{Haering04} and \citet{McConnell13}, which were used 
as reference to obtain an optimal $f_{\rm smbh}$. In the future we will explore the QSO luminosity functions as they offer independent tests to assess how realistic the BH population 
in \shark\ is.  

\subsection{The Main sequence and mass-metallicity relation}

\begin{figure}
\begin{center}
\includegraphics[trim=3mm 4mm 10mm 13mm, clip,width=0.45\textwidth]{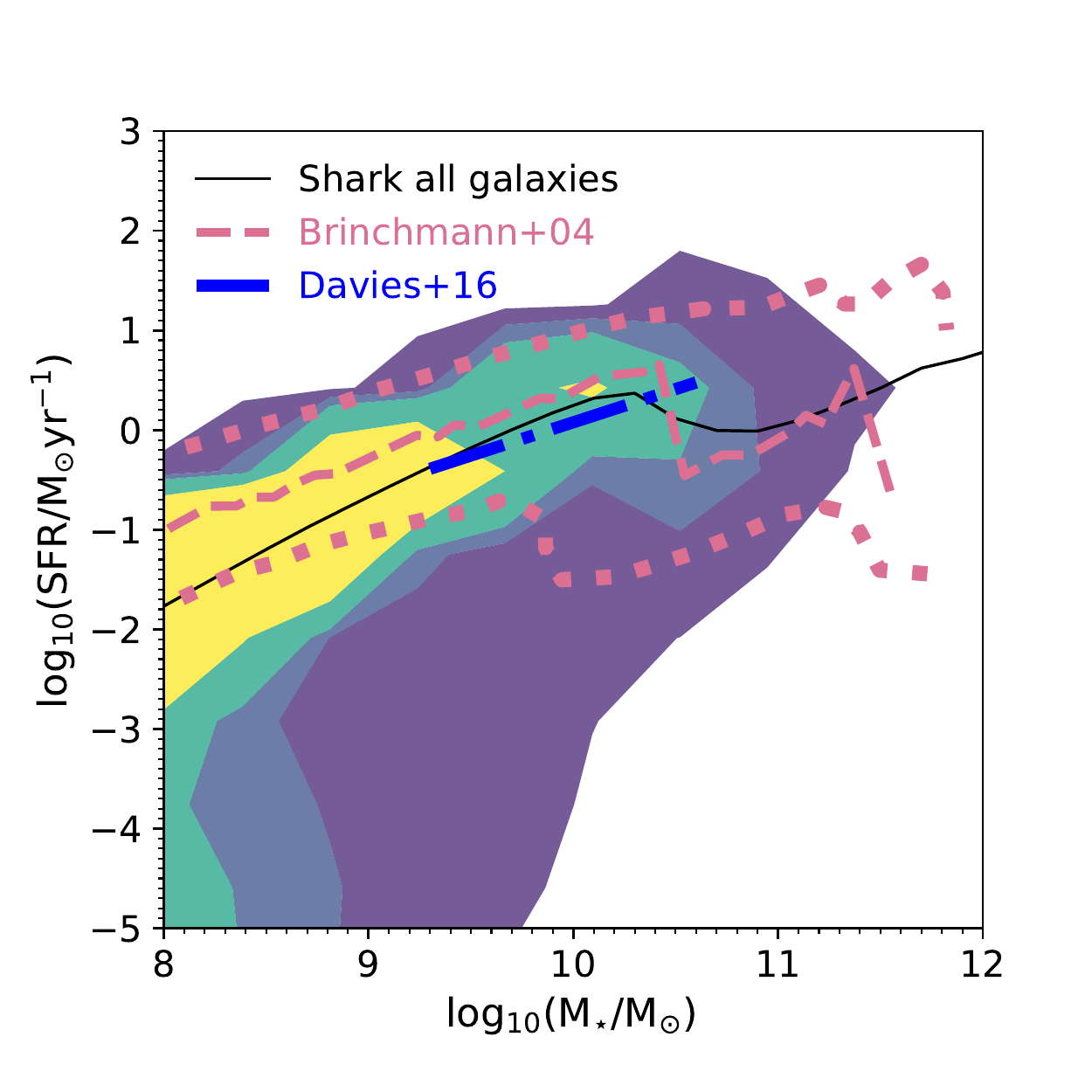}
\caption{SFR vs. stellar mass at $z=0$
for our default \shark\ model.
Here, we include 
 all the galaxies that have an $\rm SFR/M_{\star} > 10^{-3}\,\rm Gyr^{-1}$ and show 
the median only for bins with  $\ge 10$ objects.
Contours show percentiles ranges ranging from $99^{\rm th}$ to $10^{\rm th}$, from the 
outer to the inner regions. The dashed line shows the median of all galaxies.
The dashed and thick dotted lines show the median and the region of $0.02$ conditional likelihood 
of SFR given a stellar mass from  \citet{Brinchmann04}. 
The dot-dashed line shows the best fit of the GAMA main sequence reported by \citet{Davies16}.}
\label{SSFRM}
\end{center}
\end{figure}

\begin{figure}
\begin{center}
\includegraphics[trim=1mm 8mm 4mm 26.1mm, clip,width=0.45\textwidth]{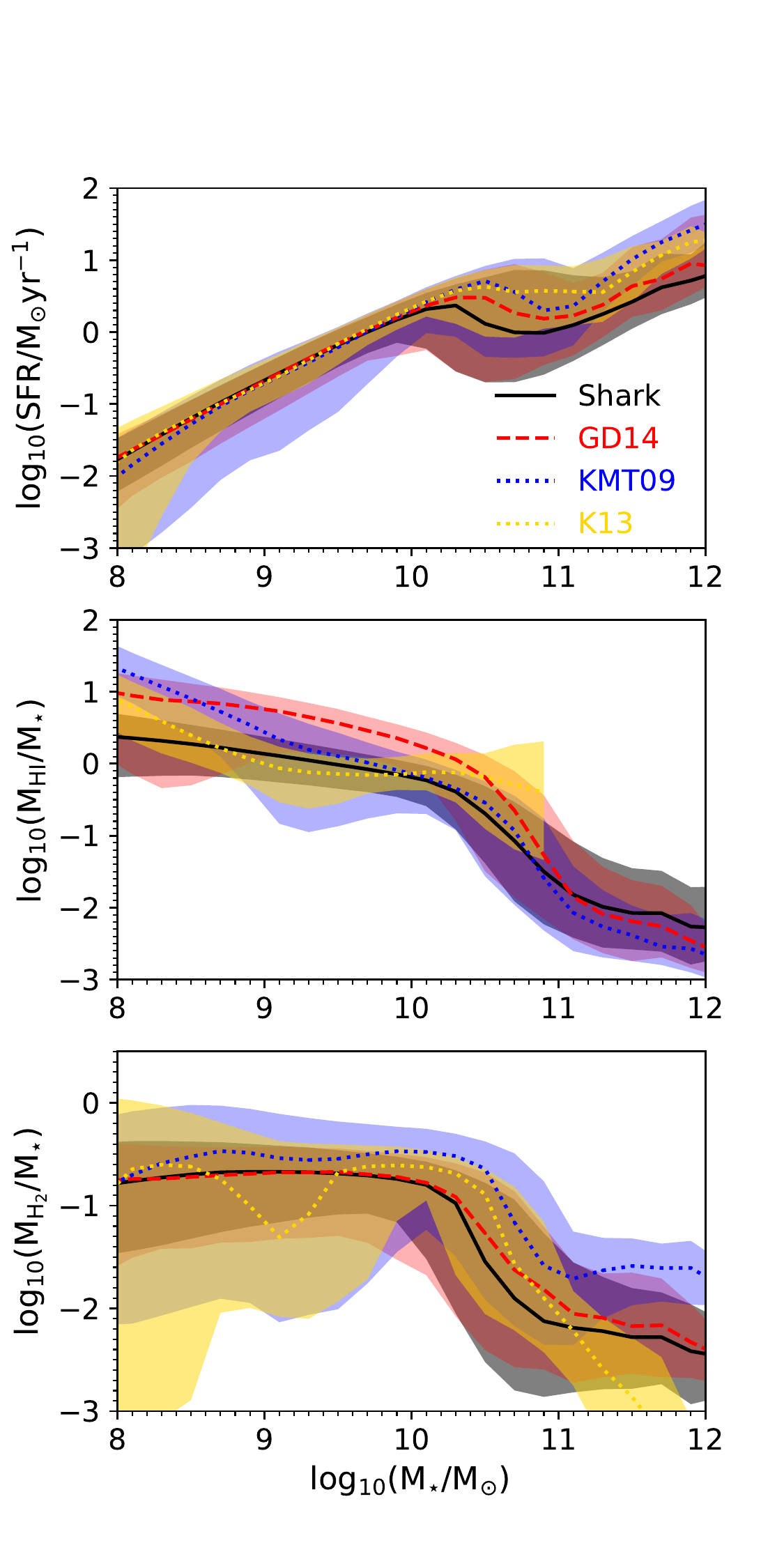}
\caption{{\it Top panel:} main sequence in the SFR vs. stellar mass at $z=0$, calculated 
with all the galaxies that have an $\rm SFR/M_{\star} > 10^{-3}\,\rm Gyr^{-1}$,   
for our default \shark\ model and 3 variants adopting the GD14, KMT09 and K13 star formation laws (see $\S$~\ref{sec:sf} for details). 
Lines show the median while shaded regions show the $16^{\rm th}$ and $84^{\rm th}$ percentiles. 
{\it Bottom panels:} the HI- (middle) and H$_2$-to-stellar mass ratio (bottom) as a function of stellar mass at $z=0$ for the same models as in the top panels. }
\label{SSFRMscatter}
\end{center}
\end{figure}

\begin{figure}
\begin{center}
\includegraphics[trim=0mm 18mm 5mm 23mm, clip,width=0.43\textwidth]{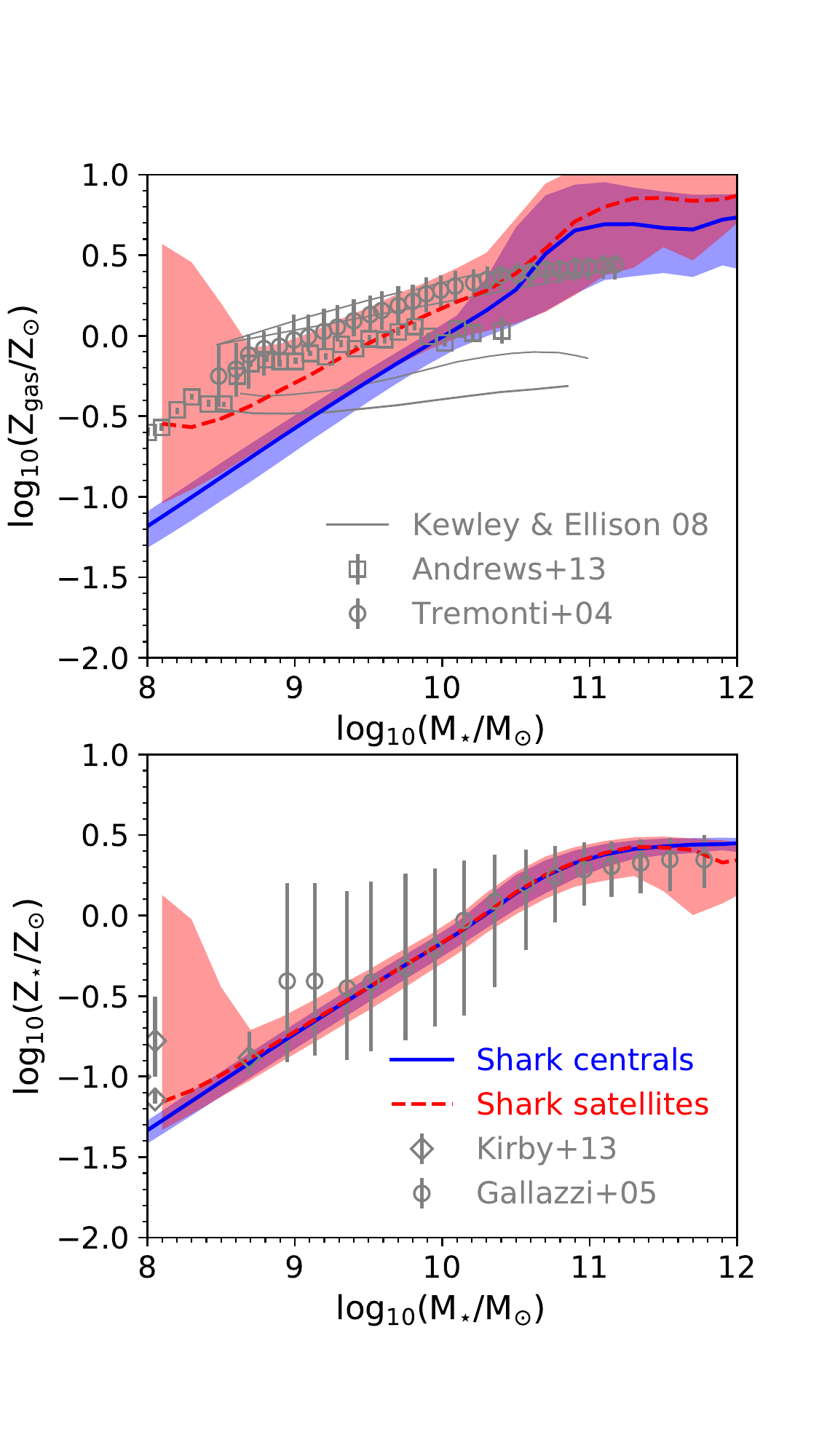}
\caption{{\it Top panel:} Gas metallicity vs. stellar mass as $z=0$, for our default \shark\ model.
Solid and dashed lines show the median of centrals and satellites, respectively. 
shaded regions show the $16^{\rm th}-84^{\rm th}$
percentile ranges. 
Here, we show all bins with $\ge 10$ objects.
Symbols show the observations of \citet{Tremonti04} and \citet{Andrews13}, as labelled, while lines 
show $4$ measurements of \citet{Kewley08}, which were produced using different methods to derive metallicities. 
The $4$ relations correspond to those giving the lowest/highest normalisation and provide a measurement of the systematic uncertainty.
{\it Bottom panel:} As in the top panel but for the stellar metallicity. Here we show observational inferences from 
\citet{Gallazzi05} and \citet{Kirby13}, as labelled. } 
\label{MZR}
\end{center}
\end{figure}

The self-regulation of galaxies, i.e. the idea of SF regulating to give an outflow rate that compensates the accretion rate, has been 
a key result of the last decade (e.g. \citealt{Schaye10, Hopkins12b, Dave12, Lagos14}). The observable consequence of this self-regulation is the formation 
of a {\it tight} relation between SFR and stellar mass, termed `main sequence of star formation'. The tightness of this relation 
has been shown to depend on how quickly galaxies self-regulate (e.g. \citealt{Lagos16a}).
In \shark\ we do not tune to get this main sequence and thus 
exploring its existence is a physical test for the model and how well it recovers the existence of self-regulation.

Fig.~\ref{SSFRM} shows the SFR-M$_{\rm star}$ plane at $z=0$ for all galaxies in our default \shark\ model together with the main sequence as reported in 
\citet{Brinchmann04} and \citet{Davies16}.
\shark\ produces a tight main sequence, with a $1\sigma$ scatter of $\approx 0.32$~dex, but the scatter increases towards both the low- and high-mass ends. The latter 
has also been observed locally by \citet{Willett15} and \citet{GuoSF15}. The \shark\ main sequence is slightly too steep compared to \citet{Brinchmann04}, 
but this may not necessarily be a worry as in SDSS the minimum SFR that would be detectable in a galaxy of $M_{\star}\approx 10^{8}\,\rm M_{\odot}$ would be $3\times 10^{-3}\,\rm M_{\odot}\,yr^{-1}$ 
(Brinchmann, priv. com.), while increasing to $10^{-2}\,\rm M_{\odot}\,yr^{-1}$ at $M_{\star}\approx 10^{10}\,\rm M_{\odot}$. This selects out an important part of the parameter 
space at low stellar masses in \shark. \shark\ also predicts a break in the main sequence at $M_{\star}\approx 10^{10.4}\,\rm M_{\odot}$, very similar to the break 
in \citet{Brinchmann04} at $M_{\star}\approx 10^{10.55}\,\rm M_{\odot}$. In \shark, galaxies with stellar masses above the break continue to form stars but at a lower rate compared to the main 
sequence, which is also hinted by the observations, though we caution that the latter is uncertain due to the confusion with low ionisation emitters at high stellar masses.

We find that there is a lot of information in the scatter of the main sequence and the gas fraction relations. This is shown in Fig.~\ref{SSFRMscatter} for our default \shark\ model 
and 3 variants adopting other star formation laws (though all are molecular gas based; see $\S$~\ref{sec:sf} for details). We find that the position of the main sequence 
(lines in the top panels of Fig.~\ref{SSFRMscatter}) is barely affected by the choice of star formation law, as expected from the existence of self-regulation. However, we find that 
the scatter around the median is quite sensitive to the choice of star formation law. At $10^{10}\,\rm M_{\odot}$, where the scatter is tightest in all the star formation laws shown 
here, we see that the K13 variant produces the tightest relation with a $1\sigma$ scatter of $0.3$~dex, while the GD14 variant produces a $1\sigma$ scatter of $0.75$~dex. Differences 
in the predicted scatter increase towards low and high masses. At $10^{9}\,\rm M_{\odot}$, the extremes are $0.55$~dex and $1.2$~dex in the K13 and KMT09 variants, respectively. 
This is also seen in the HI- and H$_2$-to-stellar mass ratio vs. stellar mass relations (shown in the middle and bottom panels of  Fig.~\ref{SSFRMscatter}), as different 
star formation laws result in very different scatter. Interestingly, the models that produce the tightest HI relation are not the same as those that produce the tightest 
H$_2$ relation. For example, in HI the tightest are our default \shark\ model, which assumes the BR06 star formation law, and the K13 variant, while in H$_2$, the BR06 and GD14 
variants produce the tightest relations. 
This shows that by studying the scatter of these relations in detail we can learn about the interplay between star formation and gas.

Metallicities were neither explored during the tuning process and thus they represent independent tests to our model. 
Fig.~\ref{MZR} shows the ISM and stellar metallicities as a function of stellar mass at $z=0$ for our default \shark\ model and 
for the observations of \citet{Tremonti04},  \citet{Kewley08}, \citet{Andrews13}, \citet{Gallazzi05} and \citet{Kirby13}.
The \citet{Kewley08} results were obtained using different metallicity diagnostics and thus provide a measurement of the 
systematic uncertainties in the measurement of gas metallicities. 
We show centrals and satellites separately. Satellites tend to have gas metallicities that are $0.3-0.5$~dex higher than centrals at fixed stellar mass. 
We find that \shark\ tends to produce a mass-gas metallicity relation for central galaxies that is too steep, resulting 
in metallicities that are too low in galaxies with $M_{\star}\lesssim 10^{9.5}\,\rm M_{\odot}$, while satellites follow the observed relation quite well. 

Stellar metallicities of \shark\ galaxies are very tightly correlated with stellar mass, and no much difference is seen between centrals and satellites.
Though there is a tendency for \shark\ galaxies to have slightly lower stellar metallicities than indicated by the observations, such offset is well within the uncertainties 
as shown by the errorbars in the observations. There is a tight relation between the details of the stellar feedback modelling and the mass-metallicity relation 
(e.g. \citealt{Xie17} and Collaccchioni et al. submitted), and those 
in \shark\ will be investigated in detail in the future. Note that one important improvement for future \shark\ versions is the inclusion of a non-instantaneous recycling approximation for the enrichment 
of gas, which can significantly modify the enrichment of galaxies \citep{Cora06,Yates13}.

\section{Discussion and conclusions}\label{conclusions}

We have introduced a new, open source, free and flexible SAM, \shark. We have presented the release of {\sc v1.1}, which includes several models for key physical processes:
gas cooling, stellar and AGN feedback, star formation and photo-ionisation feedback. \shark\ includes, in addition, at least one model for all the other important 
physical processes in galaxies that are required to obtain realistic galaxy populations, such as environmental effects, chemical enrichment, galaxy mergers and disk instabilities. 
Including these processes we can converge into an default \shark\ model that is able to reproduce a large set of observations beyond those that we used 
as primary constraints for the tuning of parameters. Below we summarise our main results.

\begin{itemize}
\item Our primary constraints to tune the free parameters are the $z=0,\,1,\,2$ SMFs, the $z=0$
the black hole-bulge mass relation and the disk (bulge) half-stellar mass size-disk (bulge) stellar mass relations. 
With the current set of physical processes included in \shark\ we are able to reproduce these observables well.
We find that the flatness and normalization of the SMF at low and high redshift is obtained due to a combination of factors including a weak redshift dependence 
of the stellar feedback efficiency, a strong dependence of the latter on the circular velocity of the galaxy and a slow reincorporation timescale of the gas ejected outside 
the halo. The high-mass end is almost solely controlled by the efficiency of AGN feedback, though in a non-linear fashion. 
The BH-bulge relation is also controlled mostly by one single parameter, which is the efficiency of gas inflowing towards BHs during starbursts. 
The latter almost linearly translates into changes in the normalization of the BH-bulge relation.
\item \shark\ obtains a very tight relation between the stellar and DM halo mass of central galaxies, with the shape being in good agreement with empirical estimates.
\item We find that our default \shark\ model reproduces reasonably well the cosmic SFR, stellar mass density and atomic/molecular hydrogen evolution.
We find that in order to get the cosmic SFR to agree with observations at $z>4$, the H$_2$ SF efficiency in starbursts needs to be $10\times$ higher than 
star formation in disks. The latter agrees with observational inferences \citep{Daddi10,Tacconi17}. Some tension arises between the cosmic SFR and the H$_2$ density, as a lower efficiency 
of H$_2$ to stars conversion allows \shark\ to reproduce better the peak of its density, at the cost of worsening the agreement with the observed cosmic SFR. 
\item \shark\ is able to reproduce out of the box the MFs of atomic and molecular hydrogen, as well as their scaling relations with 
stellar mass. We find hints that the gas content of early-type galaxies in \shark\ may be slightly too high, which is a topic that we will investigate 
in detail in the future. We also find a reasonable agreement with the gas- and stellar-metallicity vs. stellar mass relations. 
\item We find that the inclusion of the gas dissipation effect during gas-rich major mergers is key to obtain a reasonable size-mass relation for bulges.
\shark\ out of the box also reproduces quite well the fraction of stellar mass in bulges as a function of stellar mass, with a transition from disk- to bulge-dominated 
stellar budget at $\approx 10^{10.3}\,\rm M_{\odot}$ if we count bulges formed via disk instabilities, or 
at $\approx 10^{10.7}\,\rm M_{\odot}$ is we only consider bulges formed via galaxy mergers. This transition is extremely sensitive to the assumed threshold for global disk instabilities. 
\item \shark\ is able to reproduce quite well the main sequence in the SFR-stellar mass plane. We find that the scatter of the main sequence and the gas scaling relations is 
very sensitive to the assumed star formation law, suggesting that by exploring the scatter in observations we may be able to find powerful new constraints on the 
physics included in galaxy formation models.
\item We compare the baryon budget growth of \shark\ with that of the GALFORM and L-galaxies semi-analytic models in the \citet{Mitchell18} and \citet{Henriques15} variants, respectively, 
 and with the EAGLE hydrodynamical simulations \citep{Schaye14}. 
The stellar mass growth is remarkably similar between these models. 
However, the largest difference is seen in the 
amount of baryons that are locked up in halos. 
All the SAMs have a halo gas component dominating the baryon budget throughout most, and in some cases all, of the history of the universe. However, EAGLE 
has most of the baryons outside halos (i.e. the difference between the amount of baryons inside the halo and the universal baryon fraction). 
In other words, halos in EAGLE have baryon fractions that are substantially below the universal baryon fraction, while in SAMs 
they are close to that universal fraction. This means that the gas accretion rate onto halos is less efficient, or alternatively that the 
 ejection rate from halos is more efficient, in EAGLE than in SAMs. 
This abysmal difference responds to the lack of observational constraints on the abundance of gas in halos.
Smaller, but still very significant differences are seen in the abundance of ISM gas. This makes the next generation of deep HI and absorption line 
surveys, as well as blind molecular gas surveys, incredibly important, as they will be able to provide key constraints in a pretty much unexplored parameter space.
\end{itemize}

We have demonstrated the power of \shark\ as a tool designed to systematically explore the effects of different physical processes and model variants of any one of them.
We expect \shark\ to be an ever evolving community tool that feeds from new theoretical and observational developments, as well as using it 
to explore new ways of modelling key physical processes in galaxy formation. 
In the near future, we will provide public tools to create 
full spectral energy distributions from the star formation histories of \shark\ galaxies and provide several large galaxy surveys with lightcones specially designed to 
address the inherent limitations of any astronomical observational experiment.

\section*{Acknowledgements}

We thank Cedric Lacey, Carlton Baugh, Simon Driver, Luke Davies, Violeta Gonz\'alez-P\'erez, Barbara Catinella and Matthieu Schaller for fruitful discussions and suggestions.
We also thank Andrew Benson for a very constructive referee report that helped improve the clarity of this manuscript.
We also thank the theory and computational 
group at ICRAR for the useful discussions during group meetings, and 
Mark Boulton for the help using the local cluster Pleiades.
We thank Rachel Somerville, Andrew Benson, Gabriella De Lucia, Sof\'ia Cora and Cristi\'an Vega for providing us with some technical details 
on the Santa-Cruz, Galacticus, GAEA and SAG models.
CL has received funding from a Discovery Early Career Researcher Award (DE150100618) and by the ARC Centre of 
Excellence for All Sky Astrophysics in 3 Dimensions (ASTRO 3D), through project number CE170100013.
CL also thanks the MERAC Foundation for a Postdoctoral Research Award.
CL thanks CRAL-Lyon Observatory for a 2018 visiting professorship grant, which allowed close collaborations with 
Dr. Peter Mitchell, and helped the development of this manuscript.
This work was supported by resources provided by The Pawsey Supercomputing Centre with funding from the 
Australian Government and the Government of Western Australia.

\bibliographystyle{mn2e_trunc8}
\bibliography{SHArkIntro}

\label{lastpage}
\appendix

\section[]{Convergence test}\label{ConvTests}

\begin{table}
	\setlength\tabcolsep{2pt}
	\centering\footnotesize
	\caption{SURFS simulation parameters.}
	\begin{tabular}{@{\extracolsep{\fill}}l|cccc|p{0.45\textwidth}}
		\hline
		\hline
	    Name & Box size & Number of & Particle Mass & Softening Length\\
        & $L_{\rm box}$ [$\rm cMpc/h$] & Particles $N_p$ & $m_p$ [$\rm M_{\odot}/h$] & $\epsilon$ [$\rm ckpc/h$] \\
	\hline
    L40N512     & $40$  & $512^3$   & $4.13\times10^7$ & 2.6  \\
    L210N512    & $210$ & $512^3$   & $5.97\times10^9$ & 13.7 \\
    L210N1024   & $210$ & $1024^3$  & $7.47\times10^8$ & 6.8  \\
    L210N1536   & $210$ & $1536^3$  & $2.21\times10^8$ & 4.5  \\
	\end{tabular}
	\label{tab:sims}
\end{table}

\begin{figure}
\begin{center}
\includegraphics[trim=2mm 10mm 2mm 22mm, clip,clip,width=0.42\textwidth]{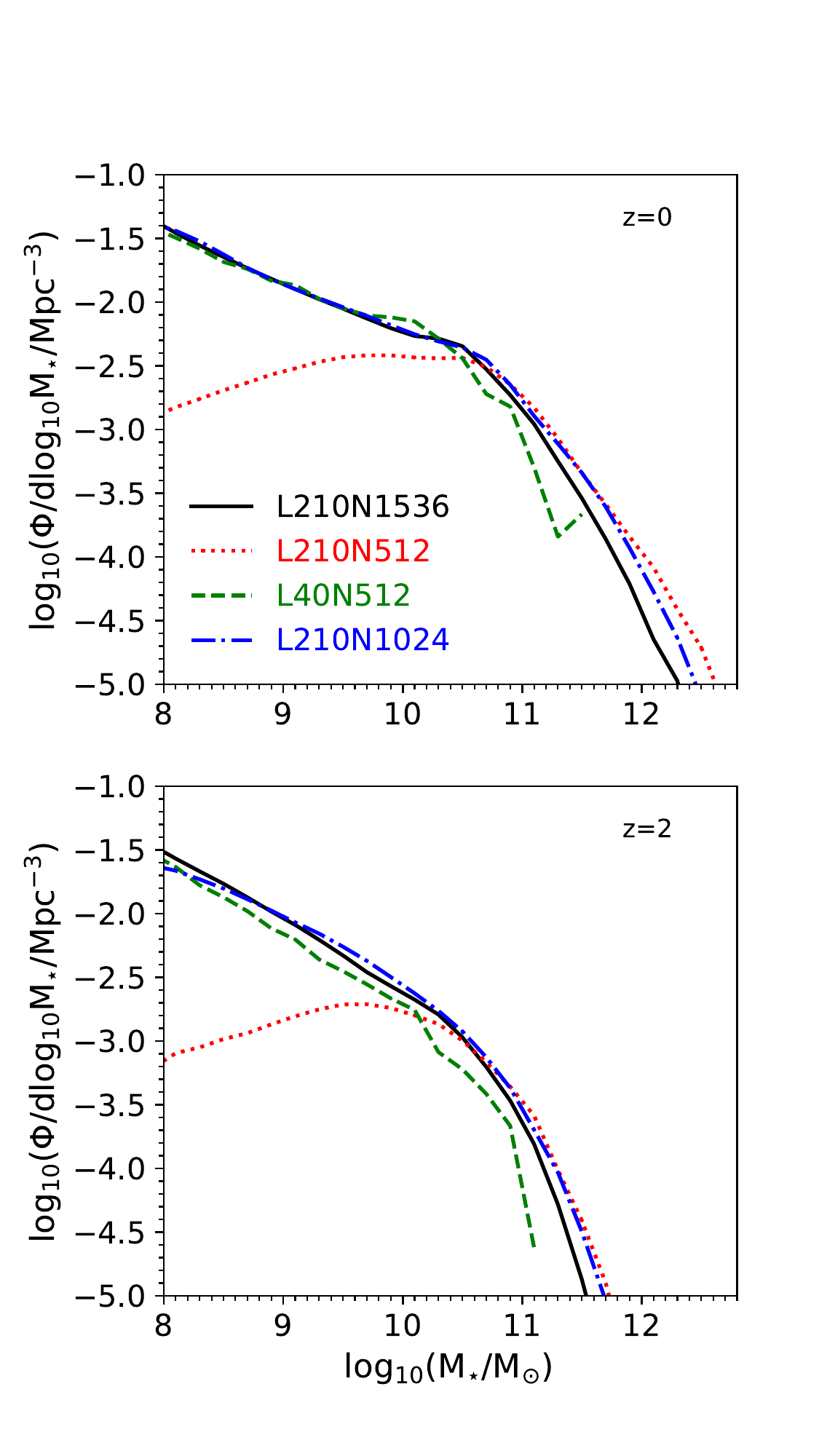}
\caption{SMF at $z=0$ and $z=0$ for our default \shark\ model, run on the $4$ simulations of Table~\ref{tab:sims}, as labelled.}
\label{SMF_resolution}
\end{center}
\end{figure}

An advantage of the SURFS simulations suite, is the possibility of testing the effect of resolution and simulated volume 
on galaxy properties. Table~\ref{tab:sims} shows the simulation parameters of all the SURFS run that were presented 
in \citet{Elahi18}. The parameters of our default \shark\ model (see values in Table~\ref{tab:parameters}) were 
chosen using the L210N1536 simulation. We fixed those parameters, and run the model on the other three 
simulations of Table~\ref{tab:sims} to assess the convergence of our results. 

Fig.~\ref{SMF_resolution} shows the SMF at two redshifts for our default \shark\ model run on all the SURFS simulations. 
The low-mass end of the SMF at $z=0$ is well converged at the resolution level of the L210N1024 simulation. However, that 
is not the case at $z=2$, as the L210N1024 produces a flatter low-mass end compared to both the L210N1536 and L40N512 
simulations. There is also a difference between L210N1536 and L40N512 in this regime, which may indicate that the 
L210N1536 is not well converged. 
At the massive end, we see that the poorer the resolution, the larger the stellar mass at fixed number density. 
This is due to the merger history of massive halos not being well converged. This affects massive galaxies more prominently 
as those are the ones whose growth is dominated by mergers \citep{Robotham14}.

\begin{figure}
\begin{center}
\includegraphics[trim=3mm 3mm 5mm 12mm, clip,width=0.42\textwidth]{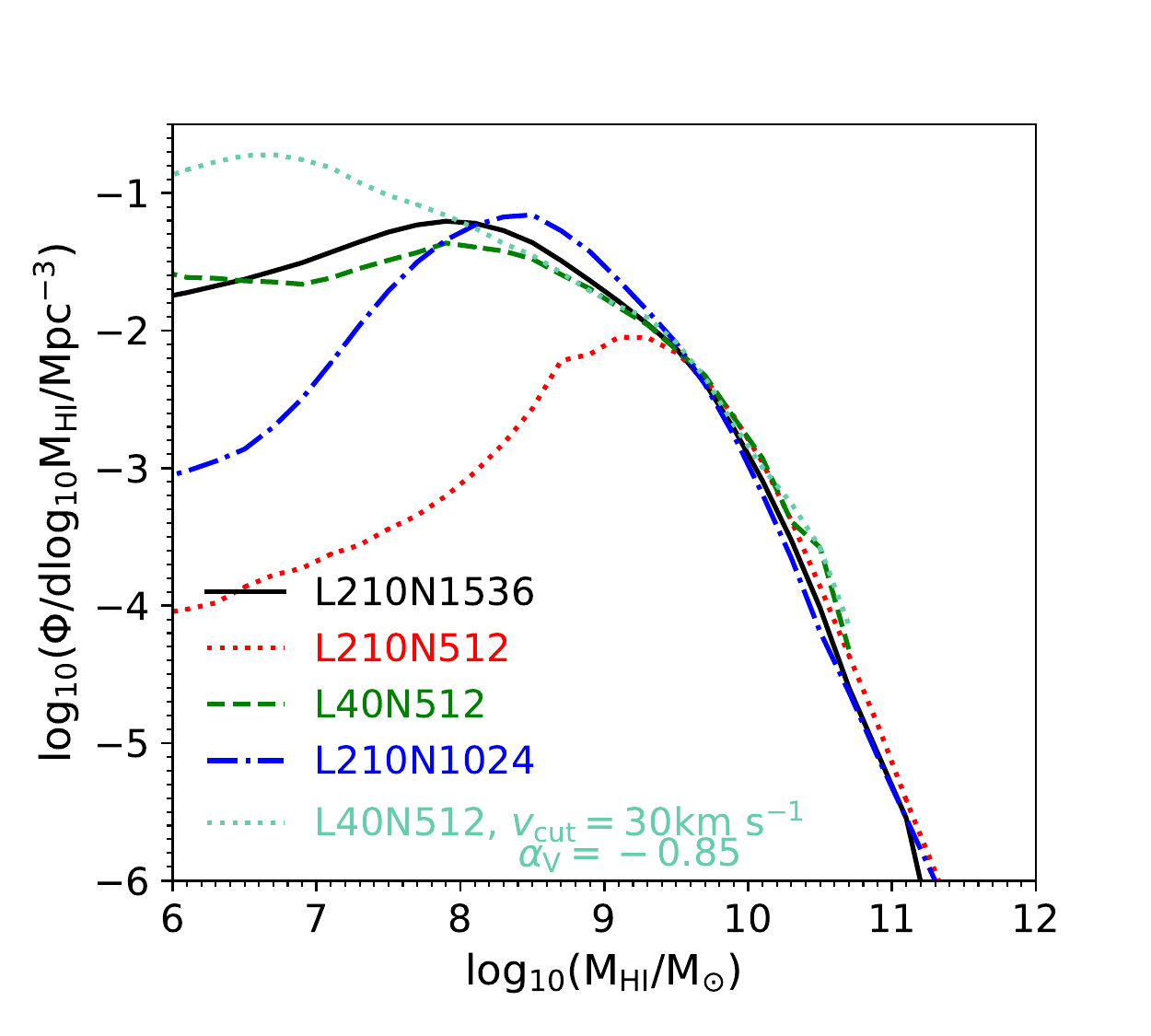}
\caption{HI MF at $z=0$ for our default \shark\ model, run on the $4$ simulations of Table~\ref{tab:sims}, as labelled.}
\label{HI_resolution}
\end{center}
\end{figure}

Fig.~\ref{HI_resolution} shows the HI Mf at $z=0$ for our default \shark\ model run on all the SURFS simulations. The effect of resolution 
at the low-mass end here is more dramatic than for the SMF. Our default L210N1536 has an HI MF converged only at $M_{\rm HI} \gtrsim 10^{8}\,\rm M_{\odot}$ 
when we compare it with the higher resolution, smaller volume L40N512. At the high-mass end we see much better convergence than 
what we obtain for the SMF. This is because the galaxies with $M_{\rm HI} \gtrsim 10^{10}\,\rm M_{\odot}$ tend to be star-forming 
galaxies with stellar masses a few times $M_{\star}\sim 10^{10}\,\rm M_{\odot}$. Those live in halos of masses 
$10^{12}\rm \, M_{\odot}$, which are well converged at the resolution of all the SURFS runs.

The drop seen in the L40N512 between $10^{7}\,\rm M_{\odot} \lesssim M_{\rm HI} \lesssim  10^{8}\,\rm M_{\odot}$ is not due to resolution,
but to the modelling of photo-ionisation feedback. To demonstrate that, we show as dotted, acqua-marine line a model variation adopting  
parameter values similar to those of \citet{Kim15} for the \citet{Sobacchi13} model. The result is that the drop disappears, and instead we see a bump at 
at $M_{\rm HI} \approx  10^{7.7}\,\rm M_{\odot}$. Thus, the shape of the HI MF at the low-mass end is the result of a complex interplay between 
photo-ionisation feedback, the star formation law (via its effect on the amount of HI in satellite galaxies) and the resolution of the 
simulation.

\end{document}